\documentclass[11pt]{article}

\usepackage[T1]{fontenc}
\usepackage[cp1251]{inputenc}
\usepackage{textcomp}
\usepackage[english]{babel}
\usepackage[centertags]{amsmath}
\usepackage[mediummath]{nccmath}
\usepackage{amsfonts}
\usepackage{amssymb}
\usepackage[hypertex]{hyperref}
\usepackage{graphicx}
\usepackage{graphbox}
\usepackage[numbers,sort&compress]{natbib}

\usepackage{paperinitial}

\paperinitialization{15mm}{15mm}{15mm}{15mm}{2pt}{10pt}

\DeclareMathOperator{\Texp}{Texp}
\DeclareMathOperator{\Sp}{Sp}

\newcommand{\lan}{\langle}
\newcommand{\ran}{\rangle}
\newcommand{\bs}{\boldsymbol}

\newcommand{\e}{\varepsilon}
\newcommand{\vf}{\varphi}

\newcommand{\s}{\sigma}
\newcommand{\Si}{\Sigma}
\newcommand{\al}{\alpha}
\newcommand{\be}{\beta}
\newcommand{\ga}{\gamma}
\newcommand{\Ga}{\Gamma}
\newcommand{\de}{\delta}
\newcommand{\De}{\Delta}

\newcommand{\la}{\lambda}

\newcommand{\spx}{\mathbf{x}}
\newcommand{\spy}{\mathbf{y}}

\newcommand{\spp}{\mathbf{p}}
\newcommand{\spk}{\mathbf{k}}
\newcommand{\spq}{\mathbf{q}}
\newcommand{\spn}{\mathbf{n}}

\begin{document}
\allowdisplaybreaks[4]
\frenchspacing

\title{{\Large\textbf{Radiation from Dirac fermions caused by a projective measurement}}}
%
\date{}
\author{P.O. Kazinski\thanks{E-mail: \texttt{kpo@phys.tsu.ru}},\;\, V.A. Ryakin\thanks{E-mail: \texttt{vlad.r.a.phys@yandex.ru}},\;\;and P.S. Shevchenko\thanks{E-mail: \texttt{psh-work@yandex.ru}}\\[0.5em]
{\normalsize Physics Faculty, Tomsk State University, Tomsk 634050, Russia}
}
\maketitle

\begin{abstract}

The theory of radiation of photons from Dirac particles caused by a projective measurement is developed. The explicit expressions for the inclusive probability to record a chain of events that the Dirac fermion had been measured in a certain state and after that the photon was recorded are derived. Stimulated and spontaneous radiations are considered. It is shown that in both cases the properties of radiation due to measurement resembles the properties of edge or transition radiation. In the case of stimulated radiation from a single particle, its wave function creates photons coherently as a charged fluid, i.e., the amplitudes of radiation from the points of the particle wave packet are summed. In the case of spontaneous radiation, the radiation of photons is incoherent, i.e., the probabilities of radiation from the points of the particle wave packet are added up. It is shown that stimulated radiation due to measurement can be used to trace the dynamics and collapse of the wave function of the Dirac particle. A systematic procedure taking into account a finiteness of the measurement time is presented. It is established that radiation due to measurement can be used as a source of hard photons, but the finiteness of the measurement time imposes an upper bound on the energy of radiated photons: the measurement time must be smaller than the radiation formation time, the latter being in inverse proportion to the photon energy. In the ultrarelativistic limit, the radiation formation time can be rather large. Several examples of radiation due to measurement of the state of free Dirac particles are investigated in detail. Namely, we scrutinize the radiation due to measurement of the spin projection, of the momentum, and of the coordinate for a general initial state of Dirac particles. The particular cases of uncorrelated and entangled particles in the beam are considered.

\end{abstract}

\section{Introduction}

A precise description of the measurement process and the dynamics of the wave function of particles undergoing the measurement is one of the urgent problems of quantum physics \cite{Neumann1932,NielsenChuang2011}. This issue is aligned with the more general one regarding the physical interpretation of the wave function. We touch these problems in the framework of quantum electrodynamics (QED) and investigate how the measurement of the intermediate state influences the properties of radiation created by the particle that has been measured. Recently, it has been shown in \cite{pra103,KazSol2022,KazSol2023,KRS2023} that there are QED processes where every point of the particle wave packet participates coherently. The outcome of these processes is such as though the wave function corresponded to some kind of a charged fluid as in the original Schr\"{o}dinger interpretation. The probabilities of these processes are determined by the modulus squared of the sum of the amplitudes from every point of the particle wave packet. These probabilities depend on the amplitude and the phase of the wave function of particles participating in the process as opposed to the more common incoherent processes where only the modulus of the wave function in the momentum representation is relevant \cite{pra103,PanGov18,MarcuseI,PMHK08,Corson2011,CorsPeat11,WCCP16,Remez19,KdGdAR21,Wong21}. The presence of such a property of coherent processes suggests that these processes can be used to recover the wave function of the particle exposed to the measurement and, in particular, to trace the dynamics of the wave function during its collapse. Stimulated radiation affected by the measurement is the example of such a coherent process and we shall study it in detail.

One of the unsettled issues in quantum theory remains the precise answer to the question when and where the collapse of the wave function of the system undergoing the measurement occurs. Namely, where is the precise boundary between the system and the detector measuring and projecting the state of this system? How long can we use the Schr\"{o}dinger equation to describe the quantum dynamics and when should we apply the projectors corresponding to the measurements? Whether does the measurement of the properties of one particle change instantaneously the properties of the second particle entangled with the first one so that the second particle emits photons?
The radiation caused by the measurement strongly depends on the dynamics of the wave function and provides the additional experimental tool that can shed light on these questions.

We shall show that the properties of radiation appearing solely due to measurement are similar to edge or transition radiation \cite{GaribYang,BazylZhev,Ginzburg,AkhShul,Bord.1,GKSSchYu1,GKSSchYu2}. This is a consequence of the fact that the measurement of the intermediate state of the radiating particles undresses this particle, the virtual photons bound to it decouple and become real \cite{Feinberg1966,AkhShul,BazylZhev,ShSSh2009}. Therefore, in order to trace the dynamics of the wave function one can employ a proper modification of the developed experimental technics for beam diagnostics based on transition and diffraction radiations \cite{BBCLS,Pbook,SukhKubPot17,SukhihDTHs}. The quantum theory of transition radiation for the initial and final particle states in the form of plane waves can be found, for example, in \cite{GaribYang,BazylZhev,Ginzburg}, where the overview of the development in this field is given. This also includes stimulated transition radiation \cite{ArutAvet1972,AvetAvetPetr1978,ZarLomNer80,IvanLomon81} that was observed experimentally in \cite{Lihn1996}. The generalization of this theory to transition radiation from particles prepared and detected in the states of a general form is presented in \cite{pra103,KazSol2022}. In fact, the paper \cite{KazSol2022} contains such a generalization for all QED radiation processes up to the second order in the coupling constant with emphasis on the coherent processes in the sense described above. Of course, there are many other papers where the influence of the form of particle wave packets on the properties of radiation was investigated for various QED processes. Just to mention a few: spontaneous radiation from a single particle \cite{PanGov18,MarcuseI,PMHK08,Corson2011,CorsPeat11,WCCP16,Remez19,KdGdAR21,PanGov19,PanGov21,BBTq1,BBTq2} and two particles \cite{Wong21,AngDiPiaz18,AngDiPiaz19,KRAK21}, stimulated radiation from a single particle \cite{PanGov21,Talebi16,GovPan18}, classical approaches to describe radiation from a wave packet of a single particle \cite{MarcuseII,IvKarl13prl,IvKarl13pra,KonPotPol,KarlZhev19,PupKarl22}.

In the present paper, we further generalize the theory developed in \cite{KazSol2022} to the case where the intermediate state of a radiating particle is measured. We obtain the general formulas for the probability of a chain of events and for the conditional probability. Then we apply the elaborated formalism to several particular examples describing the radiation from free Dirac particles undergoing a measurement. We study spontaneous and stimulated radiations due to measurement in the leading nontrivial order of perturbation theory and describe their main properties. In particular, we consider radiation due to measurement of the particle spin projection, of the particle momentum, and of the particle coordinate. The initial state of the Dirac particles is assumed to be of a general form and is described by the many-particle density matrix. The particular case of radiation due to measurement of the spin projection of one of the pair of entangled particles is also investigated.

The paper is organized as follows. In Sec. \ref{GenForm}, the problem statement is given and the general formulas for the inclusive probability of a chain of events and for the conditional probability are derived in the QED framework. In Sec. \ref{Stim_Rad_Sec}, stimulated radiation due to measurement is investigated in the leading nontrivial order of perturbation theory. Section \ref{Spont_Radiat} is devoted to spontaneous radiation due to measurement. In measuring the properties of particles in the intermediate state, i.e., the properties of virtual (bare) particles, we encounter with the problem of a finite time evolution in QED (see, e.g., Chapt. VII of \cite{BogShir1959}). In order to overcome it, we employ some kind of the procedure elaborated in \cite{Dyson1951} (see also \cite{Walhout1999}). This formal procedure allows us to take qualitatively into account a finiteness of the measurement time. In Sec. \ref{Examples_Sec}, several examples of the radiation due to measurement from free Dirac particles are presented. In Sec. \ref{Stim_Rad_Free_Part}, stimulated radiation is considered, whereas spontaneous radiation from a single Dirac particle is described in Sec. \ref{Spont_Rad_Sing_Part}. In Sec. \ref{Stim_Rad_Spin_Measur}, we consider stimulated radiation due to spin measurement. Stimulated radiation due to measurement of the momentum and of the coordinates is scrutinized in Sec. \ref{Stim_Rad_Mom_Coord}. In Sec. \ref{Spin_Meas_Entangl}, stimulated radiation due to spin measurement of one of the pair of entangled particles is investigated. In Conclusion we summarize the results. In calculating the inclusive probabilities of chains of event, it is useful to employ the Bargmann-Fock representation \cite{Bargmann1961,BerezMSQ1.4,Glaub2,KlauSud}. A summary of this formalism can be found in Appendix A of the paper \cite{ippccb}. For the reader convenience, some of the most frequently used formulas are collected in Appendix \ref{Ap_BF_represnt}. In Appendix \ref{Measurement_Proj_App}, we briefly describe the main postulates of the measurement procedure in quantum theory and define the projectors specifying the measurements used in the present paper. In Appendix \ref{Traces_App}, we provide the expression for the density matrices of Dirac particles and photons. Furthermore, in this Appendix, we obtain the traces of operators needed for evaluation of the probabilities. Appendix \ref{Beams_App} is devoted to several simple models of the density matrices for beams of Dirac particles.

We use the notation and the normalization agreements for the modes of quantum fields adopted in the papers \cite{pra103,KazSol2022,KRS2023}. The Greek indices $\al$, $\be$, $\bar{\al}$, $\bar{\be}$, $\ldots$ denote the quantum numbers of the particle states. The energy operator of particles is diagonal in the chosen basis of one-particle states. We use the star as the sign of complex conjugation. The bar over the Dirac spinor means as usual the Dirac conjugate. The summation (integration) over repeated indices is always understood unless otherwise stated. We also assume that the system is confined into a box with large volume $V$ and the states of particles are normalized to unity. Wherever it does not lead to misunderstanding, we use the matrix notation. For example,
\begin{equation}
    a^*a\equiv a^*_\al a_\al,\qquad d^*Dd\equiv d^*_\al D_{\al\bar{\al}}d_{\bar{\al}},\qquad \text{etc}.
\end{equation}
The operator acting in the Fock space are distinguished by the hats.

We use the system of units such that $\hbar=c=1$ and $e^2=4\pi\al$, where $\al$ is the fine structure constant. The Minkowski metric is taken with the mostly minus signature. The round (square) brackets at a pair of indices mean symmetrization (antisymmetrization) without the factor $1/2$. The square brackets at a pair of vectors denote the cross product of these vectors.

\section{Inclusive probabilities}\label{GenForm}

Let us investigate an influence of measurement of the intermediate state of a Dirac fermion on radiation created by it in the framework of quantum electrodynamics. We shall find the explicit expression for the probability to record the photon in a certain state at the given instant of time $t=t_{out}\rightarrow\infty$ and the Dirac fermion in a certain state at the given instant of time $t=t_0$, $t_0\in(t_{in},t_{out})$. We shall also obtain the respective conditional probability to detect a photon and the intensity of radiation recorded under the stipulation that the Dirac fermion had been detected at the instant of time $t=t_0$. The Dirac particle can be an electron or a hadron with spin $1/2$ or an atomic nucleus with spin $1/2$. For definiteness and brevity, we will frequently refer to a Dirac particle as an electron. Deliberation of general Dirac particles differs substantially only by the form of the interaction Hamiltonian, which explicitly will be given below.

Let the initial state of electrons be specified by the density operator $\hat{R}_e$ of a general form \eqref{dens_matrx_gen}. As the initial state of photons, $\hat{R}_{ph}$, we take the coherent state \eqref{R_ph_coher}. We suppose that the initial state prepared at the instant of time $t=t_{in}\rightarrow-\infty$ is such that the electrons and the photons are uncorrelated and the positrons are absent
\begin{equation}\label{init_state}
    \hat{R}=\hat{R}_{ph}\otimes \hat{R}_e\otimes |0\ran_{e^+}\lan0|_{e^+}.
\end{equation}
The creation and annihilation operators are denoted as ($\hat{c}^\dag_{\bar{\ga}}$, $\hat{c}_{\ga}$) for photons, as ($\hat{a}^\dag_{\bar{\al}}$, $\hat{a}_{\al}$) for electrons, and as ($\hat{b}^\dag_{\bar{\be}}$, $\hat{b}_{\be}$) for positrons.

The measurement is performed at the instant of time $t_0$. As a result, one of the electrons is detected in one of the states distinguished by the projector $D_e$ in the one-particle Hilbert space of electron states. The projector in the Fock space, $\hat{\Pi}_{D_e}$, describing this measurement is defined in formula \eqref{Pi_D_defn}. At the instant of time $t=t_{out}$, a single photon is recorded in one of the states singled out by the projector $D$ in the one-particle Hilbert space of photon states. The corresponding projector in the Fock space, $\hat{\Pi}_{D}$, has the form \eqref{Pi_D_defn}, where one ought to replace the creation and annihilation operators for electrons by the creation and annihilation operators for photons. In accordance with the standard postulates of quantum theory (see the main formulas in Appendix \ref{Measurement_Proj_App}), the probability of such a chain of events is
\begin{equation}\label{incl_prob_gen}
    P(\hat{\Pi}_D\leftarrow \hat{\Pi}_{D_e})=\Sp(\hat{\Pi}_D \hat{U}_{t_{out},t_0}\hat{\Pi}_{D_e}\hat{U}_{t_0,t_{in}}\hat{R}\hat{U}_{t_{in},t_0} \hat{\Pi}_{D_e}\hat{U}_{t_0,t_{out}}),
\end{equation}
where $\hat{U}_{t_2,t_1}$ is the evolution operator of QED and all the operators are given in the Schr\"{o}dinger representation. The respective conditional probability becomes
\begin{equation}\label{cond_prob_gen}
    P(\hat{\Pi}_D|\hat{\Pi}_{D_e})=P(\hat{\Pi}_D\leftarrow \hat{\Pi}_{D_e})/P(\hat{\Pi}_{D_e}),
\end{equation}
where the probability to detect the electron at the instant of time $t_0$ in the states distinguished by the projector $D_e$,
\begin{equation}\label{probab_electr_gen}
    P(\hat{\Pi}_{D_e})=\Sp(\hat{\Pi}_{D_e}\hat{U}_{t_0,t_{in}}\hat{R}\hat{U}_{t_{in},t_0}).
\end{equation}
In order to obtain the average number of radiated photons with quantum numbers specified by the projector $D$ under the condition that  the electron had been detected in one of the states singled out by the projector $D_e$ at the instant of time $t=t_0$, one formally needs to tend $D$ to zero in \eqref{cond_prob_gen} and to find the leading asymptotics linear in $D$. The conditional intensity of radiation of photons with quantum numbers $\ga_0$ is deduced from the average number of radiated photons by the replacement of $D_{\bar{\ga}\ga}$ by $\de_{\bar{\ga}\ga_0}\de_{\ga\ga_0}k_{0\ga_0}$, where $k_{0\ga_0}$ is the photon energy in the state with quantum numbers $\ga_0$. As it has been already mentioned in Introduction, we choose such a basis of one-particle photon states that the photon energy operator is diagonal in this basis.

In the interaction picture, the probabilities \eqref{incl_prob_gen} and \eqref{probab_electr_gen} can be cast into the form
\begin{subequations}\label{probab_inter_pict}
\begin{align}
    P(\hat{\Pi}_D\leftarrow \hat{\Pi}_{D_e})&=\Sp(\hat{\Pi}_{D(t_{out})}\hat{S}_{t_{out},t_0}\hat{\Pi}_{D_e(t_0)}\hat{S}_{t_0,t_{in}} \hat{R}(t_{in})\hat{S}_{t_{in},t_0}\hat{\Pi}_{D_e(t_0)}\hat{S}_{t_0,t_{out}}),\label{prob_chain}\\
    P(\hat{\Pi}_{D_e})&=\Sp(\hat{\Pi}_{D_e(t_0)}\hat{S}_{t_0,t_{in}}\hat{R}(t_{in})\hat{S}_{t_{in},t_0}),
\end{align}
\end{subequations}
where
\begin{equation}
    D^e_{\bar{\al}\al}(t)=D^e_{\bar{\al}\al}e^{i(p^0_{\bar{\al}}-p^0_{\al})t},\qquad D_{\bar{\ga}\ga}(t)=D_{\bar{\ga}\ga}e^{i(k^0_{\bar{\ga}}-k^0_{\ga})t},
\end{equation}
and $p^0_\al$ is the energy of a single electron in the state with quantum numbers $\al$. The operator
\begin{equation}\label{S_oper}
    \hat{S}_{t_2,t_1}=\Texp\Big\{-i\int_{t_1}^{t_2}dx:e\hat{A}_i\hat{\bar{\psi}}\ga^i\hat{\psi}: -i\int_{t_1}^{t_2}dt\hat{V}_{\text{Coul}}(t) \Big\}.
\end{equation}
The Coulomb term in the interaction Hamiltonian is written as
\begin{equation}\label{V_Coul}
    \hat{V}_{\text{Coul}}(t)=-\frac{e^2}2\int d\spx d\spy: (\hat{\bar{\psi}}\ga^0\hat{\psi})(t,\spx): \De^{-1}(\spx-\spy) :(\hat{\bar{\psi}}\ga^0\hat{\psi})(t,\spy):,
\end{equation}
where $\hat{A}_i(x)$ is the electromagnetic potential operator in the interaction representation, $\hat{\psi}(x)$ is the electron-positron field operator, $\De^{-1}(\spx-\spy)$ is the kernel of the operator $\De^{-1}$. Hereinafter, the Coulomb gauge is implied. To study the radiation produced not only from electrons but from general Dirac fermions, one just has to substitute
\begin{equation}
    \ga^\mu\rightarrow\Ga^\mu
\end{equation}
in the interaction Hamiltonian \eqref{S_oper}, \eqref{V_Coul}, where $\Ga^\mu$ is the nonlocal vertex containing the electric and magnetic form-factors (see, e.g., \cite{LandLifQED}).

The density operator reads
\begin{equation}
    \hat{R}(t_{in})=\hat{R}_{ph}(t_{in})\otimes \hat{R}_e(t_{in})\otimes |0\ran_{e^+}\lan0|_{e^+},
\end{equation}
where
\begin{equation}
    \hat{R}_{ph}(t_{in})=|d(t_{in})\ran\lan d^*(t_{in})|e^{-d^*d},\qquad d_\ga(t_{in})=e^{ik^0_\ga t_{in}}d_\ga,
\end{equation}
and $\hat{R}_e(t_{in})$ has the form \eqref{dens_matrx_gen} with
\begin{equation}\label{dens_matr_electr}
    \rho_{\al_N\cdots\al_1|\bar{\al}_1\cdots\bar{\al}_M}\rightarrow \rho_{\al_N\cdots\al_1|\bar{\al}_1\cdots\bar{\al}_M}(t_{in}):=\rho_{\al_N\cdots\al_1|\bar{\al}_1\cdots\bar{\al}_M} e^{i(p^0_{\al_N}+\cdots+p^0_{\al_1}-p^0_{\bar{\al}_1}-\cdots-p^0_{\bar{\al}_M})t_{in}}.
\end{equation}
Henceforth, for brevity, the arguments $t_{in}$, $t_0$, and $t_{out}$ of the initial states and the projectors $D_e$, $D$ will be omitted.

The above formulas \eqref{probab_inter_pict} are obtained under the assumptions that the measurement is carried out instantaneously as it is usually supposed in the standard formulations of postulates of quantum theory \cite{Neumann1932,NielsenChuang2011,BraginskiKhalili1992}. It is clear that the measurement implemented by an actual detector lasts some time $\tau$. The precise description of such a measurement depends on the details of the measurement procedure and on the internal structure of the detector. Nevertheless, in Sec. \ref{Spont_Radiat}, we will consider the simplest model of a finite time measurement allowing one to generalize formulas \eqref{probab_inter_pict} to the case $\tau\geqslant0$ and to estimate qualitatively the influence of finiteness of the measurement time on the observables.

\section{Stimulated radiation}\label{Stim_Rad_Sec}

Let us consider stimulated radiation of photons by electrons in the leading nontrivial order of perturbation theory. Assaying different contributions to the operator $\hat{S}_{t_2,t_1}$, it is not difficult to see that in this case the leading nontrivual contribution to the inclusive probability \eqref{probab_inter_pict} is of first order in the coupling constant $e$ (see for details \cite{KazSol2022}). Therefore, employing the notation borrowed from \cite{KazSol2022}, we can write
\begin{equation}\label{S_expanded}
    \hat{S}_{t_2,t_1}=\hat{1}+\hat{V}_{t_2,t_1}+\cdots,
\end{equation}
where
\begin{equation}
    \hat{V}_{t_2,t_1}=(V_{t_2,t_1})^{\bar{\ga}}_{\bar{\al}\al}\hat{a}^\dag_{\bar{\al}}\hat{a}_{\al}\hat{c}^\dag_{\bar{\ga}}- (V^\dag_{t_2,t_1})^{\ga}_{\bar{\al}\al}\hat{a}^\dag_{\bar{\al}}\hat{a}_{\al}\hat{c}_{\ga},
\end{equation}
and $(V^\dag_{t_2,t_1})^{\ga}_{\bar{\al}\al}=(V^*_{t_2,t_1})^{\ga}_{\al\bar{\al}}$. Due to unitarity of the operator $\hat{S}_{t_2,t_1}$, the following properties are valid in the given order of perturbation theory
\begin{equation}
    \hat{V}_{t_2,t_1}=\hat{V}^\dag_{t_1,t_2}=-\hat{V}_{t_1,t_2}.
\end{equation}
Then substituting the expansion \eqref{S_expanded} into \eqref{prob_chain} and keeping the terms at most linear in $e$, we come to
\begin{equation}\label{P_stim}
    P(\hat{\Pi}_D\leftarrow \hat{\Pi}_{D_e})=I_0+(I_1+I_2+c.c.),
\end{equation}
where
\begin{equation}\label{I_traces}
\begin{split}
    I_0:=&\,\Sp(\hat{R}_e\hat{\Pi}_{D_e})\Sp(\hat{R}_{ph}\hat{\Pi}_{D}),\\
    I_1:=&\,\Sp(\hat{R}_e\hat{\Pi}_{D_e}\hat{a}^\dag_{\bar{\al}}\hat{a}_{\al}\hat{\Pi}_{D_e}) \big[(V_{t_{out},t_0})^{\bar{\ga}}_{\bar{\al}\al} \Sp(\hat{R}_{ph}\hat{\Pi}_D\hat{c}^\dag_{\bar{\ga}}) - (V^\dag_{t_{out},t_0})^{\ga}_{\bar{\al}\al} \Sp(\hat{R}_{ph}\hat{\Pi}_D\hat{c}_{\ga})\big],\\
    I_2:=&\,\Sp(\hat{R}_e\hat{\Pi}_{D_e}\hat{a}^\dag_{\bar{\al}}\hat{a}_{\al})\big[(V_{t_0,t_{in}})^{\bar{\ga}}_{\bar{\al}\al} \Sp(\hat{R}_{ph}\hat{\Pi}_D\hat{c}^\dag_{\bar{\ga}}) - (V^\dag_{t_0,t_{in}})^{\ga}_{\bar{\al}\al} \Sp(\hat{R}_{ph}\hat{\Pi}_D\hat{c}_{\ga})\big].
\end{split}
\end{equation}
The probability to record the electron takes the form
\begin{equation}\label{P_stim1}
    P(\hat{\Pi}_{D_e})=J_0+(J_1+c.c.),
\end{equation}
where
\begin{equation}\label{J_traces}
\begin{split}
    J_0:=&\,\Sp(\hat{R}_e\hat{\Pi}_{D_e}),\\
    J_1:=&\,\Sp(\hat{R}_e\hat{\Pi}_{D_e}\hat{a}^\dag_{\bar{\al}}\hat{a}_\al)\big[(V_{t_0,t_{in}})^{\bar{\ga}}_{\bar{\al}\al} \Sp(\hat{R}_{ph}\hat{c}^\dag_{\bar{\ga}}) - (V^\dag_{t_0,t_{in}})^{\ga}_{\bar{\al}\al} \Sp(\hat{R}_{ph}\hat{c}_{\ga}) \big].
\end{split}
\end{equation}
The conditional probability in the given order of perturbation theory is written as
\begin{equation}\label{cond_prob_stim}
    P(\hat{\Pi}_D|\hat{\Pi}_{D_e})=\frac{I_0}{J_0}+\frac{1}{J_0}\big(I_1+I_2-\frac{I_0}{J_0}J_1+c.c.\big).
\end{equation}
The traces of operators appearing in the expressions above are evaluated in Appendix \ref{Traces_App}.

Substituting these expressions into \eqref{I_traces} and \eqref{J_traces}, we arrive at
\begin{equation}\label{P_stim_rad_gen}
\begin{split}
    P(\hat{\Pi}_D\leftarrow \hat{\Pi}_{D_e})=&\,(1-\rho^{(0)}_{\tilde{D}_e})(1-e^{- d^*Dd})+\Big\{ \big(\rho^{(1)} -\rho^{(1)}_{\tilde{D}_e}\big)_{\al\bar{\al}} (V_{t_{out},t_{in}})^{\bar{\ga}}_{\bar{\al}\al} (d^*D)_{\bar{\ga}} e^{-d^*Dd}+\\
    &+(\rho^{(1)}_{\tilde{D}_e}D_e)_{\al\bar{\al}} \Big[\big((d^*D)_{\bar{\ga}} +(1-e^{-d^*Dd})(d^*\tilde{D})_{\bar{\ga}} \big)(V_{t_0,t_{in}})^{\bar{\ga}}_{\bar{\al}\al} -(V^\dag_{t_0,t_{in}})^{\ga}_{\bar{\al}\al}(1-e^{-d^*Dd})d_\ga \Big]+\\
    &+(D_e\rho^{(1)}_{\tilde{D}_e}D_e)_{\al\bar{\al}} (V_{t_{out},t_0})^{\bar{\ga}}_{\bar{\al}\al} (d^*D)_{\bar{\ga}} e^{-d^*Dd} +c.c.\Big\},
\end{split}
\end{equation}
and
\begin{equation}
\begin{split}
    P(\hat{\Pi}_{D_e})&=1-\rho^{(0)}_{\tilde{D}_e} +\Big\{(\rho^{(1)}_{\tilde{D}_e}D_e)_{\al\bar{\al}} \big( (V_{t_0,t_{in}})^{\bar{\ga}}_{\bar{\al}\al} d^*_{\bar{\ga}} -(V^\dag_{t_0,t_{in}})^{\ga}_{\bar{\al}\al} d_{\ga}  \big) +c.c.\Big\}=\\
    &=1-\rho^{(0)}_{\tilde{D}_e} +[\rho^{(1)}_{\tilde{D}_e},D_e]_{\al\bar{\al}} \big( (V_{t_0,t_{in}})^{\bar{\ga}}_{\bar{\al}\al} d^*_{\bar{\ga}} -(V^\dag_{t_0,t_{in}})^{\ga}_{\bar{\al}\al} d_{\ga}  \big),
\end{split}
\end{equation}
where $\tilde{D}^e_{\al\bar{\al}}:=\de_{\al\bar{\al}}-D^e_{\al\bar{\al}}$. Consequently, the conditional probability \eqref{cond_prob_stim} becomes
\begin{equation}
\begin{split}
    P(\hat{\Pi}_D|\hat{\Pi}_{D_e})=&\,1-e^{-d^*Dd}+\frac{1}{1-\rho^{(0)}_{\tilde{D}_e}}\Big\{\big[\big(\rho^{(1)} -\rho^{(1)}_{\tilde{D}_e}\big)_{\al\bar{\al}} (V_{t_{out},t_{in}})^{\bar{\ga}}_{\bar{\al}\al}+\\ &+(D_e\rho^{(1)}_{\tilde{D}_e}D_e)_{\al\bar{\al}} (V_{t_{out},t_0})^{\bar{\ga}}_{\bar{\al}\al} +(\rho^{(1)}_{\tilde{D}_e}D_e)_{\al\bar{\al}}(V_{t_0,t_{in}})^{\bar{\ga}}_{\bar{\al}\al} \big](d^*D)_{\bar{\ga}} e^{-d^*Dd} +c.c.\Big\}.
\end{split}
\end{equation}
For $D\rightarrow0$, we have in the leading order
\begin{equation}\label{cond_prob_stim1}
\begin{split}
    P(\hat{\Pi}_D\leftarrow\hat{\Pi}_{D_e})=&\,\big\{1-\rho^{(0)}_{\tilde{D}_e} +[\rho^{(1)}_{\tilde{D}_e},D_e]_{\al\bar{\al}} \big[ (V_{t_0,t_{in}})^{\bar{\ga}}_{\bar{\al}\al} d^*_{\bar{\ga}} -(V^\dag_{t_0,t_{in}})^{\ga}_{\bar{\al}\al} d_{\ga}\big] \big\} (d^*Dd)+\\
    &+(1-\rho^{(0)}_{\tilde{D}_e})\big(\mathcal{A}^{\bar{\ga}}(d^*D)_{\bar{\ga}}+c.c.\big),\\
    P(\hat{\Pi}_D|\hat{\Pi}_{D_e})=&\,d^*Dd+\big(\mathcal{A}^{\bar{\ga}}(d^*D)_{\bar{\ga}}+c.c.\big),
\end{split}
\end{equation}
where
\begin{equation}\label{class_amplitude_gen}
    \mathcal{A}^{\bar{\ga}}:=\frac{1}{1-\rho^{(0)}_{\tilde{D}_e}}\Big[\big(\rho^{(1)} -\rho^{(1)}_{\tilde{D}_e}\big)_{\al\bar{\al}} (V_{t_{out},t_{in}})^{\bar{\ga}}_{\bar{\al}\al}+ (D_e\rho^{(1)}_{\tilde{D}_e}D_e)_{\al\bar{\al}} (V_{t_{out},t_0})^{\bar{\ga}}_{\bar{\al}\al} +(\rho^{(1)}_{\tilde{D}_e}D_e)_{\al\bar{\al}}(V_{t_0,t_{in}})^{\bar{\ga}}_{\bar{\al}\al}\Big].
\end{equation}
Then the last formula in \eqref{cond_prob_stim1} can be cast into a more suggestive form
\begin{equation}\label{cond_prob_interfer}
    P(\hat{\Pi}_D|\hat{\Pi}_{D_e})=(d^*_\ga+\mathcal{A}^*_\ga) D_{\ga\bar{\ga}}(d_{\bar{\ga}}+\mathcal{A}_{\bar{\ga}}).
\end{equation}
The term proportional to $e^2$ in this expression should be discarded since it exceeds the accuracy of derivation of formula \eqref{cond_prob_stim1}.

Let us consider several particular cases of the general formulas obtained and give the interpretation to them. In the case of one-particle initial state of electrons, the amplitude of stimulated radiation is reduced to
\begin{equation}\label{amplit_stim_onepart}
\begin{split}
    \mathcal{A}^{\bar{\ga}}=&\,\frac{1}{\rho^{(1)}_{\al_1\bar{\al}_1}D^e_{\bar{\al}_1\al_1}} \big[(D_e\rho^{(1)}D_e)_{\al\bar{\al}} (V_{t_{out},t_0})^{\bar{\ga}}_{\bar{\al}\al} +(\rho^{(1)}D_e)_{\al\bar{\al}}(V_{t_0,t_{in}})^{\bar{\ga}}_{\bar{\al}\al}\big]=\\
    =&\,\frac{1}{\rho^{(1)}_{\al_1\bar{\al}_1}D^e_{\bar{\al}_1\al_1}} \big[(D_e\rho^{(1)}D_e)_{\al\bar{\al}} (V_{t_{out},t_{in}})^{\bar{\ga}}_{\bar{\al}\al} +(\tilde{D}_e\rho^{(1)}D_e)_{\al\bar{\al}}(V_{t_0,t_{in}})^{\bar{\ga}}_{\bar{\al}\al}\big].
\end{split}
\end{equation}
If the volume of phase space singled out by the projector $D_e$ is small, then the amplitude \eqref{class_amplitude_gen} for an arbitrary initial state of electrons transforms into
\begin{equation}\label{amplit_stim_contin}
    \mathcal{A}^{\bar{\ga}}=\frac{1}{\rho^{(1)}_{\al_1\bar{\al}_1}D^e_{\bar{\al}_1\al_1}} \Big[\rho^{(2)}_{\be\al|\bar{\al}\bar{\be}} D^e_{\bar{\be}\be} (V_{t_{out},t_{in}})^{\bar{\ga}}_{\bar{\al}\al} +(\rho^{(1)}D_e)_{\al\bar{\al}}(V_{t_0,t_{in}})^{\bar{\ga}}_{\bar{\al}\al}\Big],
\end{equation}
in the leading order in $D_e$. For $D_e=1$ and, consequently, $\tilde{D}_e=0$, i.e., when the detector of electrons can record a particle in any state, it follows from formulas \eqref{cond_prob_stim1}, \eqref{class_amplitude_gen}, and \eqref{rho_tilde_Dz} that
\begin{equation}\label{prob_stim}
    \frac{P(\hat{\Pi}_D\leftarrow \hat{\Pi}_{D_e})}{1-\rho_0}=P(\hat{\Pi}_D|\hat{\Pi}_{D_e})=(d^*_\ga +\mathcal{A}^{0*}_\ga) D_{\ga\bar{\ga}}(d_{\bar{\ga}}+\mathcal{A}^0_{\bar{\ga}}),
\end{equation}
where
\begin{equation}\label{class_amplitude}
    \mathcal{A}_0^{\bar{\ga}}=\frac{1}{1-\rho_0} \rho^{(1)}_{\al\bar{\al}}(V_{t_{out},t_{in}})^{\bar{\ga}}_{\bar{\al}\al},
\end{equation}
and it is implied as in \eqref{cond_prob_interfer} that the term proportional to $e^2$ should be cast out. The quantity $\rho_0$ defines the probability to detect the system of electrons in the vacuum state for the initial state \eqref{class_amplitude_gen}. When $\rho_0=0$, the expressions \eqref{prob_stim}, \eqref{class_amplitude} coincide with expression (79) of the paper \cite{KazSol2022} describing stimulated radiation from electrons without measurement of the electron state. Recall that expression \eqref{class_amplitude} is the amplitude of radiation from the classical Dirac current, i.e., without quantum recoil, for the particle in the state with the density matrix $\rho^{(1)}_{\al\bar{\al}}$ (see \cite{KazSol2022}). In particular, in the case of a pure one-particle electron state, this amplitude equals the amplitude of radiation from the classical Dirac current for the electron wave function. It is also evident that for such a measurement the free Dirac particles do not radiate, because in this case the radiation amplitude $(V_{t_{out},t_{in}})^{\bar{\ga}}_{\bar{\al}\al}=0$ in virtue of the energy-momentum conservation law. One may say that such a measurement does not perturb the state of free Dirac particles and so they do not radiate.

In general, it follows from \eqref{I_traces} that for $(V_{t_{out},t_{in}})^{\bar{\ga}}_{\bar{\al}\al}=0$ the nontrivial contribution to the probability to record a photon is proportional to
\begin{equation}
    \Sp(\hat{\tilde{\Pi}}_{D_e}\hat{R}_e\hat{\Pi}_{D_e}\hat{a}^\dag_{\bar{\al}} \hat{a}_\al)=\Sp(\hat{\tilde{\Pi}}_{D_e}[\hat{R}_e,\hat{\Pi}_{D_e}]\hat{a}^\dag_{\bar{\al}} \hat{a}_\al).
\end{equation}
Hence, if the commutator of the projector defining the measurement and the density matrix of Dirac particles is negligible, then stimulated radiation due to measurement is small. The equality to zero is reached for the so-called nondemolition measurement \cite{BraginskiKhalili1992} when $[\hat{R}_e,\hat{\Pi}_{D_e}]=0$.

As seen from the particular cases considered above, the following interpretation can be given to the different terms in \eqref{class_amplitude_gen}. The first term in \eqref{class_amplitude_gen} has the form \eqref{class_amplitude} and defines the amplitude of radiation produced by the electrons that have not been recorded by the detector. The second term in \eqref{class_amplitude_gen} describes the amplitude of radiation created by the electron whose state has been measured by the detector. This radiation is produced on the time interval $[t_0,t_{out}]$, i.e., after the measurement, and it does not suffer the quantum recoil. The third term in \eqref{class_amplitude_gen} is the amplitude of radiation generated by the electron jumping from the initial state to the state obtained from the initial one by the action of the projector $D_e$. This radiation is produced on the time interval $[t_{in},t_0]$. The common factor at the square brackets is the normalization factor. Its inverse is equal to the probability to find the Dirac fermion in the one-particle state distinguished by the projector $D_e$ in the initial state. Notice that even in the case when the process evolves in vacuum or in a stationary external field or in a stationary dispersive medium, the energy conservation law is not fulfilled for the matrix elements $(V_{t_0,t_{in}})^{\bar{\ga}}_{\bar{\al}\al}$ and $(V_{t_{out},t_0})^{\bar{\ga}}_{\bar{\al}\al}$ for $t_{out}\rightarrow\infty$ and $t_{in}\rightarrow-\infty$. It is a consequence of semiinfinite integration limits with respect to time in these matrix elements. The measurement of the electron state gives rise to a new type of radiation similar to edge or transition radiation \cite{GaribYang,BazylZhev,Ginzburg,AkhShul,Bord.1,GKSSchYu1,GKSSchYu2}. In contrast to edge and transition radiations, the abrupt change of trajectories of charged particles in the beam occurs not on the timelike hypersurface describing, for example, the surface of the mirror or the section of the accelerator bending magnet but on the spacelike hypersurface $t=t_0$. Of course, the actual measurement is not instantaneous. The procedure allowing one to take into account a finiteness of the measurement time will be given in the next section.

\section{Spontaneous radiation}\label{Spont_Radiat}

Now we consider the influence of measurement of a single electron state on spontaneous radiation created by the beam of electrons. In this case, the initial state of the system has the form \eqref{init_state} with
\begin{equation}
    \hat{R}_{ph}=|0\ran_{ph}\lan0|_{ph}.
\end{equation}
The examination of different contributions to the probabilities \eqref{probab_inter_pict} reveals that the leading nontrivial contribution to the inclusive probability is of order $e^2$ (see for details \cite{KazSol2022}). To reproduce this contribution, it is sufficient to take the expansion of the operator $\hat{S}_{t_2,t_1}$ in the form \eqref{S_expanded}. Then keeping the terms of at most second order in the coupling constant $e$, we come to
\begin{equation}\label{probab_spont_gen}
\begin{split}
    P(\hat{\Pi}_D\leftarrow \hat{\Pi}_{D_e})&=\Sp\big[\hat{R}(\hat{V}_{t_{in},t_0}\hat{\Pi}_{D_e} +\hat{\Pi}_{D_e}\hat{V}_{t_0,t_{out}} )\hat{\Pi}_D(\hat{V}_{t_{out},t_0}\hat{\Pi}_{D_e} +\hat{\Pi}_{D_e}\hat{V}_{t_0,t_{in}})\big]=\\
    &=K_1+K_2+K^*_2+K_3,
\end{split}
\end{equation}
where
\begin{equation}\label{K_123}
\begin{split}
    K_1&:=\Sp(\hat{R}_e\hat{\Pi}_{D_e} \hat{a}^\dag_{\bar{\al}}\hat{a}_\al\hat{a}^\dag_{\bar{\be}} \hat{a}_\be\hat{\Pi}_{D_e})(V^\dag_{t_{out},t_0})^\ga_{\bar{\al}\al} (V_{t_{out},t_0})^{\bar{\ga}}_{\bar{\be}\be}\Sp(\hat{R}_{ph}\hat{c}_\ga\hat{\Pi}_D\hat{c}^\dag_{\bar{\ga}}),\\
    K_2&:=\Sp(\hat{R}_e \hat{a}^\dag_{\bar{\al}}\hat{a}_\al \hat{\Pi}_{D_e}\hat{a}^\dag_{\bar{\be}} \hat{a}_\be\hat{\Pi}_{D_e})(V^\dag_{t_0,t_{in}})^\ga_{\bar{\al}\al} (V_{t_{out},t_0})^{\bar{\ga}}_{\bar{\be}\be}\Sp(\hat{R}_{ph}\hat{c}_\ga\hat{\Pi}_D\hat{c}^\dag_{\bar{\ga}}),\\
    K^*_2&:=\Sp(\hat{R}_e \hat{\Pi}_{D_e}\hat{a}^\dag_{\bar{\al}}\hat{a}_\al \hat{\Pi}_{D_e} \hat{a}^\dag_{\bar{\be}} \hat{a}_\be)(V^\dag_{t_{out},t_0})^\ga_{\bar{\al}\al} (V_{t_0,t_{in}})^{\bar{\ga}}_{\bar{\be}\be}\Sp(\hat{R}_{ph}\hat{c}_\ga\hat{\Pi}_D\hat{c}^\dag_{\bar{\ga}}),\\
    K_3&:=\Sp(\hat{R}_e \hat{a}^\dag_{\bar{\al}}\hat{a}_\al \hat{\Pi}_{D_e}\hat{a}^\dag_{\bar{\be}} \hat{a}_\be)(V^\dag_{t_0,t_{in}})^\ga_{\bar{\al}\al} (V_{t_0,t_{in}})^{\bar{\ga}}_{\bar{\be}\be}\Sp(\hat{R}_{ph}\hat{c}_\ga\hat{\Pi}_D\hat{c}^\dag_{\bar{\ga}}).
\end{split}
\end{equation}
Obviously, the expression \eqref{probab_spont_gen} is positive-definite. The trace over the photonic degrees of freedom is readily evaluated
\begin{equation}
    \Sp(\hat{R}_{ph}\hat{c}_\ga\hat{\Pi}_D\hat{c}^\dag_{\bar{\ga}})=D_{\ga\bar{\ga}}.
\end{equation}
The traces over the electronic degrees of freedom are calculated in Appendix \ref{Traces_App} with the result presented in formula \eqref{traces_spontan}.

Substituting the explicit expressions for the traces into \eqref{K_123}, we obtain
\begin{equation}\label{probab_spont1}
\begin{split}
    P(\hat{\Pi}_D\leftarrow \hat{\Pi}_{D_e})=&\,(\rho^{(1)}_{\tilde{D}_e})_{\be\bar{\al}} D_{\ga\bar{\ga}} \big(D_eV^\dag_{t_{out,t_0}}+V^\dag_{t_0,t_{in}}D_e\big)^{\ga}_{\bar{\al}\al} \big(V_{t_{out,t_0}}D_e +D_eV_{t_0,t_{in}}\big)^{\bar{\ga}}_{\al\be}+\\
    &+\big[\de_{\al\bar{\be}}(\rho^{(1)}-\rho^{(1)}_{\tilde{D}_e})_{\be\bar{\al}} -\rho^{(2)}_{\al\be|\bar{\al}\bar{\be}}\big] D_{\ga\bar{\ga}}(V^\dag_{t_{out},t_{in}})^\ga_{\bar{\al}\al}(V_{t_{out},t_{in}})^{\bar{\ga}}_{\bar{\be}\be}+\\
    &+(\rho^{(2)}_{\tilde{D}_e})_{\al\be|\bar{\al}\bar{\be}} D_{\ga\bar{\ga}} \big[\big(V^\dag_{t_{out},t_0} +V^\dag_{t_0,t_{in}}\tilde{D}_e \big)^\ga_{\bar{\al}\al} \big(V_{t_{out},t_0} +\tilde{D}_e V_{t_0,t_{in}} \big)^{\bar{\ga}}_{\bar{\be}\be}-\\
    &-(V^\dag_{t_0,t_{in}}\tilde{D}_e)^\ga_{\bar{\al}\al} (D_eV_{t_{out},t_0}D_e)^{\bar{\ga}}_{\bar{\be}\be} - (D_eV^\dag_{t_{out},t_0}D_e)^{\ga}_{\bar{\al}\al} (\tilde{D}_eV_{t_0,t_{in}})^{\bar{\ga}}_{\bar{\be}\be}\big]-\\
    &-(\pi_2\rho^{(2)}_{\tilde{D}_e}\pi_2)_{\al\be|\bar{\al}\bar{\be}} D_{\ga\bar{\ga}} (V^\dag_{t_{out},t_0})^\ga_{\bar{\al}\al} (V_{t_{out},t_0})^{\bar{\ga}}_{\bar{\be}\be},
\end{split}
\end{equation}
where the notation introduced in Appendix \ref{Traces_App} has been used. In order to obtain the conditional probability  $P(\hat{\Pi}_D|\hat{\Pi}_{D_e})$ in the case we consider, one just needs to find the probability to detect one electron in the state distinguished by the projector $D_e$ in the leading order of perturbation theory,
\begin{equation}\label{P_spont1}
    P(\hat{\Pi}_{D_e})=\Sp(\hat{R}_e\hat{\Pi}_{D_e})=1-\rho^{(0)}_{\tilde{D}_e},
\end{equation}
and to substitute it to the general formula \eqref{cond_prob_gen}. For $D_e\rightarrow0$, the expressions are considerably simplified
\begin{equation}\label{probab_spont2}
\begin{split}
    P(\hat{\Pi}_D\leftarrow \hat{\Pi}_{D_e})=&\,\rho^{(1)}_{\be\bar{\al}} D_{\ga\bar{\ga}} D^e_{\al\bar{\be}} (V^\dag_{t_0,t_{in}})^{\ga}_{\bar{\al}\al} (V_{t_0,t_{in}})^{\bar{\ga}}_{\bar{\be}\be}+\\
    &+\big[\de_{\al\bar{\be}}\rho^{(2)}_{\be_1\be|\bar{\al}\bar{\be}_1}D^e_{\bar{\be}_1\be_1} -\rho^{(3)}_{\ga_1\al\be|\bar{\al}\bar{\be}\bar{\ga}_1} D^e_{\bar{\ga}_1\ga_1}\big] (V^\dag_{t_{out},t_{in}})^\ga_{\bar{\al}\al} (V_{t_{out},t_{in}})^{\bar{\ga}}_{\bar{\be}\be}-\\
    &-\rho^{(2)}_{\al\be|\bar{\al}\bar{\be}} D_{\ga\bar{\ga}} \big[(V^\dag_{t_{out},t_{in}})^\ga_{\bar{\al}\al} (D_eV_{t_0,t_{in}})^{\bar{\ga}}_{\bar{\be}\be} + (V^\dag_{t_0,t_{in}}D_e)^\ga_{\bar{\al}\al} (V_{t_{out},t_{in}})^{\bar{\ga}}_{\bar{\be}\be}\big],\\
    P(\hat{\Pi}_{D_e})=&\,\rho^{(1)}_{\al\bar{\al}}D^e_{\bar{\al}\al}.
\end{split}
\end{equation}
In the case of spontaneous radiation from a single electron, only the expression on the first line stays in formula \eqref{probab_spont1}, whereas in formula \eqref{probab_spont2} only the first term remains.

One can give the following interpretation to the various terms in \eqref{probab_spont1}. The terms on the first line of \eqref{probab_spont1} describe spontaneous incoherent radiation from the particle the undergone the measurement. The term on the second line of \eqref{probab_spont1} describes radiation, both coherent and incoherent, from the particles whose states have not been measured by the detector (compare with formula (40) of the paper \cite{KazSol2022}). The contribution on the last line describes coherent radiation from particles after the measurement. The contributions on the third and fourth lines of \eqref{probab_spont1} are the interference terms.

To conclude this section, we describe the procedure allowing one to take qualitatively into account a finiteness of the measurement time. Usually, in quantum field theory the states of particles are prepared at $t=t_{in}\rightarrow-\infty$ and are measured at $t=t_{out}\rightarrow\infty$. At that, it is supposed that the interaction Hamiltonian is switched on adiabatically at $t=t_{in}\rightarrow-\infty$ and is switched off at $t=t_{out}\rightarrow\infty$. The adiabatic switching of interaction results in dressing of the vacuum state and the particle states of the free Hamiltonian turning them into the vacuum state and the particle states of the total Hamiltonian with interaction.

In the expressions for the probability of a chain of events and for the conditional probability \eqref{probab_inter_pict}, the projector $\hat{\Pi}_{D_e}$ acts at a certain instant of time and specifies the measurement of states of virtual (bare) particles. The real detector performs measurement during some interval of time $\tau$ and the measured particles have time to become partly dressed. On a qualitative level, this process can be described by the replacement,
\begin{equation}\label{projector_dress}
    \hat{\Pi}_{D_e}(t_0)\rightarrow \hat{S}^{\tau}_{t_0,\infty} \hat{\Pi}_{D_e(t_0)} \hat{S}^{\tau}_{\infty,t_0},
\end{equation}
in the probabilities \eqref{probab_inter_pict}, where
\begin{equation}
    \hat{S}^{\tau}_{\infty,t_0}=\hat{S}_{\infty,t_0}\Big|_{e\rightarrow\la(t)e},\qquad \hat{S}^{\tau}_{\infty,t_0} \hat{S}^{\tau}_{t_0,\infty}=\hat{S}^{\tau}_{t_0,\infty} \hat{S}^{\tau}_{\infty,t_0}=1,
\end{equation}
and $\la(t)$ is an infinitely differentiable function that is equal to zero for $t\geqslant t_0+\tau$ and is equal to unity for $t\leqslant t_0$. If the chain of events contains a larger number of projectors, then every projector is replaced by the expression of the form \eqref{projector_dress} but with its own $\tau$ and $\la(t)$ for every detector. Such a procedure to take into account a finiteness of the measurement time possesses the properties:
\begin{enumerate}
  \item The modified expression \eqref{projector_dress} describing the measurement is a self-adjoint projector;
  \item The instantaneous measurement is reproduced for $\tau=0$, whereas the completely dressed particles are measured for $\tau\rightarrow\infty$;
  \item If the measurement is absent, viz., $\hat{\Pi}_{D_e(t_0)}=1$, then the dressed projector \eqref{projector_dress} is the identity operator;
  \item The Feynman rules are easily modified to include this measurement procedure.
\end{enumerate}

To make the substitution \eqref{projector_dress} in the expressions for probabilities \eqref{probab_inter_pict}, one just replaces the operators by the rule
\begin{equation}
\begin{aligned}
    \hat{S}_{t_0,t_{in}}&\rightarrow\tilde{S}_{t_0,t_{in}}:=\hat{S}^\tau_{\infty,t_0} \hat{S}_{t_0,t_{in}},&\qquad \hat{S}_{t_{in},t_0}&\rightarrow\tilde{S}_{t_{in},t_0}:= \hat{S}_{t_{in},t_0} \hat{S}^\tau_{t_0,\infty},\\
    \hat{S}_{t_{out},t_0}&\rightarrow\tilde{S}_{t_{out},t_0}:= \hat{S}_{t_{out},t_0} \hat{S}^\tau_{t_0,\infty},&\qquad \hat{S}_{t_0,t_{out}}&\rightarrow\tilde{S}_{t_0,t_{out}}:= \hat{S}^\tau_{\infty,t_0} \hat{S}_{t_0,t_{out}},\\
\end{aligned}
\end{equation}
Let the modified $S$-matrix be written as
\begin{equation}
    \tilde{S}_{t_0,t_{in}}=:1+\tilde{V}_{t_0,t_{in}},\qquad \tilde{S}_{t_{out},t_0}=:1+\tilde{V}_{t_{out},t_0},
\end{equation}
in the nontrivial leading order of the perturbation theory. As long as
\begin{equation}
    \tilde{S}_{t_{out},t_0}\tilde{S}_{t_0,t_{in}}=\hat{S}_{t_{out},t_{in}},
\end{equation}
we have
\begin{equation}\label{tilde_V}
    \tilde{V}_{t_{out},t_0}+\tilde{V}_{t_0,t_{in}} =\hat{V}_{t_{out},t_{in}},
\end{equation}
in the leading order of perturbation theory. Then one can verify that the formulas obtained above in this and preceding sections hold true provided one replaces everywhere
\begin{equation}
\begin{gathered}
    (V_{t_0,t_{in}})^{\bar{\ga}}_{\bar{\al}\al} \rightarrow (\tilde{V}_{t_0,t_{in}})^{\bar{\ga}}_{\bar{\al}\al},\qquad (V_{t_{out},t_0})^{\bar{\ga}}_{\bar{\al}\al}\rightarrow (\tilde{V}_{t_{out},t_0})^{\bar{\ga}}_{\bar{\al}\al},\\
    (V_{t_{out},t_{in}})^{\bar{\ga}}_{\bar{\al}\al}\rightarrow (\tilde{V}_{t_{out},t_0})^{\bar{\ga}}_{\bar{\al}\al} + (\tilde{V}_{t_0,t_{in}})^{\bar{\ga}}_{\bar{\al}\al} =(V_{t_{out},t_{in}})^{\bar{\ga}}_{\bar{\al}\al}.
\end{gathered}
\end{equation}
The last equality follows from \eqref{tilde_V}.

\section{Examples}\label{Examples_Sec}
\subsection{Stimulated radiation from free particles}\label{Stim_Rad_Free_Part}

As the simplest example of radiation due to measurement, we consider stimulated radiation from the free Dirac particles whose state is prepared at $t=t_{in}\rightarrow-\infty$ and is measured at the instant of time $t=t_0=0$. The radiated photons are recorded at $t=t_{out}\rightarrow+\infty$ and the Dirac particles are not detected at $t=t_{out}$. We assume that the measurement of the state of  Dirac particles is carried out instantaneously and provide the estimates when such an approximation is justified.

The energy-momentum conservation law for the free Dirac particles implies that the amplitude $V^{\bar{\ga}}_{t_{out},t_{in}}=0$. As a result, the amplitude of stimulated radiation \eqref{class_amplitude_gen} is reduced to
\begin{equation}\label{amplit_stim}
    \mathcal{A}^{\bar{\ga}}=\frac{1}{1-\rho^{(0)}_{\tilde{D}_e}} (\tilde{D}_e\rho^{(1)}_{\tilde{D}_e}D_e)_{\al\bar{\al}}(V_{t_0,t_{in}})^{\bar{\ga}}_{\bar{\al}\al},
\end{equation}
where the condensed notation has been used. For example,
\begin{equation}
    \sum_\al\equiv \sum_s\int\frac{Vd\spp}{(2\pi)^3}.
\end{equation}
The amplitude of photon radiation during the time interval $t\in[t_{in},t_0]$ is
\begin{equation}\label{V_1}
    (V_{t_0,t_{in}})^{\bar{\ga}}_{\bar{\al}\al}=-\frac{iem}{V} \int_{-\infty}^0 dx^0\int d\spx\frac{ \bar{u}_{\bar{\al}}\Ga^i u_\al f^*_{(\la)i}(\spk)}{\sqrt{2Vk_0p'_0p_0}} e^{i(k+p'-p)x}= -iem\frac{(2\pi)^3}{V} \frac{\de(\spk-\spp+\spp')}{\sqrt{2Vk_0p'_0p_0}} \frac{\bar{u}_{\bar{\al}}\Ga^i u_\al f^*_{(\la)i}(\spk)}{i(k_0-p_0+p_0'-i0)},
\end{equation}
where $\mathbf{f}_{(\la)}(\spk)$ is the polarization vector of photon with momentum $\spk$, $m$ is the mass of Dirac particle, $\al=(s,\spp)$, $\bar{\al}=(s',\spp')$, and \cite{KRS2023}
\begin{equation}\label{buGammaU}
    \bar{u}_{\bar{\al}}\Ga^i u_\al=\frac12\big[\de_{s's}\tilde{G}^i(\spp,\spp') +(\s_a)_{s's}\tau_a^j\tilde{Z}^{ji}(\spp,\spp')\big].
\end{equation}
The set of vectors $\tau_a^i$, $a=\overline{1,3}$, specifies the right-handed tetrad transforming the vectors on the Poincar\'{e} sphere to the vectors in the $x$-space. The explicit expressions for these vectors are presented in formula (30) of the paper \cite{KRS2023}. The explicit expressions for $\tilde{G}^i(\spp,\spp')$ and $\tilde{Z}^{ji}(\spp,\spp')$ are given in formulas (A4) and (A11) of the paper \cite{KRS2023}, where one has to make the replacement
\begin{equation}
    k_c^\mu\rightarrow p_c^\mu,\qquad q^\mu\rightarrow-q^\mu.
\end{equation}
Here $q_\mu=p_\mu-p'_\mu$ and $p_c^\mu=(p_\mu+p'_\mu)/2$.

Further, we need to substitute the amplitude of photon radiation into \eqref{amplit_stim}. Let us write the projectors and the one-particle density matrices entering into \eqref{amplit_stim} as
\begin{equation}\label{projectors}
\begin{gathered}
    \tilde{D}^e_{\al\bar{\be}}=\frac{(2\pi)^3}{V}\frac{\tilde{D}_e(\spp,\bar{\spp}_1)}{2} [1+(\bs\s\tilde{\bs\zeta}(\spp,\bar{\spp}_1))]_{s\bar{s}_1},\qquad (\rho^{(1)}_{\tilde{D}_e})_{\bar{\be}\be}=\frac{(2\pi)^3}{V} \frac{\rho^{(1)}_{\tilde{D}_e}(\bar{\spp}_1,\spp_1)}{2}[1+(\bs\s\bs\xi(\bar{\spp}_1,\spp_1))]_{\bar{s}_1s_1},\\ D^e_{\be\bar{\al}}=\frac{(2\pi)^3}{V}\frac{D_e(\spp_1,\spp')}{2} [1+(\bs\s\bs\zeta(\spp_1,\spp'))]_{s_1s'}.
\end{gathered}
\end{equation}
Substituting expressions \eqref{V_1} and \eqref{projectors} into \eqref{amplit_stim}, we obtain
\begin{equation}
    \mathcal{A}^{\bar{\ga}}=-\frac{em}{16(1-\rho^{(0)}_{\tilde{D}_e})}\frac{f^*_{(\la)i}(\spk)}{\sqrt{2Vk_0}}\int\frac{d\spp_c d\spp_1 d\bar{\spp}_1}{\sqrt{p_0p'_0}} \frac{\tilde{D}_e \rho^{(1)}_{\tilde{D}_e} D_e }{k_0-q_0} \Sp\big[ (1+(\bs\s\tilde{\bs\zeta}))(1+(\bs\s\bs\xi)) (1+(\bs\s\bs\zeta)) (\tilde{G}^i +(\bs\s\bs\tau^j)\tilde{Z}^{ji}) \big],
\end{equation}
where $\spp=\spp_c+\spk/2$, $\spp'=\spp_c-\spk/2$, and $\spq=\spk$. Evaluating the trace, we arrive at
\begin{equation}\label{amplit_stim_1}
\begin{split}
    \mathcal{A}^{\bar{\ga}}=&-\frac{em}{8(1-\rho^{(0)}_{\tilde{D}_e})}\frac{f^*_{(\la)i}(\spk)}{\sqrt{2Vk_0}}\int\frac{d\spp_c d\spp_1 d\bar{\spp}_1}{\sqrt{p_0p'_0}} \frac{ \tilde{D}_e \rho^{(1)}_{\tilde{D}_e} D_e  }{k_0-q_0} \big[ (1+(\tilde{\bs\zeta}\bs\zeta) +(\bs\xi,\bs\zeta+\tilde{\bs\zeta}) +i(\bs\xi,\bs\zeta,\tilde{\bs\zeta}))\tilde{G}^i+\\
    &+(\bs\xi+\bs\zeta+\tilde{\bs\zeta} +\bs\zeta(\tilde{\bs\zeta}\bs\xi) +\tilde{\bs\zeta}(\bs\zeta\bs\xi)-\bs\xi(\bs\zeta\tilde{\bs\zeta}) +i[\bs\xi,\bs\zeta-\tilde{\bs\zeta}] +i[\tilde{\bs\zeta},\bs\zeta])^j\tilde{Z}^{ji} \big].
\end{split}
\end{equation}
Henceforth, the vectors contracted with the tetrad $\tau^i_a$ are denoted by the same letters. For example,
\begin{equation}
    \xi_a\tau^i_a=\xi^i,\qquad \zeta_a\tau^i_a=\zeta^i.
\end{equation}
Notice that in the small quantum recoil limit, $|\spk|\ll p_0^c$, there are the approximate equalities (see \cite{pra103,KRS2023})
\begin{equation}\label{small_recoil}
\begin{gathered}
    p_0\approx p^0_c+(\bs\be_c\spk)/2,\qquad p'_0\approx p^0_c-(\bs\be_c\spk)/2,\qquad p_0p'_0\approx (p_0^c)^2=m^2+\spp_c^2,\\
    k_0-q_0\approx k_0(1-(\spn\bs\be_c)),\qquad
    \tilde{G}^i(\spp,\spp')=\frac{2p_c^i}{m}F_e(q^2)+O(\spq^2/m^2),\\
    \tilde{Z}^{ij}(\spp,\spp')=-\frac{i}{m}\Big[\e^{ijl}q_l F_m-\frac{\e^{ijl}q_0p^c_l}{p^0_c+m}F_m -\frac{p_c^j\e^{ikl}q_kp^c_l}{m(p_0^c+m)} (F_e-F_m)\Big]+O(\spq^2/m^2),
\end{gathered}
\end{equation}
where $\e^{123}=1$, $\spn:=\spk/|\spk|$, $\bs\be_c=\spp_c/p^0_c$, and $p^0_c=p^0|_{\spp=\spp_c}$. In the nonrelativistic limit, $|\spp_c|\ll p^c_0$, only the first term in the square brackets in the expression for $\tilde{Z}^{ij}$ is left, provided $|F_m|$ is not anomalously small in comparison with $|F_e|$. It is clear from the above expressions for $\tilde{G}^i$ and $\tilde{Z}^{ij}$ that the contributions proportional to $\tilde{G}^i$ describe the radiation from the distribution of charge while the contributions proportional to $\tilde{Z}^{ji}$ are related to the radiation from the magnetic moment of the Dirac particle. Recall that $F_e(q^2)$ defines the distribution of charge, $F_m(q^2)$ specifies the distribution of magnetic moment, and $F_m(q^2)-F_e(q^2)$ is responsible for the distribution of the anomalous magnetic moment \cite{LandLifQED}. As for the electrons without the anomalous magnetic moment, $F_e(0)=F_m(0)=1$.

In some cases that we will consider below, there appears to be more useful the representation of the amplitude of stimulated radiation given in \eqref{class_amplitude_gen}. This expression contains
\begin{equation}\label{amplit_stim_first_t}
    (\rho^{(1)}_{\tilde{D}_e}D_e)_{\al\bar{\al}}(V_{t_0,t_{in}})^{\bar{\ga}}_{\bar{\al}\al}=-\frac{em}{4} \frac{f^*_{(\la)i}(\spk)}{\sqrt{2Vk_0}}\int\frac{d\spp_c d\spp_1 }{\sqrt{p_0p'_0}} \frac{ \rho^{(1)}_{\tilde{D}_e}(\spp,\spp_1) D_e(\spp_1,\spp') }{k_0-q_0} \big[ (1+ (\bs\xi\bs\zeta))\tilde{G}^i+(\bs\xi+\bs\zeta +i[\bs\xi,\bs\zeta])^j\tilde{Z}^{ji} \big],
\end{equation}
where $\bs\xi\equiv\bs\xi(\spp,\spp_1)$ and the other notation is the same as in formulas \eqref{projectors} and \eqref{amplit_stim_1}. The amplitude of photon radiation during the time interval $t\in[t_0,t_{out}]$ equals
\begin{equation}\label{V_2}
    (V_{t_{out},t_0})^{\bar{\ga}}_{\bar{\al}\al}= iem\frac{(2\pi)^3}{V} \frac{\de(\spk-\spp+\spp')}{\sqrt{2Vk_0p'_0p_0}} \frac{\bar{u}_{\bar{\al}}\Ga^i u_\al f^*_{(\la)i}(\spk)}{i(k_0-p_0+p_0'-i0)}.
\end{equation}
This amplitude enters into another contribution to the amplitude of stimulated radiation,
\begin{equation}
\begin{split}
    (D_e\rho^{(1)}_{\tilde{D}_e}D_e)_{\al\bar{\al}}(V_{t_{out},t_0})^{\bar{\ga}}_{\bar{\al}\al}=&\,\frac{em}{8} \frac{f^*_{(\la)i}(\spk)}{\sqrt{2Vk_0}}\int\frac{d\spp_c d\spp_1 d\bar{\spp}_1}{\sqrt{p_0p'_0}} \frac{ D_e(\spp,\bar{\spp}_1) \rho^{(1)}_{\tilde{D}_e}(\bar{\spp}_1,\spp_1) D_e(\spp_1,\spp') }{k_0-q_0}\times\\
    &\times\big[ \big(1+(\tilde{\bs\zeta}\bs\zeta) +(\bs\xi,\bs\zeta+\tilde{\bs\zeta})+i(\bs\xi,\bs\zeta,\tilde{\bs\zeta})\big)\tilde{G}^i+\\
    &+\big(\bs\xi+\bs\zeta+\tilde{\bs\zeta} +\bs\zeta(\tilde{\bs\zeta}\bs\xi) +\tilde{\bs\zeta}(\bs\zeta\bs\xi)-\bs\xi(\bs\zeta\tilde{\bs\zeta}) +i[\bs\xi,\bs\zeta-\tilde{\bs\zeta}] +i[\tilde{\bs\zeta},\bs\zeta]\big)^j\tilde{Z}^{ji} \big],
\end{split}
\end{equation}
where $\tilde{\bs\zeta}=\bs\zeta(\spp,\bar{\spp}_1)$ and the other notation is the same as in formulas \eqref{projectors} and \eqref{amplit_stim_1}.

A more comprehensive physical interpretation of the terms in the radiation amplitude will be given below in considering the particular cases of the general formulas obtained. Nevertheless, it is seen even now that the radiation amplitude contains the factor $1/(k_0-q_0)$ which is typical for edge and transition radiations \cite{GaribYang,BazylZhev,Ginzburg,AkhShul,Bord.1,GKSSchYu1,GKSSchYu2}. For comparison, we provide here the amplitude of edge radiation created by the classical Dirac current for the one-particle state with the density matrix $\rho_{ss'}(\spp,\spp')$ of the form \eqref{projectors}, edge radiation being generated in passing through the spacelike hypersurface $t=0$ in the spacetime. This amplitude reads
\begin{equation}\label{class_trans_ampl}
    \mathcal{A}^{\bar{\ga}}_c=-\frac{em}{2} \frac{f^*_{(\la)i}(\spk)}{\sqrt{2Vk_0}}\int\frac{d\spp_c}{\sqrt{p_0p'_0}} \frac{ \rho(\spp,\spp') }{k_0-q_0} (\tilde{G}^i+\xi^j\tilde{Z}^{ji}),
\end{equation}
and is obtained from \eqref{amplit_stim_first_t} by the replacement
\begin{equation}
    \rho^{(1)}_{\tilde{D}_e}(\spp,\spp_1)\rightarrow\rho(\spp,\spp_1),\qquad D_e(\spp_1,\spp') \rightarrow 2\de(\spp_1-\spp'),\qquad \bs\zeta\rightarrow0.
\end{equation}
To provide a clearer physical interpretation to the formulas we derive, we introduce the Weyl symbol of the density matrix (the Wigner function)
\begin{equation}\label{Wigner_func}
\begin{gathered}
    \rho_{ss'}(\spx,\spp_c)\equiv\frac{\rho(\spx,\spp_c)}{2}\big[1+(\bs\s\bs\xi(\spx,\spp_c))\big]_{ss'}=\int d\spq e^{i\spq\spx} \rho_{ss'}\Big(\spp_c+\frac{\spq}{2},\spp_c-\frac{\spq}{2}\Big),\\
    \rho_{ss'}(\spp,\spp')=\int\frac{d\spx}{(2\pi)^3} e^{-i(\spp-\spp')\spx} \rho_{ss'}\Big(\spx,\frac{\spp+\spp'}{2}\Big).
\end{gathered}
\end{equation}
In that case,
\begin{equation}\label{class_trans_ampl_Wig}
    \mathcal{A}^{\bar{\ga}}_c=-\frac{em}{2}\int\frac{d\spx d\spp_c}{(2\pi)^3\sqrt{p_0p'_0}}  \frac{f^*_{(\la)i}(\spk)e^{-i\spk\spx}}{\sqrt{2Vk_0}} \frac{ \rho(\spx,\spp_c) }{k_0-q_0} (\tilde{G}^i+\xi^j\tilde{Z}^{ji}),
\end{equation}
where $\bs\xi=\bs\xi(\spx,\spp_c)$. It is supposed in \eqref{class_trans_ampl} and \eqref{class_trans_ampl_Wig} that the radiation is generated by the Dirac current of the particle that either disappears at the instant of time $t=0$ or the spatial components of the electric current produced by this particle vanish for $t\geqslant0$.

In order to see that formula \eqref{class_trans_ampl} reproduces the standard expression for the amplitude of edge radiation in the classical limit, we assume that the state of the Dirac particle is unpolarized, $\bs\xi=0$, and $|\spk|=|\spp-\spp'|\ll|\spp_c|$. For the Gaussian wave packet of the form
\begin{equation}\label{Gaussian_wp}
    \rho(\spp,\spp')=(2\pi\s^2)^{-3/2} e^{-\frac{(\spp-\spp_0)^2}{4\s^2} -\frac{(\spp'-\spp_0)^2}{4\s^2}- i\spx_0(\spp-\spp')}= (2\pi\s^2)^{-3/2} e^{-\frac{(\spp_c-\spp_0)^2}{2\s^2} -\frac{\spk^2}{8\s^2}- i\spx_0\spk},
\end{equation}
the last condition is fulfilled when $|\spp_0|\gg\s$. The Wigner function for the wave packet \eqref{Gaussian_wp} becomes
\begin{equation}
    \rho(\spx,\spp_c)=8e^{-\frac{(\spp_c-\spp_0)^2}{2\s^2}}e^{-2\s^2(\spx-\spx_0)^2}.
\end{equation}
Then employing relations \eqref{small_recoil}, the amplitude \eqref{class_trans_ampl} is reduced to
\begin{equation}
    \mathcal{A}^{\bar{\ga}}_c=-e \frac{f^*_{(\la)i}(\spk)}{\sqrt{2Vk_0^3}} e^{ -\frac{\spk^2}{8\s^2}- i\spx_0\spk} \int\frac{d\spp_c \be_c^i}{(2\pi\s^2)^{3/2}} \frac{ \exp[-\frac{(\spp_c-\spp_0)^2}{2\s^2}] }{1-(\spn\bs\be_c)}\approx \frac{e}{\sqrt{2Vk_0^3}}  \frac{ \mathbf{f}_{(\la)}^*(\spk)\bs\be_0 }{1-(\spn\bs\be_0)}  e^{ -\frac{\spk^2}{8\s^2}- i\spx_0\spk}.
\end{equation}
If $|\spk|\ll\s$, one may put $\spk^2/(8\s^2)\approx0$. As a result, the standard expression for the amplitude of edge radiation from the bunch of particles concentrated near the point $\spx_0$ at the instant of time $t=0$ with Gaussian distribution with respect to momenta $\spp_c$ is revealed.

The radiation amplitude \eqref{amplit_stim_1} includes the factor $1/(k_0-q_0)$ that declines slowly as a power with increasing the photon energy. This property is a consequence of the approximation of instantaneous measurement of the intermediate state of the Dirac particle. If one takes into account a finiteness of the measurement time as it was done at the end of the previous section, then in evaluating the amplitude \eqref{amplit_stim_1} one should make the replacement
\begin{equation}
    \int_{-\infty}^0 dte^{i(k_0-q_0)t}=\frac{-i}{k_0-q_0}\rightarrow \int_{-\infty}^{\tau} dt\la(t)e^{i(k_0-q_0)t}= \frac{i}{k_0-q_0} \int_0^\tau dt \lambda'(t)e^{i(k_0-q_0)t},
\end{equation}
where $\la(t)=1$ for $t\leqslant0$, $\la(t)=0$ for $t\geqslant\tau$, and $\la(t)$ is an infinitely differentiable function. Now it is clear that to account for a finiteness of the measurement time, it is sufficient to multiply the amplitude \eqref{amplit_stim_1} by
\begin{equation}\label{Fourier_lambda_pr}
    -\int_0^\tau dt\lambda'(t)e^{i(k_0-q_0)t},
\end{equation}
where
\begin{equation}
    -\int_0^\tau dt\lambda'(t)=1.
\end{equation}
The function $\lambda'(t)$ is infinitely differentiable and is equal to zero out of the interval $[0,\tau]$. Therefore, the Fourier transform in formula \eqref{Fourier_lambda_pr} tends to zero for large $|k_0-q_0|$ faster than any power of $1/|k_0-q_0|$ (see, e.g., \cite{GSh}). If the typical time of measurement is $\tau$, then the radiation amplitude \eqref{amplit_stim_1} accounting for a finiteness of the measurement time rapidly tends to zero when
\begin{equation}\label{formation_time}
    \tau\gg t_f,\qquad t_f:=1/(k_0-q_0)\approx1/[k_0(1-(\spn\bs\be_c))].
\end{equation}
In the opposite case $\tau\ll t_f$, one may put $\tau=0$ in calculating the radiation amplitude and so expression \eqref{amplit_stim_1} is valid. The quantity $t_f$ is the formation time of edge and transition radiations \cite{Ginzburg}. In the ultrarelativistic case, $\be_3\approx1$ and $|\bs\be_\perp|\ll1$, supposing that $n_3\approx1$ and $|\spn_\perp|\ll1$, we have
\begin{equation}\label{formation_time1}
    t_f\approx\frac{2\ga^2}{k_0(1+(\bs\be_\perp-\spn_\perp)^2\ga^2)}.
\end{equation}
Due to the factor $\ga^2$, this quantity can be rather large in the case when $|\bs\be_\perp-\spn_\perp|\lesssim1/\ga$. As a result, we deduce that for
\begin{equation}
    k_0\tau\frac{1+(\bs\be_\perp-\spn_\perp)^2\ga^2}{2\ga^2}\ll1,
\end{equation}
it is justified to use expression \eqref{amplit_stim_1} for the radiation amplitude.

In measuring the intermediate state of a Dirac particle, an abrupt change of its wave function occurs. Hence, one may expect that the high-energy photons will be radiated in such a process. Let us estimate the maximum photon energies that can be achieved in this process. The above estimates show that if the momenta in the initial state of the particle are concentrated near the value $\spp$ with the spread $|\de\spp|\sim\s$ and the momenta in the state after measurement are located near the value $\spp'$ with the spread $|\de\spp'|\sim\s'$, then the amplitude of stimulated radiation tends rapidly to zero for the radiated photon energies
\begin{equation}
    k_0\gtrsim \s+\s'+|\spp-\spp'|.
\end{equation}
Furthermore, it also vanishes rapidly when $\tau\gtrsim t_f$.

\subsubsection{Spin measurement}\label{Stim_Rad_Spin_Measur}

Consider the particular cases of the formulas obtained above. At first, we investigate the case when only the projection of the Dirac particle spin is measured at the instant of time $t=0$. In this case, it is convenient to use formula \eqref{amplit_stim_1}, where
\begin{equation}\label{zeta_zetap}
\begin{gathered}
    D_e(\spp_1,\spp')=\de(\spp_1-\spp'),\qquad\tilde{D}_e(\spp,\bar{\spp}_1)=\de(\spp-\bar{\spp}_1),\\
    \bs\zeta(\spp_1,\spp')\equiv \bs\zeta(\spp'),\qquad\tilde{\bs\zeta}(\spp,\bar{\spp}_1)\equiv\tilde{\bs\zeta}(\spp)=-\bs\zeta(\spp).
\end{gathered}
\end{equation}
For brevity, we denote $\bs\zeta(\spp)\equiv\bs\zeta$ and $\bs\zeta(\spp')\equiv\bs\zeta'$. Notice that $|\bs\zeta|=|\bs\zeta'|=1$. Substituting \eqref{Wigner_func} and \eqref{zeta_zetap} into general expression \eqref{amplit_stim_1} for the amplitude of stimulated radiation, we have
\begin{equation}\label{amplit_stim_spin}
\begin{split}
    \mathcal{A}^{\bar{\ga}}=&-\frac{em}{8(1-\rho^{(0)}_{\tilde{D}_e})} \int\frac{d\spx d\spp_c }{\sqrt{p_0p'_0}} \frac{f^*_{(\la)i}(\spk)e^{-i\spk\spx}}{\sqrt{2Vk_0}}  \frac{\rho^{(1)}_{\tilde{D}_e}(\spx,\spp_c)  }{k_0-q_0} \big[ \big(1-(\bs\zeta\bs\zeta') +(\bs\xi,\bs\zeta'-\bs\zeta) +i(\bs\xi,\bs\zeta,\bs\zeta')\big)\tilde{G}^i+\\
    &+\big(\bs\xi+\bs\zeta'-\bs\zeta -\bs\zeta'(\bs\zeta\bs\xi) -\bs\zeta(\bs\zeta'\bs\xi)+\bs\xi(\bs\zeta\bs\zeta') +i[\bs\xi,\bs\zeta'+\bs\zeta] -i[\bs\zeta,\bs\zeta']\big)^j\tilde{Z}^{ji}) \big],
\end{split}
\end{equation}
where $\bs\xi=\bs\xi(\spx,\spp_c)$. In the small recoil limit and for $\bs\zeta\approx\bs\zeta'$, we come to
\begin{equation}\label{amplit_stim_spin_a}
    \mathcal{A}^{\bar{\ga}}=-\frac{em}{4(1-\rho^{(0)}_{\tilde{D}_e})} \int\frac{d\spx d\spp_c }{p^0_c} \frac{f^*_{(\la)i}(\spk)e^{-i\spk\spx}}{\sqrt{2Vk_0^3}}  \frac{\rho^{(1)}_{\tilde{D}_e}(\spx,\spp_c)  }{1-(\spn\bs\be_c)} (\bs\kappa +i[\bs\kappa,\bs\zeta])^j\tilde{Z}^{ji},
\end{equation}
where $\bs\kappa:=\bs\xi-\bs\zeta(\bs\zeta\bs\xi)$. As expected on physical grounds, in measuring spin, the radiation is determined  only by the magnetic moment of the particles. It is seen that for $\bs\xi\parallel\bs\zeta$ the radiation amplitude $\mathcal{A}^{\bar{\ga}}=0$ in the approximation considered. This is the case of the so-called quantum nondemolition measurement \cite{BraginskiKhalili1992}. Moreover, it follows from \eqref{amplit_stim_spin_a} that, in rotating the vector $\bs\kappa$ around the vector $\bs\zeta$ by an angle of $\vf$, the amplitude of stimulated radiation \eqref{amplit_stim_spin_a} is changed only by the common phase factor $\exp(i\vf)$.

If $\bs\xi(\spx,\spp_c)$ and $\bs\zeta(\spp)$ can be approximately regarded as constant vectors, then the abovementioned property implies that the polarization of radiation described by this amplitude does not depend on the polarization vector of the measured state of the Dirac particle $\bs\xi(\spx,\spp_c)$. The component of this vector orthogonal to $\bs\zeta(\spp)$, i.e., $\bs\kappa$, determines the intensity of stimulated radiation due to measurement, whereas the direction of the vector $\bs\kappa$ in the plane orthogonal to $\bs\zeta$ affects only the interference with the incoming coherent radiation described by the complex amplitude $d_{\bar{\ga}}$ (see \eqref{cond_prob_stim1}).

Let us find the polarization of the radiation described by the amplitude \eqref{amplit_stim_spin_a} for $\bs\xi(\spx,\spp_c)\approx const$ and $\bs\zeta(\spp)\approx const$ in the nonrelativistic and ultrarelativistic limits. In the small quantum recoil approximation, the polarization of radiation is specified by the factor
\begin{equation}
    \chi_{\la}:=(\bs\kappa +i[\bs\kappa,\bs\zeta])^j\tilde{Z}^{ji}f^*_{(\la)i}(\spk)\approx -\frac{i}{m}(\bs\kappa +i[\bs\kappa,\bs\zeta])^j \Big[\e^{jil}k^l F_m -\frac{\e^{jil}p_c^l}{p_c^0+m}(\bs\be_c\spk) F_m +\frac{p^i_c\e^{jkl} k^kp^l_c}{m(p_c^0+m)} a \Big] f_{(\la)}^{*i}(\spk) ,
\end{equation}
where $a:=F_e-F_m$.

In the nonrelativistic limit, $|\spp|\ll m$, the main contribution to $\chi_{\la}$ comes from the first term in the square brackets,
\begin{equation}
    \chi_\la\approx-\frac{ik_0}{m}F_m (\bs\kappa+i[\bs\kappa,\bs\zeta],\mathbf{f}^*_{(\la)}(\spk),\spn),
\end{equation}
where $(\mathbf{a},\mathbf{b},\mathbf{c})$ means a triple product of vectors. Choosing the $z$ axis of the spherical system of coordinates to be directed along the vector $\bs\zeta$ and
\begin{equation}\label{polarization_vect}
\begin{gathered}
    \mathbf{f}_{(1)}=(\cos\phi\cos\theta,\sin\phi\cos\theta,-\sin\theta),\qquad \mathbf{f}_{(2)}=(-\sin\phi,\cos\phi,0),\\ \spn=(\sin\theta\cos\phi,\sin\theta\sin\phi,\cos\theta),\qquad\bs\kappa=|\bs\kappa|(\cos\vf,\sin\vf,0),
\end{gathered}
\end{equation}
we obtain
\begin{equation}
    \chi=-\frac{ik_0}{m} F_m |\bs\kappa|e^{i(\vf-\phi)}
    \left[
      \begin{array}{c}
        i \\
        \cos\theta \\
      \end{array}
    \right].
\end{equation}
As a result,
\begin{equation}
    \chi\chi^\dag= \frac{k^2_0 \bs\kappa^2}{2m^2} F_m^2(1+\cos^2\theta)(1+(\mathbf{b}\bs\s)),
\end{equation}
where
\begin{equation}\label{Stokes_vect_nonrel}
    b_2=-\frac{2\cos\theta}{1+\cos^2\theta},\qquad b_1=0,\qquad b_3=\frac{\sin^2\theta}{1+\cos^2\theta}.
\end{equation}
The vector $\mathbf{b}$ defines the Stokes parameters of the radiation described by the amplitude \eqref{amplit_stim_spin_a}. It is clear that $\cos\theta=(\bs\zeta\spn)$ and $\sin^2\theta=\bs\zeta_\perp^2=1-(\bs\zeta\spn)^2$. As we see, stimulated radiation due to measurement of spin projection possesses a circular polarization for $\bs\zeta\parallel\spn$. As for the other values of $\bs\zeta$, there is an admixture of linear polarization along the vector $\mathbf{f}_{(1)}$, i.e., in the plane spanned by the vectors $\bs\zeta$ and $\spn$. In the case $\bs\zeta\perp\spn$, the radiation is completely linearly polarized along the vector $\mathbf{f}_{(1)}$.

In the ultrarelativistic limit, $|\spp_c|\gg m$, we assume that the wave packet of the Dirac particle moves approximately along the $z$ axis and is sufficiently narrow with respect to the transverse momentum components. Consider the range of parameters where the main part of radiation is concentrated. In this case (see, e.g., \cite{GaribYang,BazylZhev,Ginzburg,AkhShul,Bord.1,GKSSchYu1,GKSSchYu2}),
\begin{equation}\label{ultrarel_param}
    |\spn_\perp|\lesssim\max(1/\ga,\be^c_\perp),\qquad |\bs\be^c_\perp|\ll1,\qquad (\spn\bs\be_c)\approx1-\frac{1+(\bs\be_\perp^c-\spn_\perp)^2\ga^2}{2\ga^2}, \qquad\be^c_3\approx1-\frac{1+(\be_\perp^c\ga)^2}{2\ga^2},
\end{equation}
where $\ga=p_0^c/m$. Take the $z$ axis to be along $\spn$. Then, in the leading order in $1/\ga$, we arrive at
\begin{equation}
    \chi_\la=-\frac{ik_0}{m}\big[F_m(\bs\kappa+i[\bs\kappa,\bs\zeta],\mathbf{f}^*_{(\la)}(\spk),\frac{\spn}{\ga}-\bs\be^c_\perp) +a\gamma(\bs\kappa+i[\bs\kappa,\bs\zeta],\spn,\bs\be^c_\perp)(\mathbf{f}^*_{(\la)}(\spk),\bs\be^c_\perp) \big],
\end{equation}
where $\bs\be^c_\perp=\bs\be^c-\spn(\spn\bs\be^c)$. As it has been mentioned above, the rotation of the vector $\bs\kappa$ around the vector $\bs\zeta$ by an angle of $\vf$ results in multiplication of $\chi_\la$ by the phase factor $\exp(i\vf)$. It is useful to rotate the vector $\bs\kappa$ so that
\begin{equation}
    \bs\kappa=\frac{|\bs\kappa|}{\zeta_\perp}[\bs\zeta,\spn].
\end{equation}
The vectors of linear polarization are convenient to choose as
\begin{equation}\label{linear_polar_bas}
    \mathbf{f}_{(1)}=\frac{\bs\be^c_\perp}{\be^c_\perp},\qquad\mathbf{f}_{(2)}=\frac{[\spn,\bs\be^c_\perp]}{\be^c_\perp}.
\end{equation}
Then
\begin{equation}
    \chi=-\frac{ik_0}{m\ga}\frac{|\bs\kappa|e^{i\vf}}{\zeta_\perp}
    \left[
      \begin{array}{c}
        (F_m-(\be^c_\perp\ga)^2 a) (\zeta_1+i\zeta_2\zeta_3) \\
        F_m[\zeta_2 -i(\zeta_1\zeta_3-\be_\perp^c\ga(1-\zeta_3^2))] \\
      \end{array}
    \right],
\end{equation}
where the components of the vector $\bs\zeta$ are given in the basis $\{\mathbf{f}_{(1)},\mathbf{f}_{(2)},\spn\}$. It is not difficult to find the Stokes parameters of stimulated radiation due to measurement of spin projection
\begin{equation}
    \chi\chi^\dag=\frac{k_0^2\bs\kappa^2}{2m^2\ga^2}A(1+(\mathbf{b}\bs\s)),
\end{equation}
where
\begin{equation}\label{A_b1_b2_b3}
\begin{split}
    A=&\,\big(F_m-(\be^c_\perp\ga)^2 a\big)^2(1-\zeta_2^2) +F_m^2\big(1+(\be_\perp^c\ga)^2 -(\zeta_1+\be_\perp^c\ga\zeta_3)^2\big),\\ b_2=&-\frac{2}{A}F_m(F_m-(\be^c_\perp\ga)^2a)(\zeta_3-\be^c_\perp\ga\zeta_1),\\
    b_3+ib_1=&\,\frac{1}{A}\big[\big( F_m(\zeta_1+\be^c_\perp\ga\zeta_3) +i(F_m-(\be^c_\perp\ga)^2a)\zeta_2 \big)^2  -(\be^c_\perp\ga)^2\big(F_e^2 -a^2(1+(\be^c_\perp\ga)^2)\big) \big]=\\
    =&\,\frac{1}{A}\big[F_m\big(\zeta_1+i\zeta_2 +\be^c_\perp\ga(1+\zeta_3)\big) -(\be^c_\perp\ga)^2a\frac{\zeta_1-i\zeta_2\zeta_3}{1-\zeta_3}\big]\times\\
    &\times\big[F_m\big(\zeta_1+i\zeta_2 -\be^c_\perp\ga(1-\zeta_3)\big) -(\be^c_\perp\ga)^2a\frac{\zeta_1+i\zeta_2\zeta_3}{1+\zeta_3}\big].
\end{split}
\end{equation}
In the particular case $\be^c_\perp\ga\ll1$, the Stokes vector $\mathbf{b}$ coincides with the Stokes vector \eqref{Stokes_vect_nonrel}. In the opposite limit, when $\be_\perp^c\ga\gg1$ and $\be_\perp^c\ga|a|\sqrt{1-\zeta_2^2}\gg|F_m|$, we have
\begin{equation}\label{A_b1_b2_b3_pc1}
    A\approx(\be^c_\perp\ga)^4a^2(1-\zeta_2^2),\qquad b_2\approx-\frac{2F_m\zeta_1}{\be^c_\perp\ga a(1-\zeta_2^2)},\qquad b_3+ib_1\approx1.
\end{equation}
In other words, in this case the contribution to the amplitude of stimulated radiation is completely linearly polarized in the plane spanned by the vectors $\{\bs\be^c,\mathbf{n}\}$.

In the case of negligible anomalous magnetic moment, $|a|\ll|F_m|$, we deduce
\begin{equation}\label{A_b1_b2_b3_pc2}
\begin{gathered}
    A=F_m^2\big[1+(\be^c_\perp\ga)^2\zeta_2^2+(\zeta_3-\be^c_\perp\ga\zeta_1)^2 \big],\\
    b_2=-\frac{2}{A}F^2_m(\zeta_3-\be^c_\perp\ga\zeta_1),\qquad
    b_3+ib_1=\frac{F_m^2}{A}\big(\zeta_1+i\zeta_2 +\be^c_\perp\ga(1+\zeta_3)\big) \big(\zeta_1+i\zeta_2 -\be^c_\perp\ga(1-\zeta_3)\big).
\end{gathered}
\end{equation}
For $\be_\perp^c\ga\sqrt{1-\zeta_3^2}\gg1$, the terms standing at $\be_\perp^c\ga$ dominate in $A$ and $\mathbf{b}$. In this case, the Stokes vector corresponds to the radiation with linear polarization in the plane spanned by the vectors $\{\bs\be^c,\mathbf{n}\}$. Recall that in both cases the estimate $\be^c_\perp\ll1$ should be satisfied.

In order to fuller appreciate the dependence of the amplitude of stimulated radiation due to measurement \eqref{amplit_stim_spin_a} on the state of radiating particles, we consider the $N$-particle state of uncorrelated Dirac fermions described in Appendix \ref{Beams_App_Uncor}. We stipulate that, instead of \eqref{rho_N_uncorr}, a slightly more stringent condition holds
\begin{equation}
    \int d\bar{\spp}\rho^i_{s\bar{s}}(\spp,\bar{\spp}) \rho^j_{\bar{s}'s'}(\bar{\spp},\spp')\approx0,\quad i\neq j,
\end{equation}
i.e., the one-particle states $i$ and $j$ do not overlap because of their dependence on momenta and not on the spin projection. Then
\begin{equation}
    \rho^i_{\al\bar{\al}} (D_e)_{\bar{\al}\be} \rho^j_{\be\bar{\be}}\approx0,\quad i\neq j;\qquad \rho^i_{\al\bar{\al}} (\tilde{D}_e)_{\bar{\al}\be} \rho^j_{\be\bar{\be}}\approx0,\quad i\neq j.
\end{equation}
Substituting the state \eqref{rho_N_uncorr} into the definition \eqref{tilde_rho}, we arrive at
\begin{equation}\label{rho0_rho1}
    \rho^{(0)}_{\tilde{D}_e}=\prod_{i=1}^N\rho^i_{\al\bar{\al}}\tilde{D}_{\bar{\al}\al},\qquad (\rho^{(1)}_{\tilde{D}_e})_{\al\bar{\al}}=\rho^{(0)}_{\tilde{D}_e}\sum_{i=1}^N \frac{\rho^i_{\al\bar{\al}}}{\rho^i_{\be\bar{\be}}\tilde{D}_{\bar{\be}\be}}.
\end{equation}
In particular, for $N=1$ we have
\begin{equation}
    1-\rho^{(0)}_{\tilde{D}_e}=\rho^1_{\al\bar{\al}}D_{\bar{\al}\al},\qquad (\rho^{(1)}_{\tilde{D}_e})_{\al\bar{\al}}= \rho^1_{\al\bar{\al}}.
\end{equation}
Employing the representation \eqref{projectors}, the normalization factor can be written as
\begin{equation}\label{norm_fact}
    1-\rho^{(0)}_{\tilde{D}_e}=\int d\spp\rho^1(\spp,\spp)(1+(\bs\zeta(\spp),\bs\xi(\spp,\spp)))/2= (1+\lan(\bs\zeta\bs\xi)\ran)/2,
\end{equation}
where the angle brackets denote the average with respect to the state of the particle.

For large $N$, we assume that the polarization of the particle beam with respect to the spin projection onto the direction  $\bs\zeta$ per particle,
\begin{equation}
    S_\zeta=\frac1{N}\sum_{i=1}^N\lan(\bs\zeta\bs\xi^i)\ran,
\end{equation}
belongs to the interval $(-1,1-\e)$, where $\e$ does not depend on $N$. For positive $p_i$ the following inequality holds
\begin{equation}
    \prod_{i=1}^N p_i\leqslant\Big(\frac{1}{N}\sum_{i=1}^Np_i\Big)^N.
\end{equation}
Consequently,
\begin{equation}
    \rho^{(0)}_{\tilde{D}_e}\leqslant\Big(\frac{1+S_\zeta}{2}\Big)^N.
\end{equation}
For large $N$, the quantity on the right-hand side of this inequality tends exponentially to zero. Therefore, for large $N$ and $N\e$ the normalization factor \eqref{norm_fact} is close to unity. As far as $(\rho^{(1)}_{\tilde{D}_e})_{\al\bar{\al}}$ is concerned, it follows from \eqref{rho0_rho1} that $(\rho^{(1)}_{\tilde{D}_e})_{\al\bar{\al}}$ tends exponentially to zero for large $N$ since the sum over $i$ in expression \eqref{rho0_rho1} can grow not faster than a linear function of $N$. Thus, the amplitude of stimulated radiation due to measurement of spin projection \eqref{amplit_stim_spin_a} is a tiny quantity for large $N$. This property can be qualitatively explained as follows: The measurement of the spin projection of a single particle in the beam with large number of particles changes feebly the quantum state of the particle beam and so such a measurement is close to the quantum nondemolition measurement. In order to detect stimulated radiation due to measurement of spin projection of a Dirac particle, one has to carry out the measurements in a series of experiments where a small number of Dirac particles participate in every experiment.

\subsubsection{Measurement of momentum and coordinates}\label{Stim_Rad_Mom_Coord}

Consider stimulated radiation due to measurement of the momentum of a Dirac particle. In this case,
\begin{equation}\label{D_proj_momen}
    D^e_{\be\bar{\al}}=\frac{(2\pi)^3}{V}\de(\spp_1-\spp_r)\de(\spp'-\spp_r)\de_{s_1s'} d\spp_r,
\end{equation}
where $\spp_r$ is the measured momentum. In the general formula \eqref{amplit_stim_contin} for the amplitude of stimulated radiation due to measurement of a continuous observable, the first terms vanishes in virtue of the energy-momentum conservation law. In order to evaluate the second term, it is useful to employ formula \eqref{amplit_stim_first_t}, where $\rho^{(1)}_{\tilde{D}_e}$ should be replaced by $\rho^{(1)}$. Comparing \eqref{D_proj_momen} with \eqref{projectors}, we see that it should be set $\bs\zeta=0$ in formula \eqref{amplit_stim_first_t} and
\begin{equation}\label{D_proj_momen_1}
    D_e(\spp_1,\spp')=2\de(\spp_1-\spp_r)\de(\spp'-\spp_r) d\spp_r.
\end{equation}
The normalization factor is equal to
\begin{equation}
    \rho^{(1)}_{\al_1\bar{\al}_1}D^e_{\bar{\al}_1\al_1}=\rho^{(1)}(\spp_r,\spp_r)d\spp_r.
\end{equation}
Substituting \eqref{amplit_stim_first_t} into \eqref{amplit_stim_contin}, we have a simple expression
\begin{equation}\label{amplit_stim_mom_meas}
    A^{\bar{\ga}}=-\frac{em\rho^{(1)}(\spp,\spp')}{2\rho^{(1)}(\spp_r,\spp_r)} \frac{f^*_{(\la)i}(\spk)}{\sqrt{2Vk_0p_0p'_0}}  \frac{ \tilde{G}^i +\xi^j\tilde{Z}^{ji}}{k_0-q_0} \Big|_{\spp=\spp_r+\spk, \spp'=\spp_r},
\end{equation}
where recall that $q_0=p_0-p_0'$. Writing the density matrix in terms of the Wigner function,
\begin{equation}
    \rho^{(1)}(\spp,\spp')=\int\frac{d\spx}{(2\pi)^3} e^{-i\spk\spx}\rho^{(1)}(\spx,\spp_r+\spk/2),
\end{equation}
and comparing the resulting expression with the amplitude of radiation from the classical current \eqref{class_trans_ampl_Wig}, we see that the amplitude \eqref{amplit_stim_mom_meas} can be interpreted as the normalized amplitude of edge radiation from the fraction of particles with the momentum $\spp_c=\spp_r+\spk/2$, where the distribution of particles in the phase space is specified by the Wigner function. In particular, this radiation contains the contributions from the domains of momenta where the Wigner function is negative provided, of course, that the state of the measured particle has been prepared in such a way that these domains are present. In the small recoil limit, $|\spk|\ll|\spp_r|$, and for the electrically charged particles, i.e., when $|F_e|$ is not small, we obtain in the leading order
\begin{equation}
    A^{\bar{\ga}}=\frac{eF_e}{\sqrt{2Vk^3_0}} \frac{\rho^{(1)}(\spp_r+\spk,\spp_r)}{\rho^{(1)}(\spp_r,\spp_r)} \frac{ (\bs\be_r\mathbf{f}_{(\la)}^*(\spk))}{1-(\spn\bs\be_r)},
\end{equation}
where $\bs\be_r=\spp_r/p^0_r$.

Now consider stimulated radiation due to measurement of the coordinates of a Dirac particle. In relativistic quantum mechanics, it is impossible to localize the wave function of an electron in a space point. By measuring the coordinates at the instant of time $t_0$, we understand the measurement specified by the projector
\begin{equation}\label{detector_projector}
    D^e_{\al\bar{\al}}=\frac{(2\pi)^3}{V}\vf_s(\spp)\vf^*_{s'}(\spp')e^{i(p^0_{\bar{\al}} -p^0_{\al})t_0},
\end{equation}
and $\vf_s(\spp)$ is, for example, the Gaussian
\begin{equation}
    \vf_s(\spp)=\frac{\de_{ss_0}}{(2\pi\s^2)^{3/2}} e^{-\frac{(\spp-\spp_0)^2}{4\s^2} - i\spx_0\spp},
\end{equation}
where the dispersion of momenta $\s^2$ is large in comparison with $\spp_0^2$. Further, we will not use the explicit form of the functions $\vf_s(\spp)$. The expression for the amplitude of stimulated radiation due to measurement that we will obtain will be valid for arbitrary $\vf_s(\spp)$. Moreover, in order to bring the radiation amplitude to a simpler form allowing for a transparent physical interpretation, we consider the particular case when the initial state of Dirac particles is a pure state of a single particle
\begin{equation}\label{rho_ini}
    \rho^{(1)}_{\al\be}=\psi_\al\psi^*_\be.
\end{equation}
Recall that the wave function $\psi_\al$ is defined at the instant of time $t=0$. The state of the particle at the instant of time $t=t_{in}$ is obtained from $\psi_\al$ by means of the free evolution to the past from $t=0$ to $t=t_{in}$.

The second term on the first line in formula \eqref{amplit_stim_onepart} for the radiation amplitude is written as
\begin{equation}\label{phiV1psi}
    (\rho^{(1)}D_e)_{\al\bar{\al}}(V_{t_0,t_{in}})^{\bar{\ga}}_{\bar{\al}\al}=\lan\psi|\vf\ran \vf^*_{\bar{\al}}(V_{t_0,t_{in}})^{\bar{\ga}}_{\bar{\al}\al}\psi_\al=-ie\lan\psi|\vf\ran\int_{-\infty}^0 dx^0\int d\spx\bar{\vf}(x)\Ga^i\psi(x) \frac{f^*_{(\la)i}(\spk)e^{ik_\mu x^\mu}}{\sqrt{2Vk_0}},
\end{equation}
where it is assumed in the last equality that the time when the measurement is performed is $t_0=0$. Besides,
\begin{equation}
    \lan\psi|\vf\ran=\sum_s\int d\spp\psi^*_s(\spp)\vf_s(\spp),
\end{equation}
and $\vf(x)$ and $\psi(x)$ are the solutions to the free Dirac equation
\begin{equation}
    \vf(x)=\sum_s\int\frac{d\spp}{(2\pi)^{3/2}}\sqrt{\frac{m}{p_0}} u_s(\spp)e^{-ip_\mu x^\mu} \vf_s(\spp),\qquad \psi(x)=\sum_s\int\frac{d\spp}{(2\pi)^{3/2}}\sqrt{\frac{m}{p_0}} u_s(\spp)e^{-ip_\mu x^\mu} \psi_s(\spp).
\end{equation}
In fact, $\vf(x)$ and $\psi(x)$ are the solutions to the Dirac equation in the coordinate representation that turn into $\vf_s(\spp)$ and $\psi_s(\spp)$ at the instant of time $t=0$ in the momentum representation, respectively. The first term on the first line of formula \eqref{amplit_stim_onepart} for the radiation amplitude becomes
\begin{equation}\label{phiV2phi}
\begin{split}
    (D_e\rho^{(1)}D_e)_{\al\bar{\al}}(V_{t_{out},t_0})^{\bar{\ga}}_{\bar{\al}\al}=&\,|\lan\psi|\vf\ran|^2 \vf^*_{\bar{\al}}(V_{t_{out},t_0})^{\bar{\ga}}_{\bar{\al}\al}\vf_\al=\\
    =&\,-ie|\lan\psi|\vf\ran|^2 \int_0^\infty dx^0\int d\spx \bar{\vf}(x)\Ga^i\vf(x) \frac{f^*_{(\la)i}(\spk)e^{ik_\mu x^\mu}}{\sqrt{2Vk_0}}.
\end{split}
\end{equation}
The normalization factor is equal to
\begin{equation}
    \rho^{(1)}_{\al_1\bar{\al}_1}D^e_{\al_1\bar{\al}_1}=|\lan\psi|\vf\ran|^2.
\end{equation}
As a result, the amplitude of stimulated radiation from a single Dirac particle due to measurement of its state is given by
\begin{equation}\label{amplit_stim_position}
    A^{\bar{\ga}}=-ie\Big[\int_0^\infty dx^0\int d\spx \bar{\vf}(x)\Ga^i\vf(x) \frac{f^*_{(\la)i}(\spk)e^{ik_\mu x^\mu}}{\sqrt{2Vk_0}} + \int_{-\infty}^0 dx^0\int d\spx \frac{\bar{\vf}(x)\Ga^i\psi(x)}{\lan\vf|\psi\ran} \frac{f^*_{(\la)i}(\spk)e^{ik_\mu x^\mu}}{\sqrt{2Vk_0}}\Big].
\end{equation}
The first term in this expression describes the radiation from the classical current of a Dirac particle after the reduction of the wave function of this particle. The second term in this expression defines the normalized amplitude of photon radiation due to transition from the state $\psi$ to the state $\vf$ during the time interval $t\in(-\infty,0]$. At first sight, it may seem confusing that the amplitude of photon radiation depends on the evolution of the function $\vf(x)$ for $t<0$ that characterizes the detector making the measurement at the instant of time $t=t_0=0$. However, such a form of the second term in \eqref{amplit_stim_position} is expected. This contribution is the sum (the integral) of the photon radiation amplitudes for the instants of time $t\in(-\infty,0]$ caused by the transitions from the state $\psi(t)$ to such states that will be recorded by the detector after their free evolution up to the instant of time $t=t_0$.

\subsubsection{Spin measurement of a pair of entangled particles}\label{Spin_Meas_Entangl}

In this paper, we consider the simplest scenario of radiation from a pair of free evolving entangled particles where stimulated radiation stems from the measurement of spin of one of the particles, the wave functions of these particles do not overlap in space but are correlated by the spin projections. Suppose that the initial state of two Dirac fermions has the form \eqref{entangled2part_state} and the approximate orthogonality condition \eqref{approx_orthogon} is fulfilled. The correlated quantum numbers are the spin projections while the uncorrelated ones are the momenta of particles. As long as we consider the measurement of spin projection of the first particle only, the detector is localized near this particle and does not record the properties of the second particle. Therefore,
\begin{equation}
    D^e_{\bar{b}_1\bar{b}_2,b_1b_2}\overset{2}{\vf}_{a_1b_2}\approx0.
\end{equation}
Using this equality, we derive from \eqref{rho2entangl} that
\begin{equation}\label{rho2D}
    \rho_{\be\al|\bar{\al}\bar{\be}}D^e_{\bar{\be}\be}=\frac{|k|^2}{2} \overset{1}{\vf}{}^*_{\bar{b}_1\bar{b}_2} D^e_{\bar{b}_1\bar{b}_2,b_1b_2}\overset{1}{\vf}_{b_1b_2} f_{a_1b_1} f^*_{\bar{a}_1\bar{b}_1}\overset{2}{\vf}_{a_1a_2}\overset{2}{\vf}{}^*_{\bar{a}_1\bar{a}_2},
\end{equation}
where $\al=(a_1,a_2)$, $\bar{\al}=(\bar{a}_1,\bar{a}_2)$, $\be=(b_1,b_2)$, and $\bar{\be}=(\bar{b}_2,\bar{b}_2)$.

The amplitude of stimulated radiation due to measurement \eqref{amplit_stim} contains
\begin{equation}\label{rho1tildeD}
    (\rho^{(1)}_{\tilde{D}_e})_{\al\bar{\al}}=2(\rho_{\be\al|\bar{\al}\be}-\rho_{\be\al|\bar{\al}\bar{\be}}D^e_{\bar{\be}\be}).
\end{equation}
Substituting the reduced density matrix \eqref{dens_matr_reduc} and its projection \eqref{rho2D} into this expression, we come to
\begin{equation}
    (\rho^{(1)}_{\tilde{D}_e})_{\al\bar{\al}}=|k|^2  \big(f_{b_1a_1}f^*_{b_1\bar{a}_1} (\overset{2}{\vf}{}^*,\overset{2}{\vf})_{b_1} \overset{1}{\vf}_{a_1a_2}\overset{1}{\vf}{}^*_{\bar{a}_1\bar{a}_2}  +f_{a_1b_1}f^*_{\bar{a}_1b_1} \overset{1}{\vf}{}^*_{\bar{b}_1\bar{b}_2} \tilde{D}^e_{\bar{b}_1\bar{b}_2,b_1,b_2} \overset{1}{\vf}_{b_1b_2} \overset{2}{\vf}_{a_1a_2}\overset{2}{\vf}{}^*_{\bar{a}_1\bar{a}_2}\big).
\end{equation}
The dependence of the amplitude of stimulated radiation \eqref{amplit_stim} on the state of free Dirac particles is determined by
\begin{equation}
    (\tilde{D}_e\rho^{(1)}_{\tilde{D}_e}D_e)_{\al\bar{\al}}=|k|^2 f_{b_1c_1}f^*_{b_1\bar{c}_1}  (\overset{2}{\vf}{}^*,\overset{2}{\vf})_{b_1} \tilde{D}^e_{a_1a_2,c_1c_2}\overset{1}{\vf}_{c_1c_2}\overset{1}{\vf}{}^*_{\bar{c}_1\bar{c}_2} D^e_{\bar{c}_1\bar{c}_2,\bar{a}_1\bar{a}_2}.
\end{equation}
Furthermore, the amplitude includes the normalization factor
\begin{equation}
    1-\rho^{(0)}_{\tilde{D}_e}=1-\rho_{\be\al|\bar{\al}\bar{\be}}\tilde{D}^e_{\bar{\al}\al}\tilde{D}^e_{\bar{\be}\be}= 2\rho_{\be\al|\al\bar{\be}}D^e_{\bar{\be}\be}= |k|^2\overset{1}{\vf}{}^*_{\bar{b}_1\bar{b}_2} D^e_{\bar{b}_1\bar{b}_2,b_1b_2}\overset{1}{\vf}_{b_1b_2} f_{a_1b_1} f^*_{a_1\bar{b}_1} (\overset{2}{\vf}{}^*,\overset{2}{\vf})_{a_1},
\end{equation}
where the orthogonality condition \eqref{approx_orthogon} has been used.

Now we can write the explicit expression for the amplitude of stimulated radiation due to measurement of the spin projection of one of the entangled particles. Introduce the notation
\begin{equation}
    (a_1,a_2)=(s,\spp),\quad (\bar{a}_1,\bar{a}_2)=(s',\spp'),\quad (b_1,b_2)=(s_1,\spp_1),\quad (c_1,c_2)=(s_2,\spp_2),\quad (\bar{c}_1,\bar{c}_2)=(\bar{s}_2,\bar{\spp}_2),
\end{equation}
and suppose that the particles $1$ and $2$ are correlated with respect to the spin projection so that the total spin projection is equal to zero, i.e.,
\begin{equation}
    f_{b_1c_1}=\de_{s_1,-s_2}\chi_{s_1}.
\end{equation}
Then
\begin{equation}\label{tildeDrhoD}
\begin{split}
    (\tilde{D}_e\rho^{(1)}_{\tilde{D}_e}D_e)_{\al\bar{\al}}(V_{t_0,t_{in}})^{\bar{\ga}}_{\bar{\al}\al}=&\,-em |k|^2|\chi_{s_1}|^2 (\overset{2}{\vf}{}^*,\overset{2}{\vf})_{s_1}\times\\
    &\times\int\frac{d\spp_cd\spp_2d\bar{\spp}_2}{\sqrt{2Vk_0p_0p'_0}} \frac{\bar{u}_{\bar{\al}}\Ga^iu_\al f^*_{(\la)i}(\spk) }{k_0-q_0} \tilde{D}^e_{s,-s_1}(\spp,\spp_2) \overset{1}{\vf}_{-s_1}(\spp_2)\overset{1}{\vf}{}^*_{-s_1}(\bar{\spp}_2) D^e_{-s_1,s'}(\bar{\spp}_2,\spp'),
\end{split}
\end{equation}
where
\begin{equation}
    |k|^{-2}=\sum_s|\chi_{-s}|^2(\overset{1}{\vf}{}^*,\overset{1}{\vf})_s (\overset{2}{\vf}{}^*,\overset{2}{\vf})_{-s},
\end{equation}
and
\begin{equation}
    (\overset{1}{\vf}{}^*,\overset{1}{\vf})_s=\int d\spp \overset{1}{\vf}{}^*_s(\spp)\overset{1}{\vf}_s(\spp),\qquad (\overset{2}{\vf}{}^*,\overset{2}{\vf})_s=\int d\spp \overset{2}{\vf}{}^*_s(\spp)\overset{2}{\vf}_s(\spp).
\end{equation}
In acting on the state $\overset{1}{\vf}_s(\spp)$, the projectors $D_e$ and $\tilde{D}_e$ are reduced to the spin projectors
\begin{equation}
    D^e_{ss'}=\frac{1}{2}(1+(\bs\s\bs\zeta))_{ss'}\de(\spp-\spp'),\qquad \tilde{D}^e_{ss'}=\frac{1}{2}(1-(\bs\s\bs\zeta))_{ss'}\de(\spp-\spp').
\end{equation}
As a result, introducing the effective wave functions of the particle $1$,
\begin{equation}\label{effective_wf}
    \phi_s(\spp):=|k|\chi_{-s}(\overset{2}{\vf}{}^*,\overset{2}{\vf})^{1/2}_{-s}\overset{1}{\vf}_s(\spp),\qquad \sum_s\int d\spp|\phi_s(\spp)|^2=1,
\end{equation}
we obtain
\begin{equation}\label{tildeDrhoD1}
    (\tilde{D}_e\rho^{(1)}_{\tilde{D}_e}D_e)_{\al\bar{\al}}(V_{t_0,t_{in}})^{\bar{\ga}}_{\bar{\al}\al}= -\int\frac{d\spp_c em}{\sqrt{2Vk_0p_0p_0'}} \frac{\bar{u}_{\bar{\al}}\Ga^iu_\al f^*_{(\la)i}(\spk)}{k_0-q_0} (1-(\bs\s\bs\zeta(\spp)))_{ss_1} \phi_{s_1}(\spp)\phi^*_{s_1}(\spp') (1+(\bs\s\bs\zeta(\spp')))_{s_1s'},
\end{equation}
and
\begin{equation}\label{norm_fact_ent}
    1-\rho^{(0)}_{\tilde{D}_e}=\sum_s\int d\spp(1+(\bs\s\bs\zeta(\spp)))_{ss}|\phi_s(\spp)|^2.
\end{equation}
These expressions are to be substituted into \eqref{amplit_stim}. The properties of expression \eqref{tildeDrhoD1} was analyzed in detail in Sec. \ref{Stim_Rad_Spin_Measur}. As is seen from the representation of the amplitude of stimulated radiation \eqref{amplit_stim_spin} in terms of the Wigner function and the explicit expression for the effective wave function \eqref{effective_wf}, the radiation is created near the particle that undergoes the measurement. The second particle entangled with the first one by the spin projection does not produce radiation when the spin projection of the first particle is measured.

\subsection{Spontaneous radiation from a single particle}\label{Spont_Rad_Sing_Part}

Let us consider spontaneous radiation due to measurement. To simplify the expressions, we restrict ourselves to the special case of spontaneous radiation from a single free Dirac particle. For the radiation from a single free particle, general formula \eqref{probab_spont1} is reduced to
\begin{equation}\label{P_spont_1part}
    P(\hat{\Pi}_D\leftarrow \hat{\Pi}_{D_e})=\rho_{\be\bar{\al}} D_{\ga\bar{\ga}} \big(D_eV^\dag_{t_{out,t_0}}+V^\dag_{t_0,t_{in}}D_e\big)^{\ga}_{\bar{\al}\al} \big(V_{t_{out,t_0}}D_e +D_eV_{t_0,t_{in}}\big)^{\bar{\ga}}_{\al\be},
\end{equation}
where the photon radiation amplitudes are given in formulas \eqref{V_1} and \eqref{V_2}, and $\rho_{\be\bar{\al}}$ is the one-particle density matrix.

Let us first consider spontaneous radiation due to measurement of the Dirac particle spin. The projector specifying the measurement in this case has the form \eqref{zeta_zetap}. For brevity, we denote $\bs\zeta(\spp)\equiv\bs\zeta$ and $\bs\zeta(\spp')\equiv\bs\zeta'$. Then
\begin{equation}
\begin{split}
    \big(V_{t_{out,t_0}}D_e +D_eV_{t_0,t_{in}}\big)^{\bar{\ga}}_{\al\be}=&-\frac{em}{2} \frac{(2\pi)^3}{V}\de(\spk-\spq) \frac{ f^*_{(\la)i}(\spk)}{\sqrt{2V k_0p_0p_0'}} \frac{(1+(\bs\s\bs\zeta'))\bar{u}_\al\Ga^i u_\be -\bar{u}_\al\Ga^i u_\be (1+(\bs\s\bs\zeta)) }{k_0-q_0},\\
    \big(D_eV^\dag_{t_{out,t_0}}+V^\dag_{t_0,t_{in}}D_e\big)^{\ga}_{\bar{\al}\al}=&-\frac{em}{2} \frac{(2\pi)^3}{V}\de(\spk-\spq) \frac{ f_{(\la')i'}(\spk)}{\sqrt{2V k_0p_0p_0'}} \frac{\bar{u}_{\bar{\al}}\Ga^{i'} u_\al (1+(\bs\s\bs\zeta')) -(1+(\bs\s\bs\zeta))\bar{u}_{\bar{\al}} \Ga^{i'} u_\al }{k_0-q_0},
\end{split}
\end{equation}
where $\al=(s',\spp')$, $\bar{\al}=(\tilde{s},\spp)$, and $\be=(s,\spp)$. We suppose that the photons with definite polarization and momentum $\spk$ are measured at $t=t_{out}$. Hence, $\bar{\ga}=(\la,\spk)$ and $\ga=(\la',\spk)$. Taking into account the symmetry properties \cite{KRS2023},
\begin{equation}
    \tilde{G}^\mu(\spp,\spp')= \tilde{G}^\mu(\spp',\spp),\qquad \tilde{Z}^{i\mu}(\spp,\spp')=-\tilde{Z}^{i\mu}(\spp',\spp),
\end{equation}
we deduce that the probability density to detect a photon is proportional to
\begin{equation}
\begin{split}
    \frac{1}{32}\Sp\Big\{(1+(\bs\s\bs\xi))&\big[(\tilde{G}^{i'}-(\bs\s\bs\tau^{j'})\tilde{Z}^{j'i'})(\bs\s\bs\zeta') -(\bs\s\bs\zeta) (\tilde{G}^{i'}-(\bs\s\bs\tau^{j'})\tilde{Z}^{j'i'}) \big]\times\\
    &\times \big[(\bs\s\bs\zeta')(\tilde{G}^{i}+(\bs\s\bs\tau^{j})\tilde{Z}^{ji}) - (\tilde{G}^{i}+(\bs\s\bs\tau^{j})\tilde{Z}^{ji})(\bs\s\bs\zeta) \big] \Big\},
\end{split}
\end{equation}
where $\tilde{G}^i=\tilde{G}^i(\spp,\spp')$, $\tilde{Z}^{ji}=\tilde{Z}^{ji}(\spp,\spp')$, and $\bs\xi=\bs\xi(\spp,\spp)$. Employing the properties of the $\s$-matrices and evaluating the trace entering into the expression for the probability density to detect a photon, we arrive at
\begin{equation}\label{prob_spont_1part}
\begin{split}
    dP(\hat{\Pi}_D\leftarrow \hat{\Pi}_{D_e})=&\frac{e^2}{16} f_{(\la')i'} D_{\la'\la}f^*_{(\la)i} \frac{d\spk}{2k_0(2\pi)^3} \int  \frac{d\spp'm^2}{p_0p_0'} \frac{\rho(\spp,\spp)}{(k_0-q_0)^2} \Big\{\tilde{G}^{i'}\tilde{G}^i (\bs\zeta'-\bs\zeta)^2
    -2i[\bs\zeta,\bs\zeta']^{j}\tilde{G}^{(i'} \tilde{Z}^{ji)}-\\
    &-2(\bs\xi\bs\zeta) (\bs\zeta'-\bs\zeta)^j \tilde{G}^{[i'} \tilde{Z}^{ji]} -\big[(\bs\zeta+\bs\zeta')^2\de^{j'j} -2\zeta^{(j'}\zeta'^{j)} +2i\zeta^{[j'} [\bs\zeta+\bs\zeta',\bs\xi]^{j]}+\\
    &+i(\bs\zeta+\bs\zeta')^2\xi^{k}\e^{kj'j} \big] \tilde{Z}^{j'i'}\tilde{Z}^{ji} \Big\},
\end{split}
\end{equation}
where $\spp=\spp'+\spk$ and $D_{\la'\la}$ is the projector onto a certain spin state of the photon with fixed momentum $\spk$.

We see that only the diagonal of the density matrix in the momentum space of a Dirac particle contributes to the probability of spontaneous radiation. Employing the Wigner function to express the density matrix diagonal,
\begin{equation}
    \rho(\spp,\spp)=\int\frac{d\spx}{(2\pi)^3}\rho(\spx,\spp),
\end{equation}
we conclude that the probability \eqref{prob_spont_1part} can be interpreted as a sum of probabilities of photon radiation from the different points of the wave packet in the coordinate space. This radiation is incoherent in the sense that the probabilities and not the amplitudes are added up. In particular, a periodic modulation of the wave packet in the coordinate space does not lead to the Bragg maxima in the radiation. In coherent radiation (see, e.g., \eqref{amplit_stim_spin}), every point of the wave packet in the coordinate space contributes to the radiation amplitude with the factor $\exp(-i\spk\spx)$, i.e., the amplitudes are summed and the modulus of the resulting sum is squared. The presence of a periodic modulation of the wave packet in the coordinate space leads to an increase in the probability of coherent radiation with momenta corresponding to the Bragg maxima.

Let us consider the special case of general formula \eqref{prob_spont_1part} for $\bs\zeta\approx\bs\zeta'$. This happens, for example, when $\spp'\approx\spp$, i.e., the quantum recoil is small, or when $\bs\zeta$ is independent of $\spp$. Then
\begin{equation}\label{prob_spont_1part2}
    dP(\hat{\Pi}_D\leftarrow \hat{\Pi}_{D_e})=\frac{e^2}{4} D_{\la'\la} \frac{d\spk}{2k_0(2\pi)^3} \int  \frac{d\spp'm^2}{p_0p_0'} \frac{\rho(\spp,\spp)}{(k_0-q_0)^2} \Pi_{\la\la'}(\spp',\spk),
\end{equation}
where
\begin{equation}
    \Pi_{\la\la'}(\spp',\spk)=-\big(\de^{j'j} -\zeta^{j'}\zeta^{j} +i\zeta^{[j'} [\bs\zeta,\bs\xi]^{j]}
    +i\xi^{k}\e^{kj'j} \big) \tilde{Z}^{j'i'}\tilde{Z}^{ji}f_{(\la')i'}f^*_{(\la)i}.
\end{equation}
As expected, the probability of radiation due to measurement of the particle spin projection is determined by the magnetic moment of this particle. In the limit of a small recoil, we have
\begin{equation}
    dP(\hat{\Pi}_D\leftarrow \hat{\Pi}_{D_e})=\frac{e^2}{4} D_{\la'\la} \frac{d\spk}{2k^3_0(2\pi)^3} \int  \frac{d\spp'm^2}{p'^2_0} \frac{\rho(\spp,\spp)}{(1-(\spn\bs\be'))^2} \Pi_{\la\la'}(\spp',\spk),
\end{equation}
in the leading order. The expression for $\tilde{Z}^{ji}$ in the small recoil limit is presented in \eqref{small_recoil}.

Let us find the polarization of this radiation. The polarization properties of radiation stem from the form of $\Pi_{\la\la'}$. In the nonrelativistic case, we obtain
\begin{equation}
    \Pi_{\la\la'}=\frac{k_0^2}{2m^2} A(1+(\bs\s\mathbf{b}))_{\la\la'},
\end{equation}
in the linear polarization basis \eqref{polarization_vect}, where
\begin{equation}\label{A_b1_b2_b3_spon_nonrel}
    A=F_m^2(1+\zeta_3^2),\qquad b_2=\frac{2\zeta_3(\bs\zeta\bs\xi)}{1+\zeta_3^2},\qquad b_3+ib_1=\frac{(\zeta_1+i\zeta_2)^2}{1+\zeta_3^2}.
\end{equation}
Here the components of the vector $\bs\zeta$ are written in the basis $\{\mathbf{f}_1,\mathbf{f}_2,\spn\}$. The degree of polarization of this radiation is
\begin{equation}\label{polariz_degree_nonrel}
    \mathbf{b}^2=1-\frac{4\zeta_3^2(1-(\bs\zeta\bs\xi)^2)}{(1+\zeta_3^2)^2},
\end{equation}
whence it follows that spontaneous radiation due to spin measurement is in the pure polarization state provided $\bs\zeta\perp\spn$ or $\bs\zeta\parallel\bs\xi$. If $\bs\zeta\parallel\spn$, then the polarized part of the radiation is circularly polarized. For other values of $\bs\zeta$, there is an admixture of linear polarization with the polarization plane spanned by the vectors $\{\bs\zeta,\spn\}$. For $\bs\zeta\perp\spn$ or $\bs\zeta\perp\bs\xi$, the radiation is completely linearly polarized in the plane spanned by the vectors $\{\bs\zeta,\spn\}$. Notice that the polarization of spontaneous radiation coincides with the polarization of stimulated radiation \eqref{Stokes_vect_nonrel} for $\bs\zeta=-\bs\xi$.

In the ultrarelativistic limit, $|\spp|\gg m$, as in considering stimulated radiation, we assume that the wave packet of the Dirac particle moves approximately along the $z$ axis and is sufficiently narrow with respect to the transverse momentum components. We also consider the parameter region \eqref{ultrarel_param} where the main part of radiation is concentrated. Then the estimates and approximations \eqref{ultrarel_param} are valid with the replacements $\bs\be^c\rightarrow\bs\be'$ and $p_0^c\rightarrow p_0'$. In evaluating $\Pi_{\la\la'}$, it is convenient to choose the basis of linear polarization vectors as in \eqref{linear_polar_bas}, where $\bs\be^c\rightarrow\bs\be'$. As a result,
\begin{equation}
    \Pi_{\la\la'}=\frac{k_0^2}{2m^2\ga'^2}A(1+(\mathbf{b}\bs\s))_{\la\la'},
\end{equation}
where $A$, $b_1$, and $b_3$ have the same form as in expression \eqref{A_b1_b2_b3}, whereas the Stokes parameter $b_2$ differs only by a factor and is written as
\begin{equation}\label{b_2_spont_ultra}
    b_2=\frac{2}{A}F_m(F_m-(\be'_\perp\ga')^2a)(\zeta_3-\be'_\perp\ga'\zeta_1)(\bs\zeta\bs\xi).
\end{equation}
For $\be'_\perp\ga'\ll1$, the Stokes vector $\mathbf{b}$ turns into the Stokes vector of radiation in the nonrelativistic limit \eqref{A_b1_b2_b3_spon_nonrel}. The degree of radiation polarization reads
\begin{equation}\label{polariz_degree_ultra}
    \mathbf{b}^2=1-\frac{4 }{A^2}F_m^2(F_m-(\be'_\perp\ga')^2a)^2(\zeta_3-\be'_\perp\ga'\zeta_1)^2(1-(\bs\zeta\bs\xi)^2).
\end{equation}
The degree of polarization is equal to unity when either $\bs\zeta\parallel\bs\xi$, or $b_2=0$ for $(\bs\zeta\bs\xi)\neq0$. For $\bs\zeta=-\bs\xi$, the polarization properties of the polarized part of spontaneous radiation coincide with the polarization properties of stimulated radiation \eqref{A_b1_b2_b3}. The expressions for the constant $A$ and the Stokes vector $\mathbf{b}$ derived in the special cases in formulas \eqref{A_b1_b2_b3_pc1} and \eqref{A_b1_b2_b3_pc2} are modified by multiplying $b_2$ by $(\bs\zeta\bs\xi)$ in the case of spontaneous radiation.

To calculate the conditional probability it is also necessary to know $P(\hat{\Pi}_{D_e})$. It is obvious that for a one-particle initial state of Dirac particles
\begin{equation}
    P(\hat{\Pi}_{D_e})=\rho_{\al\bar{\al}}D^e_{\bar{\al}\al}=\frac{1}{2}\int d\spp\rho(\spp,\spp)(1+\bs\xi(\spp,\spp)\bs\zeta(\spp)).
\end{equation}
This expression should be substituted into general formula \eqref{cond_prob_gen}.

Now we consider spontaneous radiation from a single Dirac particle being measured at the instant of time $t_0=0$, where the measurement is specified by the projector \eqref{detector_projector}. The initial state of the particle is assumed to be pure and of the form \eqref{rho_ini}. Then it follows from \eqref{P_spont_1part} that
\begin{equation}\label{P_spon_gen}
\begin{split}
    P(\hat{\Pi}_D\leftarrow \hat{\Pi}_{D_e})=&\,|\lan\psi|\vf\ran|^2 D_{\ga\bar{\ga}} \Big[\vf^*_\be(V^\dag_{t_{out},t_0})^\ga_{\be\al} (V_{t_{out},t_0})^{\bar{\ga}}_{\al\bar{\be}}\vf_{\bar{\be}} + \vf^*_{\bar{\al}}(V_{t_{out},t_0})^{\bar{\ga}}_{\bar{\al}\al}\vf_\al \frac{\psi^*_{\bar{\be}}(V^\dag_{t_0,t_{in}})^\ga_{\bar{\be}\be}\vf_\be }{\lan\psi|\vf\ran}  +\\
    &+\vf^*_{\bar{\be}}(V^\dag_{t_{out},t_0})^\ga_{\bar{\be}\be}\vf_\be  \frac{\vf^*_{\bar{\al}}(V_{t_0,t_{in}})^{\bar{\ga}}_{\bar{\al}\al}\psi_\al}{\lan\vf|\psi\ran} + \frac{\psi^*_{\bar{\be}}(V^\dag_{t_0,t_{in}})^\ga_{\bar{\be}\be}\vf_\be }{\lan\psi|\vf\ran}\frac{\vf^*_{\bar{\al}}(V_{t_0,t_{in}})^{\bar{\ga}}_{\bar{\al}\al}\psi_\al}{\lan\vf|\psi\ran}\Big].
\end{split}
\end{equation}
The radiation amplitudes entering into this expression arose in the analysis of stimulated radiation and are given in formulas \eqref{phiV1psi} and \eqref{phiV2phi}. The only contribution that was not present in stimulated radiation is the first term in the square brackets in \eqref{P_spon_gen}. The amplitude entering into this term can be written as
\begin{equation}
    (V_{t_{out},t_0})^{\bar{\ga}}_{\al\bar{\be}}\vf_{\bar{\be}}=\sqrt{\frac{(2\pi)^3}{V}} \frac{f^{(\la)^*}_i(\spk)}{\sqrt{2Vk_0}} \frac{em}{\sqrt{p_0p_0'}}\frac{\bar{u}_{s'}(\spp') \Ga^i u_s(\spp)}{k_0-p_0+p_0'+i0}\vf_s(\spp) \Big|_{\spp=\spp'+\spk},
\end{equation}
where $\al=(s',\spp')$ and $\bar{\ga}=(\la,\spk)$. If the detector records photons with the definite momentum $\spk$, then
\begin{equation}\label{vfVVvf}
    D_{\ga\bar{\ga}} \vf^*_\be(V^\dag_{t_{out},t_0})^\ga_{\be\al} (V_{t_{out},t_0})^{\bar{\ga}}_{\al\bar{\be}}\vf_{\bar{\be}}= e^2 \frac{f_{(\la')i'}(\spk) D_{\la'\la}f^*_{(\la)i}(\spk)d\spk}{2k_0(2\pi)^3} \int \frac{d\spp' m^2}{p_0p_0'} \frac{\bar{u}_{\bar{s}}(\spp) \Ga^{i'} u_{s'}(\spp') \bar{u}_{s'}(\spp') \Ga^i u_s(\spp) }{(k_0-q_0)^2} D^e_{s\bar{s}}(\spp,\spp),
\end{equation}
where $D_{\la'\la}$ is the projector onto a certain polarization state of the photon with definite momentum $\spk$, $\spp=\spp'+\spk$, and the summation is understood over all the repeated indices. Further, we need to substitute the explicit expressions for the matrix elements of the transition current \eqref{buGammaU} into \eqref{vfVVvf}, to represent the projector $D_e$ in terms of the spin vector of the state on which it projects, and to evaluate the sums over the indices characterizing the spin polarization of the Dirac particle. We will not give here the resulting expression. We only note that this contribution contains the diagonal of the density matrix for the state $\vf_s(\spp)$ in the momentum space and, consequently, as discussed above, this contribution describes the incoherent radiation from the wave packet after its reduction. In the small recoil limit, for electrically charged Dirac particles, when $|F_e|$ is not small and
\begin{equation}
    \bar{u}_{s'}(\spp') \Ga^i u_s(\spp)\approx \de_{s's} F_e p^i/m,
\end{equation}
this contribution takes the form of incoherent edge radiation from a beam of particles. The distribution of particles with respect to momenta and coordinates in this beam is given by the Weyl symbol of the operator with the kernel $D^e(\spp,\spp')$, i.e., by the Wigner function for the state $\vf_s(\spp)$ averaged over spin and with the shifted momentum $\spp=\spp'+\spk$.

If the detector of Dirac particles described by the projector $D_e$ records the particles in the states characterized by a continuous parameter, for example, by a coordinate or a momentum, then only the last term in the square brackets remains in formula \eqref{P_spon_gen}. In particular, for
\begin{equation}
    \vf_s(\spp)=\sqrt{\frac{(2\pi)^3}{V}}\chi_s \de(\spp-\spp'),
\end{equation}
where $\spp'$ is the particle momentum measured by the detector and $\chi_s(\spp')$ defines the spin state onto which the detector projects the state of Dirac particles, the probability density can be cast into the form
\begin{equation}\label{prob_spon_mom}
    dP(\hat{\Pi}_D\leftarrow \hat{\Pi}_{D_e})=e^2 f_{(\la')i'}(\spk) D_{\la'\la}f^*_{(\la)i}(\spk)  \frac{m^2}{p_0p_0'} \frac{\bar{u}_{\bar{s}}(\spp) \Ga^{i'} u_{\bar{s}'}(\spp') \chi_{\bar{s}'}\chi^*_{s'} \bar{u}_{s'}(\spp') \Ga^i u_s(\spp) }{(k_0-q_0)^2} \rho_{s\bar{s}}(\spp,\spp) \frac{d\spp'd\spk}{2k_0(2\pi)^3},
\end{equation}
where the same notation as in formula \eqref{vfVVvf} has been used. In the small recoil limit and for electrically charged Dirac particles, i.e., when $|F_e|$ is not small, we obtain
\begin{equation}
    dP(\hat{\Pi}_D\leftarrow \hat{\Pi}_{D_e})=e^2F_e^2   \chi^*_{s'}\rho_{s'\bar{s}'}(\spp,\spp)\chi_{\bar{s}'} D_{\la'\la}\frac{(\bs\be'\mathbf{f}^*_{(\la)}) (\bs\be'\mathbf{f}_{(\la')})}{(1-(\spn\bs\be')^2)^2} \frac{d\spp'd\spk}{2k^3_0(2\pi)^3},
\end{equation}
and
\begin{equation}
    dP(\hat{\Pi}_{D_e})=\chi^*_{s'}\rho_{s'\bar{s}'}(\spp',\spp')\chi_{\bar{s}'} d\spp'.
\end{equation}
If the detector does not record the spin projection of the Dirac particle, the expression for the probability of radiation is the same as for the probability of radiation of a photon in incoherent edge radiation from a beam of particles with distribution in the phase space given by the Wigner function $\rho(\spx,\spp'+\spk)$, only the fraction of particles possessing the momentum $\spp'+\spk$ contributing to the probability. Recall that, throughout this paper, edge radiation from the particle beam is understood as edge radiation produced on the hypersurface $x^0=0$ in the spacetime.

Formulas \eqref{prob_spont_1part} and \eqref{prob_spon_mom} show that the energy of a radiated photon in spontaneous radiation due to measurement can be rather large. In fact, as follows from these formulas the whole momentum of the Dirac particle can be transferred to the photon and the probability of such a process is suppressed only by a certain power of the radiated photon energy. However, the upper bound of the radiated photon energy comes from the requirement that the measurement time $\tau$ must be much smaller than the formation time $t_f$ of edge radiation \eqref{formation_time}, \eqref{formation_time1}. The estimates given in Sec. \ref{Stim_Rad_Spin_Measur} imply that if the energy of radiated photon is so large that $t_f\lesssim \tau$, then the probability to record such a photon tends rapidly to zero.

\section{Conclusion}

Let us sum up the results. We have obtained the explicit expressions for the inclusive probability to record a chain of events that the Dirac fermion had been measured in a certain state and after that the photon was recorded. We have considered the two cases: the photons are present in the initial state (stimulated radiation) and they are absent in the initial state (spontaneous radiation). The general formulas for the probability to detect a photon \eqref{P_stim_rad_gen}, \eqref{probab_spont1} and the intensity of radiation \eqref{cond_prob_stim1}, \eqref{probab_spont1} have been derived. They depend explicitly on the measurement of the intermediate state of the Dirac particle. We have revealed that, in general, this probability is not zero even in the case of freely moving Dirac particles. The measurement causes radiation of photons and this radiation resembles edge or transition radiation. We have developed a systematic procedure taking qualitatively into account a finiteness of the measurement time and investigated its influence on the properties of radiation.

In the case of stimulated radiation, the conditional probability to record a photon \eqref{cond_prob_interfer} has been cast into the form corresponding to the interference of the free evolving photon complex amplitude and the radiation amplitude from the Dirac particle wave packet. The physical interpretation to the different terms in the amplitude of stimulated radiation has been given. In particular, it has been shown that this amplitude vanishes for the free evolving Dirac particles undergoing a quantum nondemolition measurement of their state. Such a measurement of the Dirac particle state does not change it and so the radiation due to measurement is absent. As far as spontaneous radiation is concerned, we have obtained the general formula for the probability to record a chain of events described above and provided the interpretation to the different contributions to this probability.

As examples, we have considered stimulated and spontaneous radiation due to measurement from freely moving Dirac particles. We have investigated the influence of measurement of the spin projection, of the momentum, and of the coordinate of the Dirac particle on the the probability to record a radiated photon. The explicit formulas for the probability to record such a chain of events have been obtained. The polarization properties of radiation due to spin measurement have been analyzed and the corresponding Stokes parameters have been found \eqref{Stokes_vect_nonrel}, \eqref{A_b1_b2_b3}, \eqref{A_b1_b2_b3_spon_nonrel}, \eqref{b_2_spont_ultra}. It has been established that the polarizations of stimulated and spontaneous radiations coincide when the spin projection measured by the detector, $\bs\zeta$, tends to the value opposite to the polarization of the initial state of the Dirac particle $\bs\xi$. Of course, the probability to find the Dirac particle with such a spin projection tends to zero. For an arbitrary direction of the measured spin, the Stokes parameters $b_1$ and $b_3$ characterizing the admixture of linear polarization are the same for both spontaneous and stimulated radiations, whereas the Stokes parameter $b_2$ characterizing a degree of circular polarization of spontaneous radiation differs by the factor $-(\bs\zeta\bs\xi)$ from the same parameter for stimulated radiation. The polarization state of stimulated radiation due to spin measurement is pure, while the polarization state of spontaneous radiation due to spin measurement is mixed in general. We have found the explicit expressions \eqref{polariz_degree_nonrel}, \eqref{polariz_degree_ultra} for the degree of polarization of spontaneous radiation. We have also investigated the dependence of stimulated radiation due to spin measurement on the form of the $N$-particle density matrix. We have considered the simplest quantum model of the particle beam consisting of $N$ uncorrelated Dirac particles. It turns out that stimulated radiation due to spin measurement is extremely small for large $N$ in virtue of the fact that the measurement of the spin projection appears to be close to a quantum nondemolition measurement in this case.

As for radiation due to measurement of particle momentum, it has the form of edge radiation from a beam of particles where only the fraction of particles with a certain momentum contribute to the probability of the chain of events. In the case of stimulated radiation, radiation due to measurement is a coherent superposition of amplitudes from the different points of the particle wave packet even in the case of a single radiating particle. In the case of spontaneous radiation from a single particle, this radiation is incoherent, i.e., the radiation probabilities from the different points of the particle wave packet are added up.

We have also considered radiation due to measurement of the Dirac particle in the state of a general form, in particular, in the state with given coordinates. In the case of radiation from a single particle prepared initially in the pure state, the amplitude of stimulated radiation \eqref{amplit_stim_position} is the sum of two terms: the amplitude of photon radiation caused by transition of the Dirac particle from the initial state to the measured one and the amplitude of radiation from the classical current of a collapsed wave function of the Dirac particle. Thus, in a certain sense, the external photons that stimulate radiation allow one to trace the dynamics and collapse of the Dirac particle wave function. In the case of spontaneous radiation from a single particle prepared in the pure state, the probability of the chain of events contains the incoherent contribution \eqref{vfVVvf} describing the radiation from the Dirac particle after measurement. It is the sum of radiation probabilities from the different points of the particle wave function exposed to measurement and collapse.

Another example of stimulated radiation due to measurement we have considered is the radiation from a pair of free Dirac particles entangled by the spin projections and well separated in space. We have derived the explicit expression for the probability of the chain of events (see formulas \eqref{tildeDrhoD1}, \eqref{norm_fact_ent}). In particular, we have shown that, in measuring the spin projection of one of the particles, the photons are emitted from the region where this particle is located and measured. As expected, the second particle entangled with the first one does not create radiation in this process.

It was hypothesized in \cite{KRS2023} that radiation due to measurement can be employed for production of high energy photons. We have shown that this is indeed the case: the whole momentum of a free Dirac particle can be transferred to a radiated photon. However, if the energy of a radiated photon is such that the radiation formation time $t_f$ defined in formula \eqref{formation_time} is much smaller than the measurement time $\tau$, then the probability to detect this photon is virtually zero. Another benefit from radiation due to measurement is that it allows one to clarify the fundamental questions regarding the behavior of the particle wave function undergoing a measurement. We have seen that stimulated radiation due to measurement strongly depends on details of the evolution of the wave function during this process and provides thereby a tool to recover this evolution.

\paragraph{Acknowledgments.}

The reported study was supported by the Ministry of Education and Science of the Russian Federation, the contract FSWM-2020-0033.

\appendix
\section{Bargmann-Fock representation}\label{Ap_BF_represnt}

The states in Bargmann-Fock representation \cite{BerezMSQ1.4},
\begin{equation}
    \Phi(a^*):=\lan a^*|\Phi\ran,\qquad\lan a^*|a\ran=e^{a^*a},
\end{equation}
satisfy the normalization condition
\begin{equation}
    \int D a^*Da e^{-a^*a} \Phi^*(a^*)\Phi(a^*)=1,
\end{equation}
where the completeness relation has been used
\begin{equation}
    \int D a^*Da e^{-a^*a}|a\ran\lan a^*|=\hat{1}.
\end{equation}
The functional integral is normalized in such a way that
\begin{equation}\label{func_int_G}
\begin{split}
    \int Da^*Da \exp\Big\{-\frac{1}{2}
    \left[
      \begin{array}{cc}
        a & a^* \\
      \end{array}
    \right]
    B
    \left[
      \begin{array}{c}
        a \\
        a^* \\
      \end{array}
    \right]
    +
     \left[
      \begin{array}{cc}
        a & a^* \\
      \end{array}
    \right] F
    \Big\}&=
    \exp\big\{\tfrac{1}{2}F^T B^{-1}F\big\}
    \Big(\det\left[
      \begin{array}{cc}
        A_{21} & A_{22} \\
        A_{11} & A_{12} \\
      \end{array}
    \right]\Big)^{-1/2},\\
    \int D a^* Da\exp\Big\{\frac{1}{2}
    \left[
      \begin{array}{cc}
        a & a^* \\
      \end{array}
    \right]
    B
    \left[
      \begin{array}{c}
        a \\
        a^* \\
      \end{array}
    \right]
    +
     \left[
      \begin{array}{cc}
        a & a^* \\
      \end{array}
    \right] F
    \Big\}&=
    \exp\big\{\tfrac{1}{2}F^T B^{-1}F\big\}
    (\det B)^{1/2},
\end{split}
\end{equation}
where the first equality is true for bosons and the second one is for fermions. The sources $F$ have the same Grassmann parity as $(a,a^*)$. Besides,
\begin{equation}\label{B_matrix}
    B:=\left[
      \begin{array}{cc}
        A_{11} & A_{12} \\
        A_{21} & A_{22} \\
      \end{array}
    \right].
\end{equation}
In particular,
\begin{equation}\label{Gaussian_int}
    \int Da^*Da e^{-a^*a +a^*\eta+\eta^*a}=e^{\eta^*\eta}.
\end{equation}
The trace of an operator is evaluated according to the formula
\begin{equation}\label{func_trace}
    \Sp\hat{A}=\int D a^*Da e^{-aa^*}\lan a^*|\hat{A}|a\ran.
\end{equation}
Expressions \eqref{Gaussian_int}, \eqref{func_trace} hold for both statistics.

In addition, for fermions we have
\begin{equation}
    \int Da^*Da e^{a^*a+a^*\eta-\eta^*a}=e^{\eta^*\eta},
\end{equation}
and
\begin{equation}\label{delta_func_ferm}
    \int D a^* e^{-a^*a +a^*\eta}=\de(a-\eta),\qquad \int Da e^{-a^*a +\eta^*a}=\de(a^*-\eta^*).
\end{equation}
Moreover, in this case,
\begin{equation}
    \Sp\hat{A}=\int Da^*Da e^{a^*a}\lan a^*|\hat{A}|a\ran=\int D a^*Da e^{-a^*a}\lan a^*|\hat{A}|-a\ran,
\end{equation}
and
\begin{equation}\label{func_int_ferm1}
    \int D a^*Da e^{a^*a} a^*_{\al_1}\cdots a^*_{\al_N}a_{\bar{\al}_N}\cdots a_{\bar{\al}_1}=\sum_{\s\in S_N}(-1)^{\e(\s)}\de_{\al_1\bar{\al}_{\s(1)}}\cdots \de_{\al_N\bar{\al}_{\s(N)}},
\end{equation}
where $S_N$ is the symmetric group of $N$ elements, and $\e(\s)$ is a parity of the permutation $\s$.

\section{Measurements and projectors}\label{Measurement_Proj_App}

Let us recall some standard definitions of quantum theory used in this paper \cite{Neumann1932,NielsenChuang2011,BraginskiKhalili1992}. Let $\hat{\xi}(t)$ be a self-adjoint operator in the Schr\"{o}dinger representation and $\hat{\Pi}_a(t)$ be its corresponding spectral projectors satisfying the following property
\begin{equation}
    \sum_a \hat{\Pi}_a(t)=\hat{1}.
\end{equation}
Suppose that the system is prepared at the instant of time $t=t_0$ in the state $\hat{R}$. Then the probability that in consecutive measurements of the quantity $\hat{\xi}(t)$ at the instants of time $t_1<\ldots<t_M$, where $t_1>t_0$, its values fall into the intervals corresponding to the spectral projectors $\hat{\Pi}_{a_1}(t_1)$, \ldots, $\hat{\Pi}_{a_M}(t_M)$ is equal to
\begin{equation}\label{probab_gen}
    P(\hat{\Pi}_{a_M}(t_M)\leftarrow\cdots\leftarrow \hat{\Pi}_{a_1}(t_1)):=\Sp(\hat{\Pi}_{a_M}\hat{U}_{t_M,t_{M-1}}\cdots \hat{U}_{t_2,t_1}\hat{\Pi}_{a_1}\hat{U}_{t_1,t_0}\hat{R}\hat{U}_{t_0,t_1}\hat{\Pi}_{a_1}\hat{U}_{t_1,t_2}\cdots \hat{U}_{t_{M-1},t_M}).
\end{equation}
This formula is obviously generalized to the case when the projectors are replaced by positive operator-valued measures \cite{NielsenChuang2011,BraginskiKhalili1992}.

For brevity, we will call the sequence of measurements described above a chain and denote it by
\begin{equation}\label{chain_meas}
    C_{t_M,t_1}:=\hat{\Pi}_{a_M}(t_M)\leftarrow\cdots\leftarrow\hat{\Pi}_{a_1}(t_1).
\end{equation}
The chains can be concatenated according to the rule: if $t'_1>t_M$, then the chain is defined
\begin{equation}
    C'_{t'_M,t'_1}\leftarrow C_{t_M,t_1}:=\hat{\Pi}_{a'_M}(t'_M)\leftarrow\cdots\leftarrow\hat{\Pi}_{a'_1}(t'_1)\leftarrow \hat{\Pi}_{a_M}(t_M)\leftarrow\cdots\leftarrow\hat{\Pi}_{a_1}(t_1).
\end{equation}
It is also convenient to associate with every chain $C_{t_M,t_1}$ a class of chains $[C_{t_M,t_1}]$ that contains all the chains of the form \eqref{chain_meas} with different $a_M$, \ldots, $a_1$.

Let $C_1\equiv C_{t_M,t_1}$, $C_2\equiv C_{t'_M,t'_1}$ and $t_1>t_0$, $t'_1>t_M$. Then the probability \eqref{probab_gen} satisfies the following properties:
\begin{subequations}
\begin{align}
    \sum_{C_2\in[C_2]}P(C_2\leftarrow C_1)&=P(C_1),\\
    \sum_{C_1\in[C_1]}P(C_1)&=1.
\end{align}
\end{subequations}
Then we can define the conditional probability as
\begin{equation}
    P(C_2| C_1):=P(C_2\leftarrow C_1)/P(C_1).
\end{equation}
This probability has the following properties
\begin{subequations}
\begin{align}
    \sum_{C_3\in[C_3]}P(C_3\leftarrow C_2|C_1)&=P(C_2|C_1),\\
    \sum_{C_2\in[C_2]}P(C_2|C_1)&=1,
\end{align}
\end{subequations}
provided $C_3\leftarrow C_2\leftarrow C_1$ is defined. It is obvious that
\begin{subequations}
\begin{align}
    P(C_2\leftarrow C_1)&=P(C_2|C_1) P(C_1),\\
    P(C_3\leftarrow C_2|C_1)&=P(C_3|C_2\leftarrow C_1) P(C_2|C_1).
\end{align}
\end{subequations}
The conditional averages are calculated using the conditional probabilities. For example, let $\hat{\Pi}_a(t_{out})$ be an operator-valued spectral measure associated with the self-adjoint operator $\hat{\xi}(t_{out})$, and $t_{out}>t_M$. Then the average value of the physical quantity $\hat{\xi}(t_{out})$ under the condition that in consecutive measurements of the quantity $\hat{\xi}(t)$ at the instants of time $t_1<\ldots<t_M$ its values fall into the intervals corresponding to the spectral projectors $\hat{\Pi}_{a_1}(t_1)$, \ldots, $\hat{\Pi}_{a_M}(t_M)$, is equal to
\begin{equation}
    \lan\hat{\xi}(t_{out})\ran:=\sum_a \xi_a(t_{out}) P(\hat{\Pi}_a(t_{out})|C_{t_M,t_1}),
\end{equation}
where $\xi_a(t_{out})$ are the eigenvalues of $\hat{\xi}(t_{out})$ and $t_1>t_0$.

In evaluating the inclusive probability to detect a particle, the following self-adjoint projection operators arise
\begin{equation}\label{Pi_D_defn}
    \hat{\tilde{\Pi}}_D:=:\exp(-\hat{a}^\dag D\hat{a}):,\qquad\hat{\Pi}_D:=(\hat{1}-\tilde{\Pi}_D)\otimes \hat{1}_\text{other species of particles},
\end{equation}
where $D=D^\dag$ is a projector in the Hilbert space of one-particle states, the creation and annihilation operators $(\hat{a}^\dag_{\bar{\al}},\hat{a}_\al)$ can be either bosonic or fermionic. Since
\begin{equation}
\begin{gathered}
    (D\hat{a})_\al \hat{\tilde{\Pi}}_D= \hat{\tilde{\Pi}}_D (\hat{a}^\dag D)_\al=0,\\
    [(\tilde{D}\hat{a})_\al,\hat{\tilde{\Pi}}_D]=[(\hat{a}^\dag \tilde{D})_\al,\hat{\tilde{\Pi}}_D]=0,\qquad\hat{\tilde{\Pi}}_D|0\ran=|0\ran,
\end{gathered}
\end{equation}
where $\tilde{D}:=1-D$, the projector $\hat{\tilde{\Pi}}_D$ maps to the subspace of the Fock space that contains a vacuum state and does not contain the particles in the states selected by the projector $D$. Hence, the projector $\hat{\Pi}_D$ singles out the states in the Fock space that contain at least one particle whose state vector belongs to the subspace of the one-particle Hilbert space distinguished by the projector $D$. In particular, for $D=1$ we obtain
\begin{equation}
    \hat{\Pi}_{D=1}=(\hat{1}-|0\ran\lan0|)\otimes \hat{1}_\text{other species of particles}.
\end{equation}

The following relations take place
\begin{equation}\label{Pi_comm_rels}
\begin{aligned}
    \,[\hat{a}_\al,\hat{\Pi}_D]&=\hat{\tilde{\Pi}}_D (D\hat{a})_\al,&\qquad [\hat{\Pi}_D,\hat{a}^\dag_\al]&= (\hat{a}^\dag D)_\al\hat{\tilde{\Pi}}_D,\\
    [\hat{a}_\al,\hat{\tilde{\Pi}}_D]&=-\hat{\tilde{\Pi}}_D (D\hat{a})_\al,&\qquad [\hat{\tilde{\Pi}}_D,\hat{a}^\dag_\al]&= -(\hat{a}^\dag D)_\al\hat{\tilde{\Pi}}_D,\\
    (D\hat{a})_\al\hat{\Pi}_D&=(D\hat{a})_\al,&\qquad \hat{\Pi}_D(\hat{a}^\dag D)_\al&=(\hat{a}^\dag D)_\al.
\end{aligned}
\end{equation}
The matrix elements of the product of projectors can be cast into the form
\begin{equation}
    \lan\bar{a}|\hat{\Pi}_{D_1}\cdots\hat{\Pi}_{D_M}|a\ran=\sum_{k=0}^M\sum_{\s\in S^k_M}(-1)^k e^{\bar{a}\tilde{D}_{\s(1)}\cdots \tilde{D}_{\s(k)}a},
\end{equation}
where $S^k_M$ is a set of $k$ different ordered elements taken from the set $\{1,\ldots,M\}$.

If the detector under consideration records $N$ identical particles in the different states distinguished by the projectors $D_i$, $i=\overline{1,N}$,
\begin{equation}\label{D_i_orthogon}
    D_iD_j=\de_{ij}D_j\;\;(\text{no summation}),
\end{equation}
then such a measurement is specified by the self-adjoint projector
\begin{equation}
    \hat{\Pi}_{D_1}\cdots\hat{\Pi}_{D_N}.
\end{equation}
It is evident that $[\hat{\Pi}_{D_i},\hat{\Pi}_{D_j}]=0$ due to the properties of the projectors \eqref{D_i_orthogon}. The matrix elements of such a projector are written as
\begin{equation}
    \lan\bar{a}|\hat{\Pi}_{D_1}\cdots\hat{\Pi}_{D_N}|a\ran=e^{\bar{a}a}\prod_{k=1}^N (1-e^{-\bar{a} D_{k}a}).
\end{equation}

If the quantum numbers of identical particles recorded by the detector can be identical, to formalize the measurement process it is useful to introduce the projector (see also \cite{Loudon1983,Agarwal2013})
\begin{equation}
    \hat{\tilde{\Pi}}^{(N)}_D=:e_{N-1}^xe^{-x}:,
\end{equation}
where
\begin{equation}
    e_{N}^x=\sum_{k=0}^N \frac{x^k}{k!},
\end{equation}
and $x=\hat{a}^\dag D\hat{a}$. The following properties are fulfilled for this projector
\begin{equation}
    \hat{\tilde{\Pi}}^{(N)}_D|0\ran=|0\ran,\qquad \hat{\tilde{\Pi}}^{(N)}_D (\hat{a}^\dag D)_{\al_1}\cdots (\hat{a}^\dag D)_{\al_M}|0\ran=(\hat{a}^\dag D)_{\al_1}\cdots (\hat{a}^\dag D)_{\al_M}|0\ran,
\end{equation}
with $M=\overline{1,N-1}$. If $M\geqslant N$, then
\begin{equation}
    \hat{\tilde{\Pi}}^{(N)}_D (\hat{a}^\dag D)_{\al_1}\cdots (\hat{a}^\dag D)_{\al_M}|0\ran=0.
\end{equation}
Therefore, the projector
\begin{equation}
    \hat{\Pi}^{(N)}_D:=(\hat{1}-\hat{\tilde{\Pi}}^{(N)}_D)\otimes \hat{1}_\text{other species of particles}
\end{equation}
singles out the states in the Fock space that contain at least $N$ identical particles in states from the subspace selected by the projector $D$.

It is obvious that
\begin{equation}
    \hat{\Pi}^{(N)}_D \underset{N\rightarrow\infty}{\rightarrow}0,\qquad
    \hat{\Pi}^{(N)}_D \underset{D\rightarrow0}{\rightarrow}:\frac{(\hat{a}^\dag D\hat{a})^N}{N!}:.
\end{equation}
The asymptotics $D\rightarrow0$ appears naturally when the projector $D$ distinguishes a small region in the phase space, for example, when $D$ selects the particle states with momenta in the domain $(\spp,\spp+d\spp)$.

\section{Density matrices and traces}\label{Traces_App}

It is convenient to bring the evaluation of traces of the operators appearing in the expressions for probabilities \eqref{P_stim}, \eqref{P_stim1}, \eqref{probab_spont_gen}, and \eqref{P_spont1} to the evaluation of the trace
\begin{equation}\label{trace_basic}
    \Sp(\hat{R}_e \hat{a}^\dag_{\bar{\al}_M}\cdots \hat{a}^\dag_{\bar{\al}_1}\hat{\tilde{\Pi}}_D \hat{a}_{\al_1}\cdots \hat{a}_{\al_M} ),
\end{equation}
where $\hat{R}_e$ is the density operator of Dirac fermions,
\begin{equation}
    \hat{R}_e=\sum_{N,M=0}^\infty \frac{1}{\sqrt{N!M!}}\rho_{\al_N\cdots\al_1|\bar{\al}_1\cdots\bar{\al}_M} \hat{a}^\dag_{\al_1}\cdots \hat{a}^\dag_{\al_N}|0\ran\lan0| \hat{a}_{\bar{\al}_M}\cdots \hat{a}_{\bar{\al}_1},
\end{equation}
with the matrix elements
\begin{equation}\label{dens_matrx_gen}
    \lan a^*|\hat{R}_e|a\ran=\sum_{N,M=0}^\infty\frac{1}{\sqrt{N!M!}}\rho_{\al_N\cdots\al_1|\bar{\al}_1\cdots\bar{\al}_M} a^*_{\al_1}\cdots a^*_{\al_N} a_{\bar{\al}_M}\cdots a_{\bar{\al}_1}.
\end{equation}
The normalization condition becomes
\begin{equation}
    \sum_{N=0}^\infty \rho_{\al_N\cdots\al_1|\al_1\cdots \al_N}=1.
\end{equation}
Let us introduce the notation
\begin{equation}\label{rho_N_M}
    \rho^{(N,M)}_{\al_1\cdots\al_M|\bar{\al}_M\cdots\bar{\al}_1}:= \rho_{\al_1\cdots\al_M\al_{M+1}\cdots\al_N| \al_N\cdots \al_{M+1}\bar{\al}_M\cdots\bar{\al}_1}.
\end{equation}
By definition,
\begin{equation}
    \rho^{(N,N)}_{\al_1\cdots\al_N|\bar{\al}_N\cdots\bar{\al}_1}:= \rho_{\al_1\cdots\al_N| \bar{\al}_N\cdots \bar{\al}_1}.
\end{equation}
The $M$-particle density matrix is defined as
\begin{equation}
    \rho^{(M)}_{\al_1\cdots\al_M|\bar{\al}_M\cdots\bar{\al}_1}:=\sum_{N=M}^\infty \frac{N!}{(N-M)!}\rho^{(N,M)}_{\al_1\cdots\al_M|\bar{\al}_M\cdots\bar{\al}_1}.
\end{equation}
It is evident that
\begin{equation}\label{rho_M}
    \rho^{(M)}_{\al_1\cdots\al_M|\bar{\al}_M\cdots\bar{\al}_1}=\Sp(\hat{R}_e \hat{a}^\dag_{\bar{\al}_M}\cdots \hat{a}^\dag_{\bar{\al}_1} \hat{a}_{\al_1}\cdots \hat{a}_{\al_M}).
\end{equation}
Let us also define the projected $M$-particle density matrix
\begin{equation}\label{tilde_rho}
    (\rho^{(M)}_{\tilde{D}})_{\al_1\cdots\al_M|\bar{\al}_M\cdots\bar{\al}_1}:= \sum_{N=M}^\infty \frac{N!}{(N-M)!} \rho_{\al_1\cdots\al_M\al_{M+1}\cdots\al_N|\bar{\al}_N\cdots\bar{\al}_{M+1}\bar{\al}_M\cdots\bar{\al}_1} \tilde{D}_{\bar{\al}_N\al_N}\cdots \tilde{D}_{\bar{\al}_{M+1}\al_{M+1}}.
\end{equation}
In particular,
\begin{equation}\label{rho_tilde_Dz}
    (\rho^{(M)}_{\tilde{D}})_{\al_1\cdots\al_M|\bar{\al}_M\cdots\bar{\al}_1}\big|_{\tilde{D}=0}= M!\rho_{\al_1\cdots\al_M|\bar{\al}_M\cdots\bar{\al}_1}.
\end{equation}
For the $N$-particle state, we have
\begin{equation}
    (\rho^{(N)}_{\tilde{D}})_{\al_1\cdots\al_N|\bar{\al}_N\cdots\bar{\al}_1}=\rho^{(N)}_{\al_1\cdots\al_N|\bar{\al}_N\cdots\bar{\al}_1}.
\end{equation}

Taking the trace \eqref{trace_basic} in the Bargmann-Fock representation \eqref{func_trace}, where
\begin{equation}
    \hat{a}^\dag_{\bar{\al}}\hat{R}_e \leftrightarrow a^*_{\bar{\al}}\lan a^*|\hat{R}_e|b\ran,\qquad
    \hat{R}_e \hat{a}_{\al} \leftrightarrow \lan a^*|\hat{R}_e|b\ran\frac{\overleftarrow{\de}}{\de b_\al},
\end{equation}
and employing formula \eqref{delta_func_ferm}, we come to
\begin{equation}
    \Sp(\hat{R}_e \hat{a}^\dag_{\bar{\al}_M}\cdots \hat{a}^\dag_{\bar{\al}_1}\hat{\tilde{\Pi}}_D \hat{a}_{\al_1}\cdots \hat{a}_{\al_M} )=\int D a^* Da e^{-aa^*} \frac{\de}{\de a^*_{\al_M}}\cdots \frac{\de}{\de a^*_{\al_1}} \lan a^*|R_e|b\ran \frac{\overleftarrow{\de}}{\de b_{\bar{\al}_1}}\cdots \frac{\overleftarrow{\de}}{\de b_{\bar{\al}_M}}\Big|_{b=\tilde{D}a}.
\end{equation}
Substituting the explicit expression \eqref{dens_matrx_gen} for the matrix elements of the density operator and calculating the functional integral using formula \eqref{func_int_ferm1}, we obtain
\begin{equation}\label{rho_MD}
    (\rho^{(M)}_{\tilde{D}})_{\al_1\cdots\al_M|\bar{\al}_M\cdots\bar{\al}_1}=\Sp(\hat{R}_e \hat{a}^\dag_{\bar{\al}_M}\cdots \hat{a}^\dag_{\bar{\al}_1}\hat{\tilde{\Pi}}_D \hat{a}_{\al_1}\cdots \hat{a}_{\al_M} ).
\end{equation}
It is clear that
\begin{equation}
    \Sp(\hat{R}_e \hat{a}^\dag_{\bar{\al}_M}\cdots \hat{a}^\dag_{\bar{\al}_1}\hat{\Pi}_D \hat{a}_{\al_1}\cdots \hat{a}_{\al_M} )=\rho^{(M)}_{\al_1\cdots\al_M|\bar{\al}_M\cdots\bar{\al}_1}-(\rho^{(M)}_{\tilde{D}})_{\al_1\cdots\al_M|\bar{\al}_M\cdots\bar{\al}_1}.
\end{equation}
For $D\rightarrow0$, we have
\begin{equation}
    \Sp(\hat{R}_e \hat{a}^\dag_{\bar{\al}_M}\cdots \hat{a}^\dag_{\bar{\al}_1}\hat{\Pi}_D \hat{a}_{\al_1}\cdots \hat{a}_{\al_M} )\underset{D\rightarrow0}{\rightarrow} \rho^{(M+1)}_{\al_1\cdots\al_M\al_{M+1}|\bar{\al}_{M+1}\bar{\al}_M\cdots\bar{\al}_1}D_{\bar{\al}_{M+1}\al_{M+1}}.
\end{equation}

In evaluating the inclusive probability to record a photon in stimulated radiation, the following traces over the fermionic degrees of freedom arise
\begin{equation}
\begin{split}
    \Sp(\hat{R}_e\hat{\Pi}_{D_e})&=1-\rho^{(0)}_{\tilde{D}_e},\\
    \Sp(\hat{R}_e\hat{\Pi}_{D_e} \hat{a}^\dag_{\bar{\al}} \hat{a}_\al \hat{\Pi}_{D_e})&=\rho^{(1)}_{\al\bar{\al}}-(\rho^{(1)}_{\tilde{D}_e})_{\al\bar{\al}} +(D_e\rho^{(1)}_{\tilde{D}_e}D_e)_{\al\bar{\al}},\\
    \Sp(\hat{R}_e\hat{\Pi}_{D_e} \hat{a}^\dag_{\bar{\al}} \hat{a}_\al)&=\rho^{(1)}_{\al\bar{\al}}-(\rho^{(1)}_{\tilde{D}_e}\tilde{D}_e)_{\al\bar{\al}},\\
    \Sp(\hat{R}_e \hat{a}^\dag_{\bar{\al}} \hat{a}_\al\hat{\Pi}_{D_e})&=\rho^{(1)}_{\al\bar{\al}}-(\tilde{D}_e\rho^{(1)}_{\tilde{D}_e})_{\al\bar{\al}}.
\end{split}
\end{equation}
In order to find these traces, one can use the commutation relations \eqref{Pi_comm_rels} and bring the expression under the trace sign into the form \eqref{rho_M} or \eqref{rho_MD}. For $D_e\rightarrow0$, we have the asymptotics
\begin{equation}
\begin{split}
    \Sp(\hat{R}_e\hat{\Pi}_{D_e})&\underset{D_e\rightarrow0}{\rightarrow}\rho^{(1)}_{\al\bar{\al}} D^e_{\bar{\al}\al},\\
    \Sp(\hat{R}_e\hat{\Pi}_{D_e} \hat{a}^\dag_{\bar{\al}} \hat{a}_\al \hat{\Pi}_{D_e})&\underset{D_e\rightarrow0}{\rightarrow}\rho^{(2)}_{\al\be|\bar{\be}\bar{\al}}D^e_{\bar{\be}\be},\\
    \Sp(\hat{R}_e\hat{\Pi}_{D_e} \hat{a}^\dag_{\bar{\al}} \hat{a}_\al)&\underset{D_e\rightarrow0}{\rightarrow}\rho^{(2)}_{\al\be|\bar{\be}\bar{\al}}D^e_{\bar{\be}\be} + (\rho^{(1)} D_e)_{\al\bar{\al}},\\
    \Sp(\hat{R}_e \hat{a}^\dag_{\bar{\al}} \hat{a}_\al\hat{\Pi}_{D_e})&\underset{D_e\rightarrow0}{\rightarrow}\rho^{(2)}_{\al\be|\bar{\be}\bar{\al}}D^e_{\bar{\be}\be} +(D_e\rho^{(1)} )_{\al\bar{\al}}.
\end{split}
\end{equation}

The traces over the photon degrees of freedom were calculated in \cite{KazSol2022}. Here we present the explicit expressions for these traces where the photon density matrix is written as
\begin{equation}\label{R_ph_coher}
    \hat{R}_{ph}=|d\ran\lan d^*| e^{-d^*d},
\end{equation}
and $d_\ga$ is the complex amplitude of the coherent state. Then
\begin{equation}
\begin{split}
    \Sp(\hat{R}_{ph}\hat{\Pi}_{D})&=1-e^{-d^*Dd},\\
    \Sp(\hat{R}_{ph}\hat{\Pi}_{D}\hat{c}^\dag_{\bar{\ga}})&=(d^*D)_{\bar{\ga}}+(1-e^{-d^*Dd})(d^*\tilde{D})_{\bar{\ga}},\\
    \Sp(\hat{R}_{ph}\hat{\Pi}_{D}\hat{c}_{\ga})&=(1-e^{-d^*Dd})d_{\ga},
\end{split}
\end{equation}
where $\hat{\Pi}_D$ has the form \eqref{Pi_D_defn} with the replacement of $(\hat{a}_\al,\hat{a}^\dag_\al)$ by $(\hat{c}_\ga,\hat{c}^\dag_\ga)$. For $D\rightarrow0$, we come to
\begin{equation}
\begin{split}
    \Sp(\hat{R}_{ph}\hat{\Pi}_{D})&\underset{D\rightarrow0}{\rightarrow} d^*Dd,\\
    \Sp(\hat{R}_{ph}\hat{\Pi}_{D}\hat{c}^\dag_{\bar{\ga}})&\underset{D\rightarrow0}{\rightarrow} (d^*D)_{\bar{\ga}}+(d^*Dd) d^*_{\bar{\ga}},\\
    \Sp(\hat{R}_{ph}\hat{\Pi}_{D}\hat{c}_{\ga})&\underset{D\rightarrow0}{\rightarrow} (d^*Dd) d_{\ga}.
\end{split}
\end{equation}

In calculating the inclusive probability to record a photon in spontaneous radiation, the following traces arise
\begin{equation}\label{traces_spontan}
\begin{split}
    \Sp(\hat{R}_e \hat{a}^\dag_{\bar{\al}}\hat{a}_\al\hat{\Pi}_{D_e} \hat{a}^\dag_{\bar{\be}}\hat{a}_\be\hat{\Pi}_{D_e})=&\,\de_{\al\bar{\be}}(\rho^{(1)}-\rho^{(1)}_{\tilde{D}_e})_{\be\bar{\al}} +D^e_{\al\bar{\be}}(D_e\rho^{(1)}_{\tilde{D}_e})_{\be\bar{\al}} -\rho^{(2)}_{\al\be|\bar{\al}\bar{\be}} +\tilde{D}^e_{\al\al_1}(\rho^{(2)}_{\tilde{D}_e})_{\al_1\be|\bar{\al}\bar{\be}}-\\
    &-\tilde{D}^e_{\al\al_1} D^e_{\be\be_1} (\rho^{(2)}_{\tilde{D}_e})_{\al_1\be_1|\bar{\al}\bar{\be}_1}D^e_{\bar{\be}_1\bar{\be}},\\
    \Sp(\hat{R}_e\hat{\Pi}_{D_e} \hat{a}^\dag_{\bar{\al}}\hat{a}_\al\hat{\Pi}_{D_e} \hat{a}^\dag_{\bar{\be}}\hat{a}_\be)=&\,\de_{\al\bar{\be}}(\rho^{(1)}-\rho^{(1)}_{\tilde{D}_e})_{\be\bar{\al}} +D^e_{\al\bar{\be}}(\rho^{(1)}_{\tilde{D}_e}D_e)_{\be\bar{\al}} -\rho^{(2)}_{\al\be|\bar{\al}\bar{\be}} +(\rho^{(2)}_{\tilde{D}_e})_{\al\be|\bar{\al}\bar{\be}_1}\tilde{D}^e_{\bar{\be}_1\bar{\be}}-\\
    &- D^e_{\al\al_1} (\rho^{(2)}_{\tilde{D}_e})_{\al_1\be|\bar{\al}_1\bar{\be}_1}D^e_{\bar{\al}_1\bar{\al}}\tilde{D}^e_{\bar{\be}_1\bar{\be}},\\
    \Sp(\hat{R}_e \hat{a}^\dag_{\bar{\al}}\hat{a}_\al\hat{\Pi}_{D_e} \hat{a}^\dag_{\bar{\be}}\hat{a}_\be)=&\,\de_{\al\bar{\be}}(\rho^{(1)}-\rho^{(1)}_{\tilde{D}_e})_{\be\bar{\al}} +D^e_{\al\bar{\be}}(\rho^{(1)}_{\tilde{D}_e})_{\be\bar{\al}} -\rho^{(2)}_{\al\be|\bar{\al}\bar{\be}} +\tilde{D}^e_{\al\al_1} (\rho^{(2)}_{\tilde{D}_e})_{\al_1\be|\bar{\al}\bar{\be}_1}\tilde{D}^e_{\bar{\be}_1\bar{\be}},\\
    \Sp(\hat{R}_e\hat{\Pi}_{D_e} \hat{a}^\dag_{\bar{\al}}\hat{a}_\al \hat{a}^\dag_{\bar{\be}}\hat{a}_\be\hat{\Pi}_{D_e})=&\,\de_{\al\bar{\be}}(\rho^{(1)} -\rho^{(1)}_{\tilde{D}_e} +D_e\rho^{(1)}_{\tilde{D}_e}D_e)_{\be\bar{\al}} -\rho^{(2)}_{\al\be|\bar{\al}\bar{\be}} +(\rho^{(2)}_{\tilde{D}_e})_{\al\be|\bar{\al}\bar{\be}} -(\pi_2\rho^{(2)}_{\tilde{D}_e}\pi_2)_{\al\be|\bar{\al}\bar{\be}},
\end{split}
\end{equation}
where
\begin{equation}
    (\pi_2)_{\al\be|\bar{\al}\bar{\be}}:=\de_{\al\bar{\al}}\de_{\be\bar{\be}}-\tilde{D}^e_{\al\bar{\al}}\tilde{D}^e_{\be\bar{\be}}.
\end{equation}
The second trace is obtained from the first one by taking the complex conjugate and reassigning the indices. For $D_e\rightarrow0$, we have
\begin{equation}\label{traces_spontan1}
\begin{split}
    \Sp(\hat{R}_e \hat{a}^\dag_{\bar{\al}}\hat{a}_\al\hat{\Pi}_{D_e} \hat{a}^\dag_{\bar{\be}}\hat{a}_\be\hat{\Pi}_e)\underset{D_e\rightarrow0}{\rightarrow}&\, \de_{\al\bar{\be}} \rho^{(2)}_{\be_1\be|\bar{\al}\bar{\al}_1}D^e_{\bar{\al}_1\be_1} -\rho^{(3)}_{\al_1\al\be|\bar{\al}\bar{\be}\bar{\al}_1}D^e_{\bar{\al}_1\al_1}-D^e_{\al\al_1}\rho^{(2)}_{\al_1\be|\bar{\al}\bar{\be}},\\
    \Sp(\hat{R}_e\hat{\Pi}_{D_e} \hat{a}^\dag_{\bar{\al}}\hat{a}_\al\hat{\Pi}_{D_e} \hat{a}^\dag_{\bar{\be}}\hat{a}_\be)\underset{D_e\rightarrow0}{\rightarrow}&\,\de_{\al\bar{\be}} \rho^{(2)}_{\be_1\be|\bar{\al}\bar{\al}_1}D^e_{\bar{\al}_1\be_1} -\rho^{(3)}_{\al_1\al\be|\bar{\al}\bar{\be}\bar{\al}_1}D^e_{\bar{\al}_1\al_1} -\rho^{(2)}_{\al\be|\bar{\al}\bar{\be}_1}D^e_{\bar{\be}_1\bar{\be}},\\
    \Sp(\hat{R}_e \hat{a}^\dag_{\bar{\al}}\hat{a}_\al\hat{\Pi}_{D_e} \hat{a}^\dag_{\bar{\be}}\hat{a}_\be)\underset{D_e\rightarrow0}{\rightarrow}&\,\de_{\al\bar{\be}} \rho^{(2)}_{\be_1\be|\bar{\al}\bar{\al}_1}D^e_{\bar{\al}_1\be_1} -\rho^{(3)}_{\al_1\al\be|\bar{\al}\bar{\be}\bar{\al}_1}D^e_{\bar{\al}_1\al_1}-D^e_{\al\al_1}\rho^{(2)}_{\al_1\be|\bar{\al}\bar{\be}} -\rho^{(2)}_{\al\be|\bar{\al}\bar{\be}_1}D^e_{\bar{\be}_1\bar{\be}}+\\
    &+ D^e_{\al\bar{\be}} \rho^{(1)}_{\be\bar{\al}},\\
    \Sp(\hat{R}_e\hat{\Pi}_{D_e} \hat{a}^\dag_{\bar{\al}}\hat{a}_\al \hat{a}^\dag_{\bar{\be}}\hat{a}_\be\hat{\Pi}_{D_e})\underset{D_e\rightarrow0}{\rightarrow}&\,\de_{\al\bar{\be}} \rho^{(2)}_{\be_1\be|\bar{\al}\bar{\al}_1}D^e_{\bar{\al}_1\be_1} -\rho^{(3)}_{\al_1\al\be|\bar{\al}\bar{\be}\bar{\al}_1}D^e_{\bar{\al}_1\al_1}.
\end{split}
\end{equation}
These traces appear in studying spontaneous radiation due to measurement in Sec. \ref{Spont_Radiat}.

\section{Beams}\label{Beams_App}
\subsection{Uncorrelated particles}\label{Beams_App_Uncor}

Let us consider the simplest quantum model of a beam of $N$ identical fermions. We suppose that
\begin{enumerate}
  \item The particles in the beam are not entangled
\begin{equation}\label{rho_N_uncorr}
    \rho_{\al_N\cdots\al_1|\bar{\al}_1\cdots\bar{\al}_N}=k\sum_{\s,\s'\in S_N}(-1)^{\e(\s)+\e(\s')} \prod_{i=1}^N\rho^i_{\al_{\s(i)}\bar{\al}_{\s'(i)}},
\end{equation}
where $k$ is a normalization constant, and $\rho^i_{\al\bar{\al}}$, $i=\overline{1,N}$, are one-particle density matrices;
  \item The one-particle states are orthogonal with good accuracy
\begin{equation}\label{orthogonality_dens_mat}
    \rho^i_{\al\bar{\al}}\rho^j_{\bar{\al}\be}\approx0,\quad i\neq j;\qquad \sum_\al\rho^i_{\al\al}=1.
\end{equation}
Then the normalization constant $k=1/N!$. The condition \eqref{orthogonality_dens_mat} is satisfied when, for example, the particle wave packets are separated in the coordinate space, or in the momentum space, or in any other space of quantum numbers;
  \item The moments of deviations of the one-particle density matrices from their average value are bounded for $N\rightarrow\infty$.
\end{enumerate}
A precise formulation of the latter requirement will be given below.

Using the $N$-particle density matrix \eqref{rho_N_uncorr}, we construct the reduced density matrices \eqref{rho_N_M}. It is not difficult to see that, for example,
\begin{equation}\label{rho_N_i_uncorr}
\begin{gathered}
    \rho^{(N,1)}_{\al\bar{\al}}=\frac{1}{N}\sum_{i=1}^N\rho^i_{\al\bar{\al}}=:\bar{\rho}_{\al\bar{\al}},\qquad \rho^{(N,2)}_{\al_1\al_2|\bar{\al}_2\bar{\al}_1}=\frac{1}{N(N-1)} \sideset{}{'}\sum_{i,j=1}^N (\rho^i_{\al_1\bar{\al}_1}\rho^j_{\al_2\bar{\al}_2} -\rho^i_{\al_1\bar{\al}_2}\rho^j_{\al_2\bar{\al}_1}),\\
    \rho^{(N,3)}_{\al_1\al_2\al_3|\bar{\al}_3\bar{\al}_2\bar{\al}_1}=\frac{1}{N(N-1)(N-2)} \sideset{}{'}\sum_{i,j,k=1}^N \sum_{\s\in S_3} (-1)^{\e(\s)} \rho^i_{\al_1\bar{\al}_{\s(1)}} \rho^j_{\al_2\bar{\al}_{\s(2)}} \rho^k_{\al_3\bar{\al}_{\s(3)}}.
\end{gathered}
\end{equation}
The prime at the sum sign means that the indices are not equal to each other for any pair of them.

Let us introduce the deviation of the one-particle density matrix from its average value over the ensemble,
\begin{equation}
    \de\rho^i_{\al\bar{\al}}:=\rho^i_{\al\bar{\al}}-\bar{\rho}_{\al\bar{\al}},
\end{equation}
and rewrite the reduced density matrices \eqref{rho_N_i_uncorr} in terms of this deviation. Given that
\begin{equation}
    \sum_{i=1}^N\de\rho^i_{\al\bar{\al}}=0,
\end{equation}
we come to
\begin{equation}\label{rho_N2_N3}
\begin{split}
    \rho^{(N,2)}_{\al_1\al_2|\bar{\al}_2\bar{\al}_1}=&\,\bar{\rho}_{\al_1[\bar{\al}_1} \bar{\rho}_{\al_2\bar{\al}_2]} -\frac{1}{N-1}\Si_{\al_1[\bar{\al}_1|\al_2\bar{\al}_2]},\\
    \rho^{(N,3)}_{\al_1\al_2\al_3|\bar{\al}_3\bar{\al}_2\bar{\al}_1}=&\sum_{\s\in S_3}(-1)^{\e(\s)}\Big\{ \bar{\rho}_{\al_1\bar{\al}_{\s(1)}} \bar{\rho}_{\al_2\bar{\al}_{\s(2)}} \bar{\rho}_{\al_3\bar{\al}_{\s(3)}}-\\
    &-\frac{1}{N-1}\big[\Si_{\al_1\bar{\al}_{\s(1)}|\al_2\bar{\al}_{\s(2)}} \bar{\rho}_{\al_3\bar{\al}_{\s(3)}} +\Si_{\al_2\bar{\al}_{\s(2)}|\al_3\bar{\al}_{\s(3)}} \bar{\rho}_{\al_1\bar{\al}_{\s(1)}} +\Si_{\al_1\bar{\al}_{\s(1)}|\al_3\bar{\al}_{\s(3)}} \bar{\rho}_{\al_2\bar{\al}_{\s(2)}} \big] +\\
    &+\frac{2}{(N-1)(N-2)}\De_{\al_1\bar{\al}_{\s(1)}|\al_2\bar{\al}_{\s(2)}|\al_3\bar{\al}_{\s(3)}} \Big\},
\end{split}
\end{equation}
where the square brackets at a pair of indices denote antisymmetrization without the factor $1/2$ and
\begin{equation}\label{averages_S_D}
    \Si_{\al_1\bar{\al}_1|\al_2\bar{\al}_2}:=\frac{1}{N}\sum_{i=1}^N \de\rho^i_{\al_1\bar{\al}_1} \de\rho^i_{\al_2\bar{\al}_2},\qquad
    \De_{\al_1\bar{\al}_1|\al_2\bar{\al}_2|\al_3\bar{\al}_3}:= \frac{1}{N}\sum_{i=1}^N \de\rho^i_{\al_1\bar{\al}_1} \de\rho^i_{\al_2\bar{\al}_2} \de\rho^i_{\al_3\bar{\al}_3}.
\end{equation}
If we now assume that $\bar{\rho}_{\al\bar{\al}}$ and the other averages \eqref{averages_S_D} are finite for $N\rightarrow\infty$ (the assumption 3 above), we can neglect the terms of order $1/N$ and $1/N^2$ in the expressions \eqref{rho_N2_N3} when $N$ is large. In this case,
\begin{equation}
    \rho^{(1)}_{\al\bar{\al}}=N\bar{\rho}_{\al\bar{\al}},\qquad \rho^{(2)}_{\al_1\al_2|\bar{\al}_2\bar{\al}_1}\approx \rho^{(1)}_{\al_1[\bar{\al}_1} \rho^{(1)}_{\al_1\bar{\al}_2]},\qquad  \rho^{(3)}_{\al_1\al_2\al_3|\bar{\al}_3\bar{\al}_2\bar{\al}_1}\approx \sum_{\s\in S_3}(-1)^{\e(\s)} \rho^{(1)}_{\al_1\bar{\al}_{\s(1)}} \rho^{(1)}_{\al_2\bar{\al}_{\s(2)}} \rho^{(1)}_{\al_3\bar{\al}_{\s(3)}}.
\end{equation}
The explicit expressions for these density matrices are used in Sec. \ref{Stim_Rad_Spin_Measur}.

\subsection{Entangled two-particle states}

As an example of entangled states, consider the two-particle state of fermions
\begin{equation}\label{entangled2part_state}
    k\sum_{\al,\be} f_{b_1a_1} \overset{2}{\vf}_{b_1b_2} \overset{1}{\vf}_{a_1a_2} \hat{a}^\dag_{b_1b_2}\hat{a}^\dag_{a_1a_2}|0\ran,
\end{equation}
where $k$ is the normalization constant, $\overset{1}{\vf}_\al$ characterizes the wave function of the particle $1$, and $\overset{2}{\vf}_\be$ characterizes the wave function of the particle $2$, the quantum numbers, $\al=(a_1,a_2)$, $\be=(b_1,b_2)$, are identically split into two groups: $a_1$, $b_1$ are correlated quantum numbers of the particles $1$ and $2$, and $a_2$, $b_2$ are uncorrelated quantum numbers. We will consider the special case of correlation (entanglement) when
\begin{equation}\label{f_corr_func}
    f_{b_1a_1}=\chi_{b_1}\de_{b_1,g(a_1)},
\end{equation}
where $g(a_1)$ and $\chi_{b_1}$ are some given functions.

The matrix elements of the density matrix for such a state are written as
\begin{equation}
\begin{split}
    \lan a^*|\hat{R}| a\ran=&\,\frac{|k|^2}{4} \sum_{\al,\bar{\al},\be,\bar{\be}} \big(f_{b_1a_1}\overset{2}{\vf}_{b_1b_2} \overset{1}{\vf}_{a_1a_2} -f_{a_1b_1} \overset{2}{\vf}_{a_1a_2} \overset{1}{\vf}_{b_1b_2}\big) \big(f^*_{\bar{b}_1\bar{a}_1} \overset{2}{\vf}{}^*_{\bar{b}_1\bar{b}_2} \overset{1}{\vf}{}^*_{\bar{a}_1\bar{a}_2} -f^*_{\bar{a}_1\bar{b}_1} \overset{2}{\vf}{}^*_{\bar{a}_1\bar{a}_2} \overset{1}{\vf}{}^*_{\bar{b}_1\bar{b}_2}\big)\times\\
    &\times\bar{a}_{b_1b_2}\bar{a}_{a_1a_2} a_{\bar{a}_1\bar{a}_2} a_{\bar{b}_1\bar{b}_2},
\end{split}
\end{equation}
where $\bar{\al}=(\bar{a}_1,\bar{a}_2)$, $\bar{\be}=(\bar{b}_1,\bar{b}_2)$. Whence we have
\begin{equation}\label{rho2entangl}
    \rho_{\be\al|\bar{\al}\bar{\be}}=\frac{|k|^2}{2} \big(f_{b_1a_1}\overset{2}{\vf}_{b_1b_2} \overset{1}{\vf}_{a_1a_2} -f_{a_1b_1} \overset{2}{\vf}_{a_1a_2} \overset{1}{\vf}_{b_1b_2}\big) \big(f^*_{\bar{b}_1\bar{a}_1} \overset{2}{\vf}{}^*_{\bar{b}_1\bar{b}_2} \overset{1}{\vf}{}^*_{\bar{a}_1\bar{a}_2} -f^*_{\bar{a}_1\bar{b}_1} \overset{2}{\vf}{}^*_{\bar{a}_1\bar{a}_2} \overset{1}{\vf}{}^*_{\bar{b}_1\bar{b}_2}\big).
\end{equation}
The reduced density matrix is
\begin{equation}
\begin{split}
    \rho_{\be\al|\bar{\al}\be}=&\,\frac{|k|^2}{2} \sum_{b_1} \big(f_{b_1a_1}f^*_{b_1\bar{a}_1} (\overset{2}{\vf}{}^*,\overset{2}{\vf})_{b_1} \overset{1}{\vf}_{a_1a_2}\overset{1}{\vf}{}^*_{a_1\bar{a}_2}  +f_{a_1b_1}f^*_{\bar{a}_1b_1}(\overset{1}{\vf}{}^*,\overset{1}{\vf})_{b_1} \overset{2}{\vf}_{a_1a_2}\overset{2}{\vf}{}^*_{a_1\bar{a}_2}  -\\
    &-f_{b_1a_1} f^*_{\bar{a}_1b_1} (\overset{1}{\vf}{}^*,\overset{2}{\vf})_{b_1} \overset{1}{\vf}_{a_1a_2}\overset{2}{\vf}{}^*_{\bar{a}_1\bar{a}_2} -f_{a_1b_1} f^*_{b_1\bar{a}_1} (\overset{2}{\vf}{}^*,\overset{1}{\vf})_{b_1} \overset{2}{\vf}_{a_1a_2}\overset{1}{\vf}{}^*_{\bar{a}_1\bar{a}_2}\big),
\end{split}
\end{equation}
where
\begin{equation}
    (\overset{1,2}{\vf}{}^*,\overset{1,2}{\vf})_{b_1}:=\sum_{b_2}\overset{1,2}{\vf}{}^*_{b_1b_2}\overset{1,2}{\vf}_{b_1b_2}.
\end{equation}

We suppose that the wave functions of the particles $1$ and $2$ are orthogonal with respect to the uncorrelated quantum numbers, i.e.,
\begin{equation}\label{approx_orthogon}
    \sum_{b_2} \overset{1}{\vf}{}^*_{a_1b_2}\overset{2}{\vf}_{b_1b_2}\approx0.
\end{equation}
For example, if the correlated quantum numbers are the spin variables and the uncorrelated quantum numbers are the momenta of particles, then the condition \eqref{approx_orthogon} is fulfilled when the particles $1$ and $2$ are spaced far enough apart. Then the reduced density matrix becomes
\begin{equation}\label{dens_matr_reduc}
    \rho_{\be\al|\bar{\al}\be}\approx \frac{|k|^2}{2} \sum_{b_1} \big(f_{b_1a_1}f^*_{b_1\bar{a}_1} (\overset{2}{\vf}{}^*,\overset{2}{\vf})_{b_1} \overset{1}{\vf}_{a_1a_2}\overset{1}{\vf}{}^*_{a_1\bar{a}_2}  +f_{a_1b_1}f^*_{\bar{a}_1b_1}(\overset{1}{\vf}{}^*,\overset{1}{\vf})_{b_1} \overset{2}{\vf}_{a_1a_2}\overset{2}{\vf}{}^*_{a_1\bar{a}_2}\big).
\end{equation}
It follows from the form of the correlation function \eqref{f_corr_func} that the reduced density matrix \eqref{dens_matr_reduc} is diagonal with respect to the correlated quantum numbers $a_1$, $\bar{a}_1$. The normalization constant in this case has the form
\begin{equation}
    |k|^{-2}\approx\sum_{a_1,b_1}|f_{b_1a_1}|^2(\overset{1}{\vf}{}^*,\overset{1}{\vf})_{a_1} (\overset{2}{\vf}{}^*,\overset{2}{\vf})_{b_1}.
\end{equation}
We use these expressions in Sec. \ref{Spin_Meas_Entangl}.



\begin{thebibliography}{999}


\bibitem{Neumann1932}
J. von Neumann,
\textsl{Mathematical Foundations of Quantum Mechanics}
(Princeton University Press, Princeton, 1932).

\bibitem{NielsenChuang2011}
M.~A. Nielsen, I.~L. Chuang,
\textsl{Quantum Computation and Quantum Information}
(Cambridge University Press, New York, 2011).

\bibitem{pra103}
P.~O. Kazinski, G.~Yu. Lazarenko,
Transition radiation from a Dirac-particle wave packet traversing a mirror,
Phys. Rev. A \textbf{103}, 012216 (2021).

\bibitem{KazSol2022}
P.~O. Kazinski, T.~V. Solovyev,
Coherent radiation of photons by particle wave packets,
Eur. Phys. J. C \textbf{82}, 790 (2022).

\bibitem{KazSol2023}
P.~O. Kazinski, T.~V. Solovyev,
Susceptibility of a single photon wave packet,
Phys. Rev. D \textbf{108}, 016004 (2023).

\bibitem{KRS2023}
P.~O. Kazinski, D.~I. Rubtsova, A.~A. Sokolov,
Inclusive probability to record an electron in elastic electromagnetic scattering by a spin one-half hadron wave packet,
Phys. Rev. D \textbf{108}, 096011 (2023).


\bibitem{PanGov18}
Y. Pan, A. Gover,
Spontaneous and stimulated radiative emission of modulated free electron quantum wavepackets -- semiclassical analysis,
J. Phys. Commun. \textbf{2}, 115026 (2018).

\bibitem{MarcuseI}
D. Marcuse,
Emission of radiation from a modulated electron beam,
J. Appl. Phys. \textbf{42}, 2255 (1971).

\bibitem{PMHK08}
J. Peatross, C. M\"{u}ller, K.~Z. Hatsagortsyan, C.~H. Keitel,
Photoemission of a single-electron wave packet in a strong laser field,
Phys. Rev. Lett. \textbf{100}, 153601 (2008).

\bibitem{Corson2011}
J.~P. Corson, J. Peatross, C. M\"{u}ller, K.~Z. Hatsagortsyan,
Scattering of intense laser radiation by a single-electron wave packet,
Phys. Rev. A \textbf{84}, 053831 (2011).

\bibitem{CorsPeat11}
J.~P. Corson, J. Peatross,
Quantum-electrodynamic treatment of photoemission by a single-electron wave packet,
Phys. Rev. A \textbf{84}, 053832 (2011).

\bibitem{WCCP16}
M. Ware, E. Cunningham, C. Coburn, J. Peatross,
Measured photoemission from electron wave packets in a strong laser field,
Opt. Lett. \textbf{41}, 689 (2016).

\bibitem{Remez19}
R. Remez \textit{et al}.,
Observing the quantum wave nature of free electrons through spontaneous emission,
Phys. Rev. Lett. \textbf{123}, 060401 (2019).

\bibitem{KdGdAR21}
O. Kfir, V. DiGiulio, F.~J.~G. de Abajo, C. Ropers,
Optical coherence transfer mediated by free electrons,
Sci. Adv. \textbf{7}, eabf6380 (2021).

\bibitem{Wong21}
L.~J. Wong \textit{et al}.,
Control of quantum electrodynamical processes by shaping electron wavepackets,
Nat. Commun. \textbf{12}, 1700 (2021).

\bibitem{GaribYang}
G.~M. Garibyan, C. Yang,
\textsl{X-Ray Transition Radiation}
(Armenian Academy of Science, Yerevan, 1983)
[in Russian].

\bibitem{BazylZhev}
V.~A. Bazylev, N.~K. Zhevago,
\textsl{Radiation from Fast Particles in a Medium and External Fields}
(Nauka, Moscow, 1987)
[in Russian].

\bibitem{Ginzburg}
V.~L. Ginzburg, V.~N. Tsytovich,
\textsl{Transition Radiation and Transition Scattering}
(Hilger, Bristol, 1990).

\bibitem{AkhShul}
A.~I. Akhiezer, N.~F. Shulga,
\textsl{High-Energy Electrodynamics in Matter}
(Gordon and Breach, New York, 1996).

\bibitem{Bord.1}
V.~G. Bagrov, G.~S. Bisnovatyi-Kogan, V.~A. Bordovitsyn, A.~V. Borisov, O.~F. Dorofeev, V.~Ya. Epp, V.~S. Gushchina, V.~C. Zhukovskii,
\textsl{Synchrotron Radiation Theory and its Development}
(World Scientific, Singapore, 1999).

\bibitem{GKSSchYu1}
G. Geloni, V. Kocharyan, E. Saldin, E. Schneidmiller, M. Yurkov,
Theory of edge radiation. Part I: Foundations and basic applications,
Nucl. Instrum. Methods Phys. Res., Sect. A \textbf{605}, 409 (2009).

\bibitem{GKSSchYu2}
G. Geloni, V. Kocharyan, E. Saldin, E. Schneidmiller, M. Yurkov,
Theory of edge radiation. Part II: Advanced applications,
Nucl. Instrum. Methods Phys. Res., Sect. A \textbf{607}, 470 (2009).

\bibitem{Feinberg1966}
E.~L. Feinberg,
High energy successive interactions,
J. Exptl. Theoret. Phys. (U.S.S.R.) \textbf{50}, 202 (1966)
[Sov. Phys. JETP \textbf{32}, 132 (1966)].

\bibitem{ShSSh2009}
N.~F. Shul'ga, V.~V. Syshchenko, S.~N. Shul'ga,
On the motion of high-energy wave packets and the transition radiation by ``half-bare'' electron,
Phys. Lett. A \textbf{374}, 331 (2009).

\bibitem{BBCLS}
C. Bal, E. Bravin, E. Chevallay, T. Lef\`{e}vre, G. Suberlucq,
OTR from non-relativistic electrons,
in Proceedings of the 6th European Workshop (DIPAC 2003), Mainz, Germany, May 5–7, 2003,
edited by A. Peters, V. Schaa
(JACoW, Geneva, Switzerland, 2013),
p. 95.

\bibitem{Pbook}
A.~P. Potylitsyn, M.~I. Ryazanov, M.~N. Strikhanov, A.~A. Tishchenko,
\textsl{Diffraction Radiation from Relativistic Particles}. Springer Tracts in Modern Physics, Vol. \textbf{239}
(Springer, Berlin, 2010).

\bibitem{SukhKubPot17}
L.~G. Sukhikh, G. Kube, A.~P. Potylitsyn,
Simulation of transition radiation based beam imaging from tilted targets,
Phys. Rev. Accel. Beams \textbf{20}, 032802 (2017).

\bibitem{SukhihDTHs}
L.~G. Sukhikh,
Measuring dimensions of high energy micrometer electron beams with the aid of transition radiation,
Doctor thesis, Tomsk Polytechnic University, 2018.

\bibitem{ArutAvet1972}
V.~M. Arutyunyan, G.~K. Avetisyan,
Emission by charged particles in the field of a plane electromagnetic wave in a medium,
Zh. Eksp. Teor. Fiz. \textbf{62}, 1639 (1972)
[Sov. Phys. JETP \textbf{35}, 854 (1972)].

\bibitem{AvetAvetPetr1978}
G.~K. Avetisyan, A.~K. Avetisyan, R.~G. Petrosyan,
Stimulated interaction of charged particles with electromagnetic radiation in a medium with nonstationary properties,
Zh. Eksp. Teor. Fiz. \textbf{75}, 382 (1978)
[Sov. Phys. JETP \textbf{48}, 192 (1978)].


\bibitem{ZarLomNer80}
L.~F. Zaretski{\u{\i}}, V.~V. Lomonosov, {\'{E}}.~A. Nersesov,
Stimulated emission from particles crossing the interface between two media,
Kvantovaya Elektronika \textbf{7}, 2367 (1980)
[Sov. J. Quantum Electron. \textbf{10}, 1379 (1980)].

\bibitem{IvanLomon81}
I.~G. Ivanter, V.~V. Lomonosov,
Polarization and angular distribution of Cherenkov and transition radiations in the field of a strong electromagnetic wave,
Zh. Eksp. Teor. Fiz. \textbf{80}, 879 (1981)
[Sov. Phys. JETP \textbf{53}, 447 (1981)].

\bibitem{Lihn1996}
H. Lihn, P. Kung, C. Settakorn, H. Wiedemann,
Observation of stimulated transition radiation,
Phys. Rev. Lett. \textbf{76}, 4163 (1996).

\bibitem{BBTq1}
V.~G. Bagrov, V.~V. Belov, A.~Yu. Trifonov,
Theory of spontaneous radiation by electrons in a trajectory-coherent approximation,
J. Phys. A: Math. Gen. \textbf{26}, 6431 (1993).

\bibitem{BBTq2}
V.~V. Belov, D.~V. Boltovskiy, A.~Yu. Trifonov,
Theory of spontaneous radiation by bosons in quasi-classical trajectory-coherent approximation,
Int. J. Mod. Phys. B \textbf{8}, 2503 (1994).

\bibitem{PanGov19}
Y. Pan, A. Gover,
Spontaneous and stimulated emissions of a preformed quantum free-electron wave function,
Phys. Rev. A \textbf{99}, 052107 (2019).

\bibitem{PanGov21}
Y. Pan, A. Gover,
Beyond Fermi's golden rule in free-electron quantum electrodynamics: acceleration/radiation correspondence,
New J. Phys. \textbf{23}, 063070 (2021).

\bibitem{AngDiPiaz18}
A. Angioi, A. Di Piazza,
Quantum limitation to the coherent emission of accelerated charges,
Phys. Rev. Lett. \textbf{121}, 010402 (2018).

\bibitem{AngDiPiaz19}
A. Angioi, A. Di Piazza,
On quantum electrodynamic processes in plasmas interacting with strong lasers,
Rend. Fis. Acc. Lincei \textbf{30}, 17 (2019).

\bibitem{KRAK21}
A. Karnieli, N. Rivera, A. Arie, I. Kaminer,
Superradiance and subradiance due to quantum interference of entangled free electrons,
Phys. Rev. Lett. \textbf{127}, 060403 (2021).


\bibitem{Talebi16}
N. Talebi,
Schr\"{o}dinger electrons interacting with optical gratings: quantum mechanical study of the inverse Smith-Purcell effect,
New J. Phys. \textbf{18}, 123006 (2016).

\bibitem{GovPan18}
A. Gover, Y. Pan,
Dimension-dependent stimulated radiative interaction of a single electron quantum wavepacket,
Phys. Lett. A \textbf{382}, 1550 (2018).

\bibitem{MarcuseII}
D. Marcuse,
Transition radiation from a modulated electron beam,
J. Appl. Phys. \textbf{42}, 2259 (1971).

\bibitem{IvKarl13prl}
I.~P. Ivanov, D.~V. Karlovets,
Detecting transition radiation from a magnetic moment,
Phys. Rev. Lett. \textbf{110}, 264801 (2013).

\bibitem{IvKarl13pra}
I.~P. Ivanov, D.~V. Karlovets,
Polarization radiation of vortex electrons with large orbital angular momentum,
Phys. Rev. A \textbf{88}, 043840 (2013).

\bibitem{KonPotPol}
A.~S. Konkov, A.~P. Potylitsyn, M.~S. Polonskaya,
Transition radiation of electrons with a nonzero orbital angular momentum,
JETP Lett. \textbf{100}, 421 (2014).

\bibitem{KarlZhev19}
D. Karlovets, A. Zhevlakov,
Intrinsic multipole moments of non-Gaussian wave packets,
Phys. Rev. A \textbf{99}, 022103 (2019).

\bibitem{PupKarl22}
A. Pupasov-Maksimov, D. Karlovets,
Passage of a vortex electron over an inclined grating,
Phys. Rev. A \textbf{105}, 042206 (2022).

\bibitem{BogShir1959}
N.~N. Bogoliubov, D.~V. Shirkov,
\textsl{Introduction to the theory of quantized fields}
(Wiley, New York, 1980).

\bibitem{Dyson1951}
F.~J. Dyson,
The Schr\"{o}dinger equation in quantum electrodynamics,
Phys. Rev. \textbf{83}, 1207 (1951).

\bibitem{Walhout1999}
T.~S. Walhout,
Similarity renormalization, Hamiltonian flow equations, and Dyson's intermediate representation,
Phys. Rev. D \textbf{59}, 065009 (1999).

\bibitem{Bargmann1961}
V. Bargmann,
On a Hilbert space of analytic functions and an associated integral transform,
Commun. Pure Appl. Mathem. \textbf{14}, 187 (1961).

\bibitem{BerezMSQ1.4}
F.~A. Berezin,
\textsl{Method of Second Quantization}
(Academic Press, New York, 1966).

\bibitem{Glaub2}
R.~J. Glauber,
Coherent and incoherent states of the radiation field,
Phys. Rev. \textbf{131}, 2766 (1963).

\bibitem{KlauSud}
J.~R. Klauder, E.~C.~G. Sudarshan,
\textsl{Fundamentals of Quantum Optics}
(Benjamin, New York, 1968).

\bibitem{ippccb}
P.~O. Kazinski,
Inclusive probability of particle creation on classical backgrounds,
Eur. Phys. J. C \textbf{80}, 734 (2020).

\bibitem{LandLifQED}
V.~B. Berestetskii, E.~M. Lifshitz, L.~P. Pitaevskii,
\textsl{Quantum Electrodynamics}
(Butterworth-Heinemann, Oxford, 1982).

\bibitem{BraginskiKhalili1992}
V.~B. Braginsky, F.~Ya. Khalili,
\textsl{Quantum Measurement}
(Cambridge University Press, Cambridge, 1992).

\bibitem{GSh}
I.~M. Gel'fand, G.~E. Shilov,
\textsl{Generalized Functions}, Vol. I: \textsl{Properties and Operations}
(Academic Press, New York, 1964).

\bibitem{Loudon1983}
R. Loudon,
\textsl{The Quantum Theory of Light}
(Clarendon Press, Oxford, 1983).

\bibitem{Agarwal2013}
G.~S. Agarwal,
\textsl{Quantum Optics}
(Cambridge University Press, Cambridge, 2013).

























%











%
%

%











%
%
%
%
%
%
%
%
%
%
%

%
%
%
%
%


%














%
%
%



%
%
%
%

%
%


%
%
%



%
%
%
%
%




%


%

%

%
%




%
%




%
%
%

%
%
%
%
%
%




%

%
%
%



%
%
%
%
%
%
%
%



%
%




\end{thebibliography}
\end{document}